\documentclass[epj]{svjour}
\usepackage{amsfonts}
\usepackage{amsmath}
\usepackage{graphicx}
\usepackage{subfigure}
\usepackage{dcolumn}
\usepackage{bm}
\usepackage{booktabs}
\usepackage{color}

\newcommand{\figcaption}{\def\@captype{figure}\caption}
\newcommand{\tabcaption}{\def\@captype{table}\caption}

\newcommand{\Rmnum}[1]{\expandafter\@slowromancap\romannumeral #1@}
\def\hlinewd#1{%
  \noalign{\ifnum0=`}\fi\hrule \@height #1 \futurelet
   \reserved@a\@xhline}
\makeatother
\def\dab{\int^{\alpha_{max}}_{\alpha_{min}}d\alpha\int^{\beta_{max}}_{\beta_{min}}d\beta}
\def\qq{\langle\bar qq\rangle}
\def\GGa{\langle GG\rangle}
\def\GGb{\langle g_s^2GG\rangle}

\def\qGqa{\langle\bar qGq\rangle}
\def\qGqb{\langle\bar qg_s\sigma\cdot Gq\rangle}
\def\FF(s){\left[(\alpha+\beta)m_c^2-\alpha\beta s\right]}
\def\HH(s){\left[m_c^2-\alpha(1-\alpha) s\right]}
\def\non{\\ \nonumber}
\allowdisplaybreaks[4]

\usepackage{ulem}

\usepackage{graphics}
\allowdisplaybreaks[4]

\begin{document}

\title{QCD sum rule study of hidden-charm pentaquarks}

\author{Hua-Xing Chen\inst{1} \and Er-Liang Cui\inst{1} \and Wei Chen\inst{2} \and Xiang Liu\inst{3,4} \and T. G. Steele\inst{2} \and Shi-Lin Zhu\inst{5,6,7}
}                     
\offprints{}          
\institute{
School of Physics and Beijing Key Laboratory of Advanced Nuclear Materials and Physics, Beihang University, Beijing 100191, China
\and
Department of Physics and Engineering Physics, University of Saskatchewan, Saskatoon, Saskatchewan, S7N 5E2, Canada
\and
School of Physical Science and Technology, Lanzhou University, Lanzhou 730000, China
\and
Research Center for Hadron and CSR Physics, Lanzhou University and Institute of Modern Physics of CAS, Lanzhou 730000, China
\and
School of Physics and State Key Laboratory of Nuclear Physics and Technology, Peking University, Beijing 100871, China
\and
Collaborative Innovation Center of Quantum Matter, Beijing 100871, China
\and
Center of High Energy Physics, Peking University, Beijing 100871, China}
\date{Received: date / Revised version: date}
%
\abstract{
We study the mass spectra of hidden-charm pentaquarks having spin $J = {1\over2}/{3\over2}/{5\over2}$ and quark contents $uud c \bar c$. We systematically construct all the relevant local hidden-charm pentaquark currents, and select some of them to perform QCD sum rule analyses. We find that the $P_c(4380)$ and $P_c(4450)$ can be identified as hidden-charm pentaquark states composed of an anti-charmed meson and a charmed baryon. Besides them, we also find a) the lowest-lying hidden-charm pentaquark state of $J^P = 1/2^-$ has the mass $4.33^{+0.17}_{-0.13}$ GeV, while the one of $J^P = 1/2^+$ is significantly higher, that is around $4.7-4.9$ GeV; b) the lowest-lying hidden-charm pentaquark state of $J^P = 3/2^-$ has the mass $4.37^{+0.18}_{-0.13}$ GeV, consistent with the $P_c(4380)$ of $J^P = 3/2^-$, while the one of $J^P = 3/2^+$ is also significantly higher, that is above $4.6$ GeV; c) the hidden-charm pentaquark state of $J^P = 5/2^-$ has a mass around $4.5-4.6$ GeV, slightly larger than the $P_c(4450)$ of $J^P = 5/2^+$.
\PACS{
      {12.39.Mk}{Glueball and nonstandard multi-quark/gluon states} \and
      {12.38.Lg}{Other nonperturbative calculations} \and
      {11.40.-q}{Currents and their properties}
     } 
} 
\maketitle

\section{Introduction}
\label{sec:intro}

Since the discovery of the $X(3872)$~\cite{Choi:2003ue}, dozens of charmonium-like $XYZ$ states have been reported~\cite{Agashe:2014kda}.
They are good candidates of tetraquark states, which are made of two quarks and two antiquarks.
In the past year, the LHCb Collaboration observed two hidden-charm pentaquark resonances, $P_c(4380)$ and
$P_c(4450)$, in the $J/\psi p$ invariant mass spectrum~\cite{lhcb}. They are good candidates of pentaquark states,
which are made of four quarks and one antiquark. They all belong to the exotic states, which can not be
explained in the traditional quark model, and are of particular importance to understand
the low energy behaviours of Quantum Chromodynamics (QCD)~\cite{Chen:2016qju}.

Before the LHCb's observation of the $P_c(4380)$ and $P_c(4450)$~\cite{lhcb}, there had been extensive theoretical discussions on the existence of hidden-charm pentaquark states~\cite{Wu:2010jy,Yang:2011wz,Xiao:2013yca,Uchino:2015uha,Karliner:2015ina,Garzon:2015zva,Wang:2011rga,Yuan:2012wz,Huang:2013mua}. Yet, this experimental observation triggered more
studies to explain their nature, such as meson-baryon molecules~\cite{Chen:2015loa,Roca:2015dva,He:2015cea,Huang:2015uda,Meissner:2015mza,Xiao:2015fia,Chen:2016heh}, diquark-diquark-antiquark pentaquarks~\cite{Maiani:2015vwa,Anisovich:2015cia,Ghosh:2015ksa,Wang:2015epa,Wang:2015ixb}, compact diquark-triquark pentaquarks~\cite{Lebed:2015tna,Zhu:2015bba}, the topological soliton model~\cite{Scoccola:2015nia}, genuine multiquark states other than molecules~\cite{Mironov:2015ica}, and kinematical effects related to the triangle singularity~\cite{Guo:2015umn,Liu:2015fea,Mikhasenko:2015vca}, etc.
Their productions and decay properties are also interesting, and have been studied in Refs.~\cite{Li:2015gta,Cheng:2015cca,Wang:2015jsa,Kubarovsky:2015aaa,Karliner:2015voa,Lu:2015fva,Hsiao:2015nna,Wang:2015qlf,Feijoo:2015kts,Wang:2016vxa,Schmidt:2016cmd}, etc. For more extensive discussions, see Refs.~\cite{Chen:2016qju,Burns:2015dwa,Oset:2016lyh}.

In this paper we use the method of QCD sum rule to study the mass spectrum of hidden-charm pentaquarks having spin $J = {1\over2}/{3\over2}/{5\over2}$ and quark contents $uud c \bar c$.
We shall investigate the possibility of interpreting the $P_c(4380)$ and $P_c(4450)$ as hidden-charm pentaquark states.
We shall also investigate other possible hidden-charm pentaquark states. The present discussion is an extension of our recent work shortly reported
in Ref.~\cite{Chen:2015moa}. In the calculation we need the resonance parameters of the $P_c(4380)$ and
$P_c(4450)$ measured in the LHCb experiment~\cite{lhcb}:
\begin{eqnarray}
\nonumber M_{P_c(4380)}&=&4380\pm 8\pm 29\, \mathrm{MeV} \, ,
\\ \nonumber \Gamma_{P_c(4380)}&=&205\pm18\pm86\, \mathrm{MeV} \, ,
\\ \nonumber M_{P_c(4450)}&=&4449.8\pm 1.7\pm 2.5 \,\mathrm{MeV} \, ,
\\ \nonumber \Gamma_{P_c(4450)}&=&39\pm5\pm19\, \mathrm{MeV} \, ,
\end{eqnarray}
as well as the preferred spin-parity assignments $(3/2^-, 5/2^+)$ for the $P_c(4380)$ and $P_c(4450)$, respectively~\cite{lhcb}.

This paper organized as follows. After this Introduction, we systematically construct the local pentaquark interpolating currents having spin $J = 1/2$ and quark contents $uud c \bar c$ in Sec.~\ref{sec:current}.
The currents having spin $J = 3/2$ and $J = 5/2$ are similarly constructed in Appendixes~\ref{app:spin32} and \ref{app:spin52}, respectively.
These currents are used to perform QCD sum rule analyses in Sec.~\ref{sec:sumrule}, and then are used to perform numerical analyses in Sec.~\ref{sec:numerical}. The results are discussed and summarized in Sec.~\ref{sec:summary}.
An example applying the Fierz transformation and the color rearrangement is given in Appendix~\ref{app:example}. This paper has a supplementary file ``OPE.nb'' containing all the spectral densities.

\section{Local Pentaquark Currents of Spin 1/2}
\label{sec:current}

In this section we systematically construct local pentaquark interpolating currents having spin $J = 1/2$ and quark contents $uud c \bar c$. There are two possible color configurations, $[\bar c_d c_d][\epsilon^{abc}q_a q_b q_c]$ and $[\bar c_d q_d][\epsilon^{abc}c_a q_b q_c]$, where $a \cdots d$
are color indices, $q$ represents the up, down or strange quark, and $c$ represents the charm quark. These two configurations, if they are local, can be related by the Fierz transformation as well as the color rearrangement:
\begin{eqnarray}
\delta^{de} \epsilon^{abc} &=& \delta^{da} \epsilon^{ebc} + \delta^{db} \epsilon^{aec} + \delta^{dc} \epsilon^{abe} \, .
\label{eq:cr}
\end{eqnarray}
With this relation, the color configurations $[\bar c^d c^d][\epsilon_{abc}q_1^a q_2^b q_3^c]$, $[\bar c^d q_1^d][\epsilon_{abc}c^a q_2^b q_3^c]$, $[\bar c^d q_2^d][\epsilon_{abc}c^a q_1^b q_3^c]$ and $[\bar c^d q_3^d][\epsilon_{abc}c^a q_1^b q_2^c]$ can actually be related, where $q_{1,2,3}$ represent three light quark fields. There are several formulae related to the Fierz transformation, some of which were given in Refs.~\cite{Chen:2006hy,Chen:2008qv}. In this paper we also need to use the product of two Dirac matrices with two symmetric Lorentz indices:
\begin{tiny}
\begin{eqnarray}
\nonumber && \left (
\begin{array}{l}
g_{\mu\nu} \mathbf{1} \otimes \mathbf{1}
\\ g_{\mu\nu} \gamma_\rho \otimes \gamma^\rho
\\ g_{\mu\nu} \sigma_{\rho\sigma} \otimes \sigma^{\rho\sigma}
\\ g_{\mu\nu} \gamma_{\rho} \gamma_5 \otimes \gamma^{\rho} \gamma_5
\\ g_{\mu\nu} \gamma_5 \otimes \gamma_5
\\ \gamma_\mu \otimes \gamma_\nu + (\mu \leftrightarrow \nu)
\\ \gamma_\mu \gamma_5 \otimes \gamma_\nu \gamma_5 + (\mu \leftrightarrow \nu)
\\ \sigma_{\mu \rho} \otimes \sigma_{\nu \rho} + (\mu \leftrightarrow \nu)
\end{array} \right )_{a b, c d}
\\ && ~~~~~~~~ = \left (
\begin{array}{llllllll}
{1\over4} & {1\over4} & {1\over8} & -{1\over4} & {1\over4} & 0 & 0 &
0
\\ 1 & -{1\over2} & 0 & -{1\over2} & -1 & 0 & 0 & 0
\\ 3 & 0 & -{1\over2} & 0 & 3 & 0 & 0 & 0
\\ -1 & -{1\over2} & 0 & -{1\over2} & 1 & 0 & 0 & 0
\\ {1\over4} & -{1\over4} & {1\over8} & {1\over4} & {1\over4} & 0 & 0 & 0
\\ {1\over2} & -{1\over2} & {1\over4} & -{1\over2} & -{1\over2} & {1\over2} & {1\over2} & -{1\over2}
\\ -{1\over2} & -{1\over2} & -{1\over4} & -{1\over2} & {1\over2} & {1\over2} & {1\over2} & {1\over2}
\\ {3\over2} & {1\over2} & -{1\over4} & -{1\over2} & {3\over2} & -1 & 1 & 0
\end{array} \right )
\left (
\begin{array}{l}
g_{\mu\nu} \mathbf{1} \otimes \mathbf{1}
\\ g_{\mu\nu} \gamma_\rho \otimes \gamma^\rho
\\ g_{\mu\nu} \sigma_{\rho\sigma} \otimes \sigma^{\rho\sigma}
\\ g_{\mu\nu} \gamma_{\rho} \gamma_5 \otimes \gamma^{\rho} \gamma_5
\\ g_{\mu\nu} \gamma_5 \otimes \gamma_5
\\ \gamma_\mu \otimes \gamma_\nu + (\mu \leftrightarrow \nu)
\\ \gamma_\mu \gamma_5 \otimes \gamma_\nu \gamma_5 + (\mu \leftrightarrow \nu)
\\ \sigma_{\mu \rho} \otimes \sigma_{\nu \rho} + (\mu \leftrightarrow \nu)
\end{array} \right )_{a d, b c}.
\end{eqnarray}
\end{tiny}
However, the detailed relations between $[\bar c_d c_d][\epsilon^{abc}q_a q_b q_c]$ and $[\bar c_d q_d][\epsilon^{abc}c_a q_b q_c]$ can not be easily obtained. We just show one example in Appendix~\ref{app:example}. We also systematically construct local pentaquark interpolating currents having spin $J = 3/2$ and $J = 5/2$,
and list the results in Appendixes~\ref{app:spin32} and \ref{app:spin52}, respectively.

\subsection{Currents of $[\bar c_d c_d][\epsilon^{abc}u_a d_b u_c]$}

In this subsection, we construct the currents of the color configuration $[\bar c_d c_d][\epsilon^{abc}u_a d_b u_c]$.
We only investigate the currents of the following type
\begin{eqnarray}
\eta &=& [\epsilon^{abc} (u^T_a C \Gamma_i d_b) \Gamma_j u_c] [\bar c_d \Gamma_k c_d] \, ,
\end{eqnarray}
where $\Gamma_{i,j,k}$ are various Dirac matrices. The currents of the other types $[\bar c_d \Gamma_k c_d][\epsilon^{abc} (u_a^T C \Gamma_i u_b) \Gamma_j d_c]$ and \\ $[\bar c_d \Gamma_k c_d][\epsilon^{abc} (d_a^T C \Gamma_i u_b) \Gamma_j u_c]$ can be
related to these currents by using the Fierz transformation. We can easily construct them based on the results of Ref.~\cite{Chen:2008qv} that there are three independent local light baryon fields of flavor-octet and having the positive parity:
\begin{eqnarray}
\nonumber N^N_1 &=& \epsilon_{abc} \epsilon^{ABD} \lambda_{DC}^N (q_A^{aT} C q_B^b) \gamma_5 q_C^c \, ,
\\ N^N_2 &=& \epsilon_{abc} \epsilon^{ABD} \lambda_{DC}^N (q_A^{aT} C \gamma_5 q_B^b) q_C^c \, ,
\label{eq:baryon}
\\ \nonumber N^N_{3\mu} &=& \epsilon_{abc} \epsilon^{ABD} \lambda_{DC}^N (q_A^{aT} C \gamma_\mu \gamma_5 q_B^b) \gamma_5 q_C^c \, ,
\end{eqnarray}
where $A \cdots D$ are flavor indices, and $q_{A}=(u\, ,d\, ,s)$ is the light quark field of flavor-triplet. Together with light baryon fields having the negative parity, $\gamma_5 N^N_{1,2}$ and $\gamma_5 N^N_{3\mu}$, and the charmonium fields:
\begin{eqnarray}
&\bar c_d c_d \, [0^+] \, , \bar c_d \gamma_5 c_d \, [0^-] \, ,&
\\ \nonumber &\bar c_d \gamma_\mu c_d \, [1^-] \, , \bar c_d \gamma_\mu \gamma_5 c_d \, [1^+] \, , \bar c_d \sigma_{\mu\nu} c_d \, [1^\pm] \, ,&
\end{eqnarray}
we can construct the following currents having $J^P=1/2^+$ and quark contents $uud c \bar c$:
\begin{eqnarray}
\nonumber \eta_1 &=& [\epsilon^{abc} (u^T_a C d_b) \gamma_5 u_c] [\bar c_d c_d] \, ,
\\ \nonumber \eta_2 &=& [\epsilon^{abc} (u^T_a C d_b) u_c] [\bar c_d \gamma_5 c_d] \, ,
\\ \nonumber \eta_3 &=& [\epsilon^{abc} (u^T_a C \gamma_5 d_b) u_c] [\bar c_d c_d] \, ,
\\ \nonumber \eta_4 &=& [\epsilon^{abc} (u^T_a C \gamma_5 d_b) \gamma_5 u_c] [\bar c_d \gamma_5 c_d] \, ,
\\ \nonumber \eta_5 &=& [\epsilon^{abc} (u^T_a C d_b) \gamma_\mu \gamma_5 u_c] [\bar c_d \gamma_\mu c_d] \, ,
\\ \nonumber \eta_6 &=& [\epsilon^{abc} (u^T_a C d_b) \gamma_\mu u_c] [\bar c_d \gamma_\mu \gamma_5 c_d] \, ,
\\ \nonumber \eta_7 &=& [\epsilon^{abc} (u^T_a C \gamma_5 d_b) \gamma_\mu u_c] [\bar c_d \gamma_\mu c_d] \, ,
\\ \eta_8 &=& [\epsilon^{abc} (u^T_a C \gamma_5 d_b) \gamma_\mu \gamma_5 u_c] [\bar c_d \gamma_\mu \gamma_5 c_d] \, ,
\\ \nonumber \eta_9 &=& [\epsilon^{abc} (u^T_a C d_b) \sigma_{\mu\nu} \gamma_5 u_c] [\bar c_d \sigma_{\mu\nu} c_d] \, ,
\\ \nonumber \eta_{10} &=& [\epsilon^{abc} (u^T_a C d_b) \sigma_{\mu\nu} u_c] [\bar c_d \sigma_{\mu\nu} \gamma_5 c_d] \, ,
\\ \nonumber \eta_{11} &=& [\epsilon^{abc} (u^T_a C \gamma_5 d_b) \sigma_{\mu\nu} u_c] [\bar c_d \sigma_{\mu\nu} c_d] \, ,
\\ \nonumber \eta_{12} &=& [\epsilon^{abc} (u^T_a C \gamma_5 d_b) \sigma_{\mu\nu} \gamma_5 u_c] [\bar c_d \sigma_{\mu\nu} \gamma_5 c_d] \, .
\\ \nonumber \eta_{13} &=& [\epsilon^{abc} (u^T_a C \gamma_\mu \gamma_5 d_b) u_c] [\bar c_d \gamma_\mu c_d] \, ,
\\ \nonumber \eta_{14} &=& [\epsilon^{abc} (u^T_a C \gamma_\mu \gamma_5 d_b) \gamma_5 u_c] [\bar c_d \gamma_\mu \gamma_5 c_d] \, ,
\\ \nonumber \eta_{15} &=& [\epsilon^{abc} (u^T_a C \gamma_\mu \gamma_5 d_b) \gamma_\nu u_c] [\bar c_d \sigma_{\mu\nu} c_d] \, ,
\\ \nonumber \eta_{16} &=& [\epsilon^{abc} (u^T_a C \gamma_\mu \gamma_5 d_b) \gamma_\nu \gamma_5 u_c] [\bar c_d \sigma_{\mu\nu} \gamma_5 c_d] \, .   \label{def:scalaretacurrent}
\end{eqnarray}
We can verify the following relations
\begin{eqnarray}
\nonumber \eta_{9} &=& \eta_{10} \, ,
\\ \eta_{11} &=& \eta_{12} \, ,
\\ \nonumber \eta_{11} &=& \eta_{9} + 2i \eta_{15} - 2i \eta_{16} \, .
\end{eqnarray}
Hence, only 13 currents are independent in Eq. \eqref{def:scalaretacurrent}. All of them have $J^P = 1/2^+$, while their partner currents, $\gamma_5 \eta_i$, have $J^P = 1/2^-$.
We shall not use all of them to perform QCD sum rule analyses, but select those containing pseudoscalar ($\bar c_d \gamma_5 c_d$) and vector ($\bar c_d \gamma_\mu c_d$) components
\begin{eqnarray}
\eta_2 - \eta_4 &=& [\epsilon^{abc} (u^T_a C d_b) u_c] [\bar c_d \gamma_5 c_d]
\label{def:eta24}
\\ \nonumber && ~~~~~~~~~~ - [\epsilon^{abc} (u^T_a C \gamma_5 d_b) \gamma_5 u_c] [\bar c_d \gamma_5 c_d] \, ,
\\ \eta_5 - \eta_7 &=& [\epsilon^{abc} (u^T_a C d_b) \gamma_\mu \gamma_5 u_c] [\bar c_d \gamma_\mu c_d]
\label{def:eta57}
\\ \nonumber && ~~~~~~~~~~ - [\epsilon^{abc} (u^T_a C \gamma_5 d_b) \gamma_\mu u_c] [\bar c_d \gamma_\mu c_d] \, ,
\\ \eta_{13} &=& [\epsilon^{abc} (u^T_a C \gamma_\mu \gamma_5 d_b) u_c] [\bar c_d \gamma_\mu c_d] \, .
\label{def:eta13}
\end{eqnarray}
Their internal structures are quite simple, suggesting that they well couple to the $[p \eta_c]$, $[p J/\psi]$, and $[N^* J/\psi]$ channels, respectively.
Especially, $\eta_2 - \eta_4$ and $\eta_5 - \eta_7$ both contain the ``Ioffe's baryon current'', which couples strongly to the lowest-lying nucleon state~\cite{Belyaev:1982sa,Espriu:1983hu}.

\subsection{Currents of $[\bar c_d u_d][\epsilon^{abc} u_a d_b c_c]$}

In this subsection, we construct the currents of the type $[\bar c_d \Gamma_k u_d][\epsilon^{abc} (u_a^T C \Gamma_i d_b) \Gamma_j c_c]$.
The currents of the other types $[\bar c_d \Gamma_k u_d][\epsilon^{abc} (u_a^T C \Gamma_i c_b) \Gamma_j d_c]$ and $[\bar c_d \Gamma_k u_d][\epsilon^{abc} (c_a^T C \Gamma_i d_b) \Gamma_j u_c]$, etc. can be
related to these currents by using the Fierz transformation.
We find the following currents having $J^P=1/2^+$ and quark contents $uud c \bar c$:
\begin{eqnarray}
\nonumber \xi_1 &=& [\epsilon^{abc} (u^T_a C d_b) \gamma_5 c_c] [\bar c_d u_d] \, ,
\\ \nonumber \xi_2 &=& [\epsilon^{abc} (u^T_a C d_b) c_c] [\bar c_d \gamma_5 u_d] \, ,
\\ \nonumber \xi_3 &=& [\epsilon^{abc} (u^T_a C \gamma_5 d_b) c_c] [\bar c_d u_d] \, ,
\\ \nonumber \xi_4 &=& [\epsilon^{abc} (u^T_a C \gamma_5 d_b) \gamma_5 c_c] [\bar c_d \gamma_5 u_d] \, ,
\\ \nonumber \xi_5 &=& [\epsilon^{abc} (u^T_a C d_b) \gamma_\mu \gamma_5 c_c] [\bar c_d \gamma_\mu u_d] \, ,
\\ \nonumber \xi_6 &=& [\epsilon^{abc} (u^T_a C d_b) \gamma_\mu c_c] [\bar c_d \gamma_\mu \gamma_5 u_d] \, ,
\\ \nonumber \xi_7 &=& [\epsilon^{abc} (u^T_a C \gamma_5 d_b) \gamma_\mu c_c] [\bar c_d \gamma_\mu u_d] \, ,
\\ \nonumber \xi_8 &=& [\epsilon^{abc} (u^T_a C \gamma_5 d_b) \gamma_\mu \gamma_5 c_c] [\bar c_d \gamma_\mu \gamma_5 u_d] \, ,
\\ \nonumber \xi_9 &=& [\epsilon^{abc} (u^T_a C d_b) \sigma_{\mu\nu} \gamma_5 c_c] [\bar c_d \sigma_{\mu\nu} u_d] \, ,
\\ \nonumber \xi_{10} &=& [\epsilon^{abc} (u^T_a C d_b) \sigma_{\mu\nu} c_c] [\bar c_d \sigma_{\mu\nu} \gamma_5 u_d] \, ,
\\ \nonumber \xi_{11} &=& [\epsilon^{abc} (u^T_a C \gamma_5 d_b) \sigma_{\mu\nu} c_c] [\bar c_d \sigma_{\mu\nu} u_d] \, ,
\\ \nonumber \xi_{12} &=& [\epsilon^{abc} (u^T_a C \gamma_5 d_b) \sigma_{\mu\nu} \gamma_5 c_c] [\bar c_d \sigma_{\mu\nu} \gamma_5 u_d] \, ,
\\ \nonumber \xi_{13} &=& [\epsilon^{abc} (u^T_a C \gamma_\mu d_b) \gamma_\mu \gamma_5 c_c] [\bar c_d u_d] \, ,
\\ \nonumber \xi_{14} &=& [\epsilon^{abc} (u^T_a C \gamma_\mu d_b) \gamma_\mu c_c] [\bar c_d \gamma_5 u_d] \, ,
\\ \nonumber \xi_{15} &=& [\epsilon^{abc} (u^T_a C \gamma_\mu \gamma_5 d_b) \gamma_\mu c_c] [\bar c_d u_d] \, ,
\\ \nonumber \xi_{16} &=& [\epsilon^{abc} (u^T_a C \gamma_\mu \gamma_5 d_b) \gamma_\mu \gamma_5 c_c] [\bar c_d \gamma_5 u_d] \, ,
\\ \nonumber \xi_{17} &=& [\epsilon^{abc} (u^T_a C \gamma_\mu d_b) \gamma_5 c_c] [\bar c_d \gamma_\mu u_d] \, ,
\\ \nonumber \xi_{18} &=& [\epsilon^{abc} (u^T_a C \gamma_\mu d_b) c_c] [\bar c_d \gamma_\mu \gamma_5 u_d] \, ,
\\ \nonumber \xi_{19} &=& [\epsilon^{abc} (u^T_a C \gamma_\mu \gamma_5 d_b) c_c] [\bar c_d \gamma_\mu u_d] \, ,
\\ \nonumber \xi_{20} &=& [\epsilon^{abc} (u^T_a C \gamma_\mu \gamma_5 d_b) \gamma_5 c_c] [\bar c_d \gamma_\mu \gamma_5 u_d] \, ,
\\ \nonumber \xi_{21} &=& [\epsilon^{abc} (u^T_a C \gamma_\mu d_b) \sigma_{\mu\nu} \gamma_5 c_c] [\bar c_d \gamma_\nu u_d] \, ,
\\ \xi_{22} &=& [\epsilon^{abc} (u^T_a C \gamma_\mu d_b) \sigma_{\mu\nu} c_c] [\bar c_d \gamma_\nu \gamma_5 u_d] \, ,
\\ \nonumber \xi_{23} &=& [\epsilon^{abc} (u^T_a C \gamma_\mu \gamma_5 d_b) \sigma_{\mu\nu} c_c] [\bar c_d \gamma_\nu u_d] \, ,
\\ \nonumber \xi_{24} &=& [\epsilon^{abc} (u^T_a C \gamma_\mu \gamma_5 d_b) \sigma_{\mu\nu} \gamma_5 c_c] [\bar c_d \gamma_\nu \gamma_5 u_d] \, ,
\\ \nonumber \xi_{25} &=& [\epsilon^{abc} (u^T_a C \gamma_\mu d_b) \gamma_\nu \gamma_5 c_c] [\bar c_d \sigma_{\mu\nu} u_d] \, ,
\\ \nonumber \xi_{26} &=& [\epsilon^{abc} (u^T_a C \gamma_\mu d_b) \gamma_\nu c_c] [\bar c_d \sigma_{\mu\nu} \gamma_5 u_d] \, ,
\\ \nonumber \xi_{27} &=& [\epsilon^{abc} (u^T_a C \gamma_\mu \gamma_5 d_b) \gamma_\nu c_c] [\bar c_d \sigma_{\mu\nu} u_d] \, ,
\\ \nonumber \xi_{28} &=& [\epsilon^{abc} (u^T_a C \gamma_\mu \gamma_5 d_b) \gamma_\nu \gamma_5 c_c] [\bar c_d \sigma_{\mu\nu} \gamma_5 u_d] \, ,
\\ \nonumber \xi_{29} &=& [\epsilon^{abc} (u^T_a C \sigma_{\mu\nu} d_b) \sigma_{\mu\nu} \gamma_5 c_c] [\bar c_d u_d] \, ,
\\ \nonumber \xi_{30} &=& [\epsilon^{abc} (u^T_a C \sigma_{\mu\nu} d_b) \sigma_{\mu\nu} c_c] [\bar c_d \gamma_5 u_d] \, ,
\\ \nonumber \xi_{31} &=& [\epsilon^{abc} (u^T_a C \sigma_{\mu\nu} \gamma_5 d_b) \sigma_{\mu\nu} c_c] [\bar c_d u_d] \, ,
\\ \nonumber \xi_{32} &=& [\epsilon^{abc} (u^T_a C \sigma_{\mu\nu} \gamma_5 d_b) \sigma_{\mu\nu} \gamma_5 c_c] [\bar c_d \gamma_5 u_d] \, ,
\\ \nonumber \xi_{33} &=& [\epsilon^{abc} (u^T_a C \sigma_{\mu\nu} d_b) \gamma_\mu \gamma_5 c_c] [\bar c_d \gamma_\nu u_d] \, ,
\\ \nonumber \xi_{34} &=& [\epsilon^{abc} (u^T_a C \sigma_{\mu\nu} d_b) \gamma_\mu c_c] [\bar c_d \gamma_\nu \gamma_5 u_d] \, ,
\\ \nonumber \xi_{35} &=& [\epsilon^{abc} (u^T_a C \sigma_{\mu\nu} \gamma_5 d_b) \gamma_\mu c_c] [\bar c_d \gamma_\nu u_d] \, ,
\\ \nonumber \xi_{36} &=& [\epsilon^{abc} (u^T_a C \sigma_{\mu\nu} \gamma_5 d_b) \gamma_\mu \gamma_5 c_c] [\bar c_d \gamma_\nu \gamma_5 u_d] \, ,
\\ \nonumber \xi_{37} &=& [\epsilon^{abc} (u^T_a C \sigma_{\mu\nu} d_b) \gamma_5 c_c] [\bar c_d \sigma_{\mu\nu} u_d] \, ,
\\ \nonumber \xi_{38} &=& [\epsilon^{abc} (u^T_a C \sigma_{\mu\nu} d_b) c_c] [\bar c_d \sigma_{\mu\nu} \gamma_5 u_d] \, ,
\\ \nonumber \xi_{39} &=& [\epsilon^{abc} (u^T_a C \sigma_{\mu\nu} \gamma_5 d_b) c_c] [\bar c_d \sigma_{\mu\nu} u_d] \, ,
\\ \nonumber \xi_{40} &=& [\epsilon^{abc} (u^T_a C \sigma_{\mu\nu} \gamma_5 d_b) \gamma_5 c_c] [\bar c_d \sigma_{\mu\nu} \gamma_5 u_d] \, ,
\\ \nonumber \xi_{41} &=& [\epsilon^{abc} (u^T_a C \sigma_{\mu\rho} d_b) \sigma_{\mu\nu} \gamma_5 c_c] [\bar c_d \sigma_{\nu\rho} u_d] \, ,
\\ \nonumber \xi_{42} &=& [\epsilon^{abc} (u^T_a C \sigma_{\mu\rho} d_b) \sigma_{\mu\nu} c_c] [\bar c_d \sigma_{\nu\rho} \gamma_5 u_d] \, ,
\\ \nonumber \xi_{43} &=& [\epsilon^{abc} (u^T_a C \sigma_{\mu\rho} \gamma_5 d_b) \sigma_{\mu\nu} c_c] [\bar c_d \sigma_{\nu\rho} u_d] \, ,
\\ \nonumber \xi_{44} &=& [\epsilon^{abc} (u^T_a C \sigma_{\mu\rho} \gamma_5 d_b) \sigma_{\mu\nu} \gamma_5 c_c] [\bar c_d \sigma_{\nu\rho} \gamma_5 u_d] \, .
\end{eqnarray}
We can verify the following relations
\begin{eqnarray}
\nonumber \xi_{9} &=& \xi_{10} \, ,
\\ \nonumber \xi_{11} &=& \xi_{12} \, ,
\\ \xi_{29} &=& \xi_{31} \, ,
\\ \nonumber \xi_{30} &=& \xi_{32} \, ,
\\ \nonumber \xi_{37} &=& \xi_{40} \, ,
\\ \nonumber \xi_{38} &=& \xi_{39} \, .
\end{eqnarray}
Then there are only 38 independent currents left.
To perform QCD sum rule analyses, we shall use
\begin{eqnarray}
\xi_2 - \xi_4 &=& [\epsilon^{abc} (u^T_a C d_b) c_c] [\bar c_d \gamma_5 u_d]
\label{def:xi24}
\\ \nonumber && ~~~~~~~~~~ - [\epsilon^{abc} (u^T_a C \gamma_5 d_b) \gamma_5 c_c] [\bar c_d \gamma_5 u_d] \, ,
\\ \xi_5 - \xi_7 &=& [\epsilon^{abc} (u^T_a C d_b) \gamma_\mu \gamma_5 c_c] [\bar c_d \gamma_\mu u_d]
\label{def:xi57}
\\ \nonumber && ~~~~~~~~~~ - [\epsilon^{abc} (u^T_a C \gamma_5 d_b) \gamma_\mu c_c] [\bar c_d \gamma_\mu u_d] \, ,
\\ \xi_{14} &=& [\epsilon^{abc} (u^T_a C \gamma_\mu d_b) \gamma_\mu c_c] [\bar c_d \gamma_5 u_d] \, ,
\label{def:xi14}
\\ \xi_{16} &=& [\epsilon^{abc} (u^T_a C \gamma_\mu \gamma_5 d_b) \gamma_\mu \gamma_5 c_c] [\bar c_d \gamma_5 u_d] \, ,
\label{def:xi16}
\\ \xi_{17} &=& [\epsilon^{abc} (u^T_a C \gamma_\mu d_b) \gamma_5 c_c] [\bar c_d \gamma_\mu u_d] \, ,
\label{def:xi17}
\\ \xi_{19} &=& [\epsilon^{abc} (u^T_a C \gamma_\mu \gamma_5 d_b) c_c] [\bar c_d \gamma_\mu u_d] \, ,
\label{def:xi19}
\end{eqnarray}
which well couple to the $[\Lambda_c \bar D]$, $[\Lambda_c \bar D^*]$, $[\Sigma_c \bar D]$, $[\Lambda^*_c \bar D]$, $[\Sigma_c^* \bar D^*]$ and $[\Lambda^*_c \bar D^*]$ channels, respectively.

\subsection{Currents of $[\bar c_d d_d][\epsilon^{abc} u_a u_b c_c]$}

In this subsection, we construct the currents of the type $[\bar c_d \Gamma_k d_d][\epsilon^{abc} (u_a^T C \Gamma_i u_b) \Gamma_j c_c]$.
The currents of the other types $[\bar c_d \Gamma_k d_d][\epsilon^{abc} (u_a^T C \Gamma_i c_b) \Gamma_j u_c]$ and $[\bar c_d \Gamma_k d_d][\epsilon^{abc} (c_a^T C \Gamma_i u_b) \Gamma_j u_c]$ can be
related to these currents by using the Fierz transformation.
We find the following currents having $J^P=1/2^+$ and quark contents $uud c \bar c$:
\begin{eqnarray}
\nonumber \psi_1 &=& [\epsilon^{abc} (u^T_a C \gamma_\mu u_b) \gamma_\mu \gamma_5 c_c] [\bar c_d d_d] \, ,
\\ \nonumber \psi_2 &=& [\epsilon^{abc} (u^T_a C \gamma_\mu u_b) \gamma_\mu c_c] [\bar c_d \gamma_5 d_d] \, ,
\\ \nonumber \psi_3 &=& [\epsilon^{abc} (u^T_a C \gamma_\mu u_b) \gamma_5 c_c] [\bar c_d \gamma_\mu d_d] \, ,
\\ \nonumber \psi_4 &=& [\epsilon^{abc} (u^T_a C \gamma_\mu u_b) c_c] [\bar c_d \gamma_\mu \gamma_5 d_d] \, ,
\\ \nonumber \psi_5 &=& [\epsilon^{abc} (u^T_a C \gamma_\mu u_b) \sigma_{\mu\nu} \gamma_5 c_c] [\bar c_d \gamma_\nu d_d] \, ,
\\ \nonumber \psi_6 &=& [\epsilon^{abc} (u^T_a C \gamma_\mu u_b) \sigma_{\mu\nu} c_c] [\bar c_d \gamma_\nu \gamma_5 d_d] \, ,
\\ \nonumber \psi_7 &=& [\epsilon^{abc} (u^T_a C \gamma_\mu u_b) \gamma_\nu \gamma_5 c_c] [\bar c_d \sigma_{\mu\nu} d_d] \, ,
\\ \nonumber \psi_8 &=& [\epsilon^{abc} (u^T_a C \gamma_\mu u_b) \gamma_\nu c_c] [\bar c_d \sigma_{\mu\nu} \gamma_5 d_d] \, ,
\\ \nonumber \psi_9 &=& [\epsilon^{abc} (u^T_a C \sigma_{\mu\nu} u_b) \sigma_{\mu\nu} \gamma_5 c_c] [\bar c_d d_d] \, ,
\\ \nonumber \psi_{10} &=& [\epsilon^{abc} (u^T_a C \sigma_{\mu\nu} u_b) \sigma_{\mu\nu} c_c] [\bar c_d \gamma_5 d_d] \, ,
\\ \nonumber \psi_{11} &=& [\epsilon^{abc} (u^T_a C \sigma_{\mu\nu} \gamma_5 u_b) \sigma_{\mu\nu} c_c] [\bar c_d d_d] \, ,
\\ \psi_{12} &=& [\epsilon^{abc} (u^T_a C \sigma_{\mu\nu} \gamma_5 u_b) \sigma_{\mu\nu} \gamma_5 c_c] [\bar c_d \gamma_5 d_d] \, ,
\\ \nonumber \psi_{13} &=& [\epsilon^{abc} (u^T_a C \sigma_{\mu\nu} u_b) \gamma_\mu \gamma_5 c_c] [\bar c_d \gamma_\nu d_d] \, ,
\\ \nonumber \psi_{14} &=& [\epsilon^{abc} (u^T_a C \sigma_{\mu\nu} u_b) \gamma_\mu c_c] [\bar c_d \gamma_\nu \gamma_5 d_d] \, ,
\\ \nonumber \psi_{15} &=& [\epsilon^{abc} (u^T_a C \sigma_{\mu\nu} \gamma_5 u_b) \gamma_\mu c_c] [\bar c_d \gamma_\nu d_d] \, ,
\\ \nonumber \psi_{16} &=& [\epsilon^{abc} (u^T_a C \sigma_{\mu\nu} \gamma_5 u_b) \gamma_\mu \gamma_5 c_c] [\bar c_d \gamma_\nu \gamma_5 d_d] \, ,
\\ \nonumber \psi_{17} &=& [\epsilon^{abc} (u^T_a C \sigma_{\mu\nu} u_b) \gamma_5 c_c] [\bar c_d \sigma_{\mu\nu} d_d] \, ,
\\ \nonumber \psi_{18} &=& [\epsilon^{abc} (u^T_a C \sigma_{\mu\nu} u_b) c_c] [\bar c_d \sigma_{\mu\nu} \gamma_5 d_d] \, ,
\\ \nonumber \psi_{19} &=& [\epsilon^{abc} (u^T_a C \sigma_{\mu\nu} \gamma_5 u_b) c_c] [\bar c_d \sigma_{\mu\nu} d_d] \, ,
\\ \nonumber \psi_{20} &=& [\epsilon^{abc} (u^T_a C \sigma_{\mu\nu} \gamma_5 u_b) \gamma_5 c_c] [\bar c_d \sigma_{\mu\nu} \gamma_5 d_d] \, ,
\\ \nonumber \psi_{21} &=& [\epsilon^{abc} (u^T_a C \sigma_{\mu\rho} u_b) \sigma_{\mu\nu} \gamma_5 c_c] [\bar c_d \sigma_{\nu\rho} d_d] \, ,
\\ \nonumber \psi_{22} &=& [\epsilon^{abc} (u^T_a C \sigma_{\mu\rho} u_b) \sigma_{\mu\nu} c_c] [\bar c_d \sigma_{\nu\rho} \gamma_5 d_d] \, ,
\\ \nonumber \psi_{23} &=& [\epsilon^{abc} (u^T_a C \sigma_{\mu\rho} \gamma_5 u_b) \sigma_{\mu\nu} c_c] [\bar c_d \sigma_{\nu\rho} d_d] \, ,
\\ \nonumber \psi_{24} &=& [\epsilon^{abc} (u^T_a C \sigma_{\mu\rho} \gamma_5 u_b) \sigma_{\mu\nu} \gamma_5 c_c] [\bar c_d \sigma_{\nu\rho} \gamma_5 d_d] \, .
\end{eqnarray}
We can verify the following relations
\begin{eqnarray}
\nonumber \psi_9 &=& \psi_{11} \, ,
\\ \psi_{10} &=& \psi_{12} \, ,
\\ \nonumber \psi_{18} &=& \psi_{19} \, ,
\\ \nonumber \psi_{17} &=& \psi_{20} \, .
\end{eqnarray}
Then there are 20 independent currents left.
To perform QCD sum rule analyses, we shall use
\begin{eqnarray}
\psi_2 &=& [\epsilon^{abc} (u^T_a C \gamma_\mu u_b) \gamma_\mu c_c] [\bar c_d \gamma_5 d_d] \, ,
\label{def:psi2}
\\ \psi_3 &=& [\epsilon^{abc} (u^T_a C \gamma_\mu u_b) \gamma_5 c_c] [\bar c_d \gamma_\mu d_d] \, ,
\label{def:psi3}
\end{eqnarray}
which well couple to the $[\Sigma^*_c \bar D]$ and $[\Sigma_c^* \bar D^*]$ channels, respectively.

\subsection{A short summary}

In the previous subsections, we systematically construct all local pentaquark interpolating currents having $J^P=1/2^+$ and quark contents $uud c \bar c$.
We find13 independent currents of the color configuration $[\bar c_d c_d][\epsilon^{abc}u_a d_b u_c]$, 38 independent currents
of the color configuration $[\bar c_d u_d][\epsilon^{abc} u_a d_b c_c]$, and 20 independent currents of the color configuration $[\bar c_d d_d][\epsilon^{abc} u_a u_b c_c]$.
One may wonder that there are altogether 71 independent currents. Actually, some of these currents can be
related by applying the color rearrangement in Eq.~(\ref{eq:cr}). Considering there are 71 currents in all, it will
be too complicate to perform thorough transformation one by one. However, we can give an estimation that
at least one third of these 71 currents are not independent and can be expressed by other currents.

We shall not discuss this any more, but start to perform QCD sum rule analyses using the currents selected in this section as well as those selected in Appendixes~\ref{app:spin32} and \ref{app:spin52}.

\section{QCD sum rules analyses}
\label{sec:sumrule}

In the following, we shall use the method of QCD sum rules~\cite{Shifman:1978bx,Reinders:1984sr,Nielsen:2009uh,Chen:2010ze,Chen:2015ata} to investigate the currents selected in Sec.~\ref{sec:current} and in Appendixes~\ref{app:spin32} and \ref{app:spin52}. The results obtained using the $J^P=1/2^+$ currents $\eta_2 - \eta_4$, $\eta_5 - \eta_7$, $\eta_{13}$, $\xi_2 - \xi_4$, $\xi_5 - \xi_7$, $\xi_{14}$, $\xi_{16}$, $\xi_{17}$, $\xi_{19}$, $\psi_2$, and $\psi_3$ are listed in Table~\ref{tab:spin12}, those obtained using the $J^P=3/2^-$ currents $\eta_{5\mu} - \eta_{7\mu}$, $\eta_{18\mu}$, $\eta_{19\mu}$, $\xi_{5\mu} - \xi_{7\mu}$, $\xi_{18\mu}$, $\xi_{20\mu}$, $\xi_{25\mu}$, $\xi_{27\mu}$, $\xi_{33\mu}$, $\xi_{35\mu}$, $\psi_{2\mu}$, $\psi_{5\mu}$, and $\psi_{9\mu}$ are listed in Table~\ref{tab:spin32}, and those obtained using the $J^P=5/2^+$ currents $\eta_{11\mu\nu}$, $\xi_{13\mu\nu}$, $\xi_{15\mu\nu}$, and $\psi_{3\mu\nu}$ are listed in Table~\ref{tab:spin52}.

We use $J$, $J_{\mu}$, and $J_{\mu\nu}$ to denote the currents having spin $J=1/2$, $3/2$, and $5/2$, respectively, and assume
they couple to the physical states $X$ through
\begin{eqnarray}
\nonumber \langle 0 | J | X_{1/2} \rangle &=& f_X u (p) \, ,
\\ \label{eq:gamma0} \langle 0 | J_{\mu} | X_{3/2} \rangle &=& f_X u_\mu (p) \, ,
\\ \nonumber \langle 0 | J_{\mu\nu} | X_{5/2} \rangle &=& f_X u_{\mu\nu} (p) \, .
\end{eqnarray}
The two-point correlation functions obtained using these currents can be written as:
\begin{eqnarray}
\label{pi:spin12} \Pi\left(q^2\right) &=& i \int d^4x e^{iq\cdot x} \langle 0 | T\left[J(x) \bar J(0)\right] | 0 \rangle
\\ \nonumber &=& (q\!\!\!\slash~ + M_X) \Pi^{1/2}\left(q^2\right) \, ,
\\ \label{pi:spin32} \Pi_{\mu \nu}\left(q^2\right) &=& i \int d^4x e^{iq\cdot x} \langle 0 | T\left[J_{\mu}(x) \bar J_{\nu}(0)\right] | 0 \rangle
\\ \nonumber &=& \left(\frac{q_\mu q_\nu}{q^2}-g_{\mu\nu}\right) (q\!\!\!\slash~ + M_X) \Pi^{3/2}\left(q^2\right) + \cdots \, ,
\\ \label{pi:spin52} \Pi_{\mu \nu \rho \sigma}\left(q^2\right) &=& i \int d^4x e^{iq\cdot x} \langle 0 | T\left[J_{\mu\nu}(x) \bar J_{\rho\sigma}(0)\right] | 0 \rangle
\\ \nonumber &=& \left(g_{\mu\rho}g_{\nu\sigma} + g_{\mu\sigma} g_{\nu\rho} \right) (q\!\!\!\slash~ + M_X) \Pi^{5/2}\left(q^2\right) + \cdots \, ,
\end{eqnarray}
where $\cdots$ in Eq.~(\ref{pi:spin32}) contains the spin $1/2$ components of $J_{\mu}$, and $\cdots$ in Eq.~(\ref{pi:spin52}) contains the spin $1/2$ and $3/2$ components of $J_{\mu\nu}$.

We note that we have assumed that $X$ has the same parity as $J$, and used the non-$\gamma_5$ coupling in Eq.~(\ref{eq:gamma0}). While, we can also use the $\gamma_5$ coupling
\begin{eqnarray}
\label{eq:gamma51} \langle 0 | J | X^\prime \rangle &=& f_{X^\prime} \gamma_5 u (p) \, ,
\end{eqnarray}
when $X^\prime$ has the opposite parity from $J$. Or we can use the partner of the current $\gamma_5 J$ having the opposite parity
\begin{eqnarray}
\label{eq:gamma52} \langle 0 | \gamma_5 J | X \rangle &=& f_X \gamma_5 u (p) \, .
\end{eqnarray}
See also discussions in Refs.~\cite{Chung:1981cc,Jido:1996ia,Kondo:2005ur,Ohtani:2012ps}.
These two assumptions both lead to the two-point correlation functions which are similar to Eqs.~(\ref{pi:spin12})--(\ref{pi:spin52}), but with $(q\!\!\!\slash~ + M_{X^{(\prime)}})$ replaced by $(- q\!\!\!\slash~ + M_{X^{(\prime)}})$.
This difference would tell us the parity of the hadron $X^{(\prime)}$. Technically, in the following analyses we use the terms proportional to $\mathbf{1}$, $\mathbf{1} \times g_{\mu\nu}$ and $\mathbf{1} \times g_{\mu\rho} g_{\nu\sigma}$ to evaluate the mass of $X$, which are then compared with those proportional to $q\!\!\!\slash~$, $q\!\!\!\slash~ \times g_{\mu\nu}$ and $q\!\!\!\slash~ \times g_{\mu\rho} g_{\nu\sigma}$ to determine its parity.

We can calculate the two-point correlation functions (\ref{pi:spin12})--(\ref{pi:spin52}) in the QCD operator product expansion (OPE) up to certain order in the expansion, which is then matched with a hadronic parametrization to extract information about hadron properties.
At the hadron level, it can be written as
%
\begin{equation}
\Pi(q^2)={\frac{1}{\pi}}\int^\infty_{s_<}\frac{{\rm Im} \Pi(s)}{s-q^2-i\varepsilon}ds \, ,
\label{eq:disper}
\end{equation}
%
where we have used the form of the dispersion relation, and $s_<$ denotes the physical threshold.
The imaginary part of the correlation function is defined as the spectral function,
which is usually evaluated at the hadron level by inserting intermediate hadron states $\sum_n|n\rangle\langle n|$
%
\begin{eqnarray}
\nonumber \rho(s) \equiv \frac{1}{\pi}{\rm Im}\Pi(s)
&=& \sum_n\delta(s-M^2_n)\langle 0|\eta|n\rangle\langle n|{\eta^\dagger}|0\rangle
\\ &=& f_X^2\delta(s-m_X^2)+ \mbox{continuum}\, ,
\label{eq:rho}
\end{eqnarray}
%
where we have adopted the usual parametrization of one-pole dominance for the ground state $X$ and a continuum contribution.
The spectral density $\rho(s)$ can also be evaluated at the quark and gluon level via the QCD operator product expansion.
After performing the Borel transform at both the hadron and QCD levels, the two-point correlation function can be expressed as
%
\begin{equation}
\Pi^{(all)}(M_B^2)\equiv\mathcal{B}_{M_B^2}\Pi(p^2) = \int^\infty_{s_<} e^{-s/M_B^2} \rho(s) ds \, .
\label{eq:borel}
\end{equation}
%
Finally, we assume that the contribution from the continuum states can be approximated well by the OPE spectral density above a threshold value $s_0$ (duality),
and arrive at the sum rule relation which can be used to perform numerical analyses:
%
\begin{eqnarray}
M^2_X(s_0, M_B) 
&=& {\int^{s_0}_{s_<} e^{-s/M_B^2} \rho(s) s ds \over \int^{s_0}_{s_<} e^{-s/M_B^2} \rho(s) ds} \, .
\label{eq:mass}
\end{eqnarray}
%

\newcounter{mytempeqncnt1}
\begin{figure*}[!hbt]
\small
\hrulefill
\begin{eqnarray}
\rho^{[N^* J/\psi]}_{\mu\nu\rho\sigma}(s) &=& \mathbf{1} \times \big( g_{\mu\rho}g_{\nu\sigma} + g_{\mu\sigma}g_{\nu\rho} \big) \times
\left ( \rho^{pert}_1(s)+\rho^{\qq}_1(s)+\rho^{\GGa}_1(s)+\rho^{\qq^2}_1(s)+\rho^{\qGqa}_1(s)+\rho^{\qq\qGqa}_1(s) \right )
\label{ope:eta11}
\non
&+& q\!\!\!\slash~\times \big( g_{\mu\rho}g_{\nu\sigma} + g_{\mu\sigma}g_{\nu\rho} \big) \times
\left ( \rho^{pert}_2(s)+\rho^{\qq}_2(s)+\rho^{\GGa}_2(s)+\rho^{\qq^2}_2(s)+\rho^{\qGqa}_2(s)+\rho^{\qq\qGqa}_2(s) \right ) + \cdots \, .
\non
\rho^{pert}_1(s) &=& 0 \, ,
\non
\rho^{\qq}_1(s) &=& -\frac{\qq}{49152\pi^6} \dab \Bigg\{ \FF(s)^4 \times \frac{(1-\alpha-\beta)(11 + 11\alpha + 11\beta + 8\alpha^2 + 16\alpha\beta + 8\beta^2)}{\alpha^3\beta^3}
\non
&& - m_c^2\FF(s)^3 \times \frac{4(1-\alpha-\beta)(11 - 7\alpha - 7\beta - 4\alpha^2 - 8\alpha\beta - 4\beta^2)}{\alpha^3\beta^3} \Bigg\} \, ,
\non
\rho^{\GGa}_1(s) &=& 0 \, ,
\non
\rho^{\qGqa}_1(s) &=& \frac{\qGqb}{4096\pi^6} \dab \Bigg\{ \FF(s)^3 \times \frac{\alpha + \beta + 2\alpha^2  + 4\alpha\beta + 2\beta^2}{\alpha^2\beta^2}
\non
&& - m_c^2\FF(s)^2 \times \frac{3(2 - \alpha - \beta - \alpha^2 - 2\alpha\beta - \beta^2)}{\alpha^2\beta^2} \Bigg\} \, ,
\non
\rho^{\qq^2}_1(s)&=& 0 \, ,
\non
\rho^{\qq\qGqa}_1(s)&=& 0 \, ,
\non
\rho^{pert}_2(s) &=& - \frac{1}{6553600\pi^8}\dab \Bigg\{ \FF(s)^5 \times \frac{(1-\alpha-\beta)^3(57 + 31\alpha + 31\beta + 12\alpha^2 + 24\alpha\beta + 12\beta^2)}{\alpha^4\beta^4}
\non
&& - m_c^2 \FF(s)^4 \times \frac{5(1-\alpha-\beta)^3(19 - 13\alpha - 13\beta - 6\alpha^2 - 12\alpha\beta - 6\beta^2)}{\alpha^4\beta^4} \Bigg\} \, ,
\non
\rho^{\qq}_2(s) &=& 0 \, ,
\non
\rho^{\GGa}_2(s)&=&\frac{\GGb}{23592960 \pi^8}\dab\Bigg\{\FF(s)^3 \times \Big( \frac{3(1-\alpha-\beta)(11 + 11\alpha + 11\beta - 4\alpha^2 - 8\alpha\beta - 4\beta^2)}{\alpha^2\beta^2}
\non
&& + \frac{(1-\alpha-\beta)^3(83 + 29\alpha + 29\beta - 12\alpha^2 - 24\alpha\beta - 12\beta^2)}{\alpha^3\beta^3} \Big)
\non
&& + m_c^2 \FF(s)^2 \times \Big( - \frac{3(1-\alpha-\beta)^3(57+31\alpha+31\beta+12\alpha^2+24\alpha\beta+12\beta^2)}{\alpha\beta^4}
\non
&& - \frac{9(1-\alpha-\beta)(11-13\alpha-13\beta+2\alpha^2+4\alpha\beta+2\beta^2)}{\alpha^2\beta^2} + \frac{9(1-\alpha-\beta)^4(19+6\alpha+6\beta)}{\alpha^2\beta^4}
\non
&& - \frac{3(1-\alpha-\beta)^3(57+31\alpha+31\beta+12\alpha^2+24\alpha\beta+12\beta^2)}{\alpha^4\beta} + \frac{9(1-\alpha-\beta)^4(19+6\alpha+6\beta)}{\alpha^4\beta^2} \Big)
\non
&& + m_c^4 \FF(s) \times \Big( \frac{6(1-\alpha-\beta)^3(19-13\alpha-13\beta-6\alpha^2-12\alpha\beta-6\beta^2)}{\alpha\beta^4}
\non
&& + \frac{6(1-\alpha-\beta)^3(19-13\alpha-13\beta-6\alpha^2-12\alpha\beta-6\beta^2)}{\alpha^4\beta}
\Big) \Bigg\}\, ,
\non
\rho^{\qGqa}_2(s) &=& 0 \, ,
\non
\rho^{\qq^2}_2(s)&=& -\frac{\qq^2}{512\pi^4} \dab \Bigg\{ \FF(s)^2 \times \frac{5(\alpha + \beta)}{\alpha\beta}
- m_c^2 \FF(s) \times \frac{10(1 - \alpha - \beta)}{\alpha\beta} \Bigg\} \, ,
\non
\rho^{\qq\qGqa}_2(s)&=& \frac{11\qq\qGqb}{1024\pi^4} \times \int^{\alpha_{max}}_{\alpha_{min}}d\alpha \Bigg\{ \HH(s) - \int^{\beta_{max}}_{\beta_{min}}d\beta \Big (\FF(s)+ m_c^2\Big) \Bigg\} \, .
\end{eqnarray}
\hrulefill
\vspace*{4pt}
\end{figure*}

\newcounter{mytempeqncnt2}
\begin{figure*}[!hbt]
\small
\hrulefill
\begin{eqnarray}
\rho^{[\Sigma_c \bar D^*]}_{\mu\nu}(s) &=& \mathbf{1} \times g_{\mu\nu} \times \left ( \rho^{pert}_3(s)+\rho^{\qq}_3(s)+\rho^{\GGa}_3(s)+\rho^{\qq^2}_3(s)+\rho^{\qGqa}_3(s)+\rho^{\qq\qGqa}_3(s) \right )
\label{ope:psi9}
\non
&+& q\!\!\!\slash~\times g_{\mu\nu} \times \left ( \rho^{pert}_4(s)+\rho^{\qq}_4(s)+\rho^{\GGa}_4(s)+\rho^{\qq^2}_4(s)+\rho^{\qGqa}_4(s)+\rho^{\qq\qGqa}_4(s) \right )
+ \cdots \, .
\non
\rho^{pert}_3(s)&=&\frac{m_c}{163840\pi^8}\dab \FF(s)^5 \times \frac{(1-\alpha-\beta)^3(3+\alpha+\beta)}{\alpha^5\beta^4}\, ,
\non
\rho^{\qq}_3(s)&=&-\frac{m_c^2\qq}{512\pi^6}\dab \FF(s)^3 \times \frac{(1-\alpha-\beta)^2}{\alpha^3\beta^3}\, ,
\non
\rho^{\GGa}_3(s)&=&\frac{m_c\GGb}{16384\pi^8}\dab\Bigg\{\FF(s)^3 \times \Big( - \frac{(1-\alpha-\beta)(1+\alpha+\beta)}{2\alpha^3\beta^2}
\non
&& - \frac{(1-\alpha-\beta)^2(4-\alpha-\beta)}{9\alpha^3\beta^3} + \frac{(1-\alpha-\beta)^2(2+\alpha+\beta)}{3\alpha^4\beta^2} + \frac{(1-\alpha-\beta)^3(3+\alpha+\beta)}{12\alpha^5\beta^2} \Big)
\non
&& + m_c^2 \FF(s)^2 \times \Big( \frac{(1-\alpha-\beta)^3(3+\alpha+\beta)}{12\alpha^2\beta^4} + \frac{(1-\alpha-\beta)^3(3+\alpha+\beta)}{12\alpha^5\beta} \Big) \Bigg\}\, ,
\non
\rho^{\qGqa}_3(s)&=&\frac{m_c^2\qGqb}{1024\pi^6}\dab \FF(s)^2 \times \frac{3(1-\alpha-\beta)}{\alpha^2\beta^2}\, ,
\non
\rho^{\qq^2}_3(s)&=& -\frac{m_c \qq^2}{32\pi^4} \dab \FF(s)^2 \times \frac{(\alpha+\beta)}{\alpha^2\beta}\, ,
\non
\rho^{\qq\qGqa}_3(s)&=& \frac{m_c\qq\qGqb}{64\pi^4} \times \int^{\alpha_{max}}_{\alpha_{min}}d\alpha \Bigg\{ \HH(s) \times \frac{2}{\alpha} - \int^{\beta_{max}}_{\beta_{min}}d\beta \FF(s) \times \frac{(3\alpha+\beta)}{\alpha^2} \Bigg\} \, ,
\non
\rho^{pert}_4(s)&=&\frac{1}{81920\pi^8}\dab \FF(s)^5 \times \frac{(1-\alpha-\beta)^3(3+\alpha+\beta)}{\alpha^4\beta^4}\, ,
\non
\rho^{\qq}_4(s)&=&-\frac{m_c\qq}{256\pi^6}\dab \FF(s)^3 \times \frac{(1-\alpha-\beta)^2}{\alpha^2\beta^3}\, ,
\non
\rho^{\GGa}_4(s)&=&\frac{\GGb}{8192\pi^8}\dab\Bigg\{
 \FF(s)^3 \times \Big( - \frac{(1-\alpha-\beta)^2(4-\alpha-\beta)}{9\alpha^2\beta^3} + \frac{(1-\alpha-\beta)^2(2+\alpha+\beta)}{6\alpha^3\beta^2}  \Big)
\non
&& + m_c^2 \FF(s)^2 \times \Big( \frac{(1-\alpha-\beta)^3(3+\alpha+\beta)}{12\alpha\beta^4} + \frac{(1-\alpha-\beta)^3(3+\alpha+\beta)}{12\alpha^4\beta} \Big) \Bigg\}\, ,
\non
\rho^{\qGqa}_4(s)&=&\frac{m_c\qGqb}{512\pi^6}\dab \FF(s)^2 \times \frac{3(1-\alpha-\beta)}{\alpha\beta^2}\, ,
\non
\rho^{\qq^2}_4(s)&=& -\frac{\qq^2}{64\pi^4} \dab \FF(s)^2 \times \frac{(\alpha+\beta)}{\alpha\beta}\, ,
\non
\rho^{\qq\qGqa}_4(s)&=& \frac{\qq\qGqb}{128\pi^4} \times \int^{\alpha_{max}}_{\alpha_{min}}d\alpha \Bigg\{ 2 \HH(s) - \int^{\beta_{max}}_{\beta_{min}}d\beta \FF(s) \times \frac{(3\alpha+\beta)}{\alpha} \Bigg\} \, .
\end{eqnarray}
\hrulefill
\vspace*{4pt}
\end{figure*}

In this paper we evaluate the QCD spectral density $\rho(s)$ at the leading order on $\alpha_s$ and up to dimension eight, including the perturbative term, the quark condensate $\langle \bar q q \rangle$, the gluon condensate $\langle g_s^2 GG \rangle$, the quark-gluon mixed condensate $\langle g_s \bar q \sigma G q \rangle$, and their combinations $\langle \bar q q \rangle^2$ and
$\langle \bar q q \rangle\langle g_s \bar q \sigma G q \rangle$.
The results of these spectral densities are too length, so we list them in the supplementary file ``OPE.nb''.
In the calculations we ignore the chirally suppressed terms with the light quark mass and adopt the factorization assumption
of vacuum saturation for higher dimensional condensates ($D=6$ and $D=8$).
We shall find that the $D=3$ quark condensate $\qq$ and the $D=5$ mixed condensate $\qGqb$ are both multiplied by the charm quark mass $m_c$, which are thus
important power corrections to the correlation functions.

To illustrate our numerical analysis, we use the current $\eta_{11\mu\nu}$ defined in Eq.~(\ref{def:eta11munu}) as an example. It has the quantum number $J^P = 5/2^+$ and couples to the $[N^* J/\psi]$ channel (or the $P$-wave $[p J/\psi]$ channel). Its spectral density $\rho^{[N^* J/\psi]}_{\mu\nu\rho\sigma}(s)$ are listed in Eqs.~(\ref{ope:eta11}), where $m_c$ is the heavy quark mass, and the integration limits are $\alpha_{min}=\frac{1-\sqrt{1-4m_c^2/s}}{2}$, $\alpha_{max}=\frac{1+\sqrt{1-4m_c^2/s}}{2}$, $\beta_{min}=\frac{\alpha m_c^2}{\alpha s-m_c^2}$, $\beta_{max}=1-\alpha$.
We only list the terms proportional to $\mathbf{1} \times \big( g_{\mu\rho}g_{\nu\sigma} + g_{\mu\sigma}g_{\nu\rho} \big)$ and $q\!\!\!\slash~ \times \big( g_{\mu\rho}g_{\nu\sigma} + g_{\mu\sigma}g_{\nu\rho} \big)$, and $\cdots$ denotes other Lorentz structures, such as $\mathbf{1} \times g_{\mu\rho} \sigma_{\nu\sigma}$, etc..
We find that the results are not useful, because many terms vanish in this spectral density: its $q\!\!\!\slash~ \times \big( g_{\mu\rho}g_{\nu\sigma} + g_{\mu\sigma}g_{\nu\rho} \big)$ part only contains the perturbative term, $\langle g_s^2 GG \rangle$, $\langle \bar q q \rangle^2$ and $\langle \bar q q \rangle\langle g_s \bar q \sigma G q \rangle$, but its $\mathbf{1} \times \big( g_{\mu\rho}g_{\nu\sigma} + g_{\mu\sigma}g_{\nu\rho} \big)$ part only contains $\langle \bar q q \rangle$ and $\langle g_s \bar q \sigma G q \rangle$. This makes bad OPE convergence and leads to unreliable results. Moreover, the parity can not be determined because these two parts are quite different.
We shall not use these currents to perform QCD sum rule analyses from which the parity can not be determined.

We use the current $\psi_{9\mu}$ defined in Eq.~(\ref{def:psi9mu}) as another example. It has the quantum number $J^P = 3/2^-$ and couples to the $[\Sigma_c \bar D^*]$ channel.
Its spectral density $\rho^{[\Sigma_c \bar D^*]}_{\mu\nu}(s)$ is listed in Eqs.~(\ref{ope:psi9}).
We find that the terms proportional to $\mathbf{1} \times g_{\mu\nu}$ are almost the same as those proportional to $q\!\!\!\slash~\times g_{\mu\nu}$.
Hence, the parity of $X$ can be well determined to be negative, the same as $\psi_{9\mu}$.
In the next section we shall use the terms proportional to $\mathbf{1} \times g_{\mu\nu}$ to evaluate the mass of $X$.

\section{Numerical Analysis}
\label{sec:numerical}

In this section we still use the current $\psi_{9\mu}$ as an example to perform the numerical analysis.
We use the following QCD parameters of quark masses and various condensates in our
analysis~\cite{Yang:1993bp,Agashe:2014kda,Eidemuller:2000rc,Narison:2002pw,Gimenez:2005nt,Jamin:2002ev,Ioffe:2002be,Ovchinnikov:1988gk,colangelo}:
\begin{eqnarray}
\nonumber && \langle \bar qq \rangle = - (0.24 \pm 0.01)^3 \mbox{ GeV}^3 \, ,
\\ \nonumber &&\langle g_s^2GG\rangle =(0.48 \pm 0.14) \mbox{ GeV}^4\, ,
\\ \label{paramaters} && \langle g_s \bar q \sigma G q \rangle = M_0^2 \times \langle \bar qq \rangle\, ,
\\ \nonumber && M_0^2 = - 0.8 \mbox{ GeV}^2\, ,
\\ \nonumber && m_c = 1.23 \pm 0.09 \mbox{ GeV} \, ,
\end{eqnarray}
where the running mass in the $\overline{MS}$ scheme is used for the charm quark.

\begin{figure*}[hbt]
\begin{center}
\scalebox{0.6}{\includegraphics{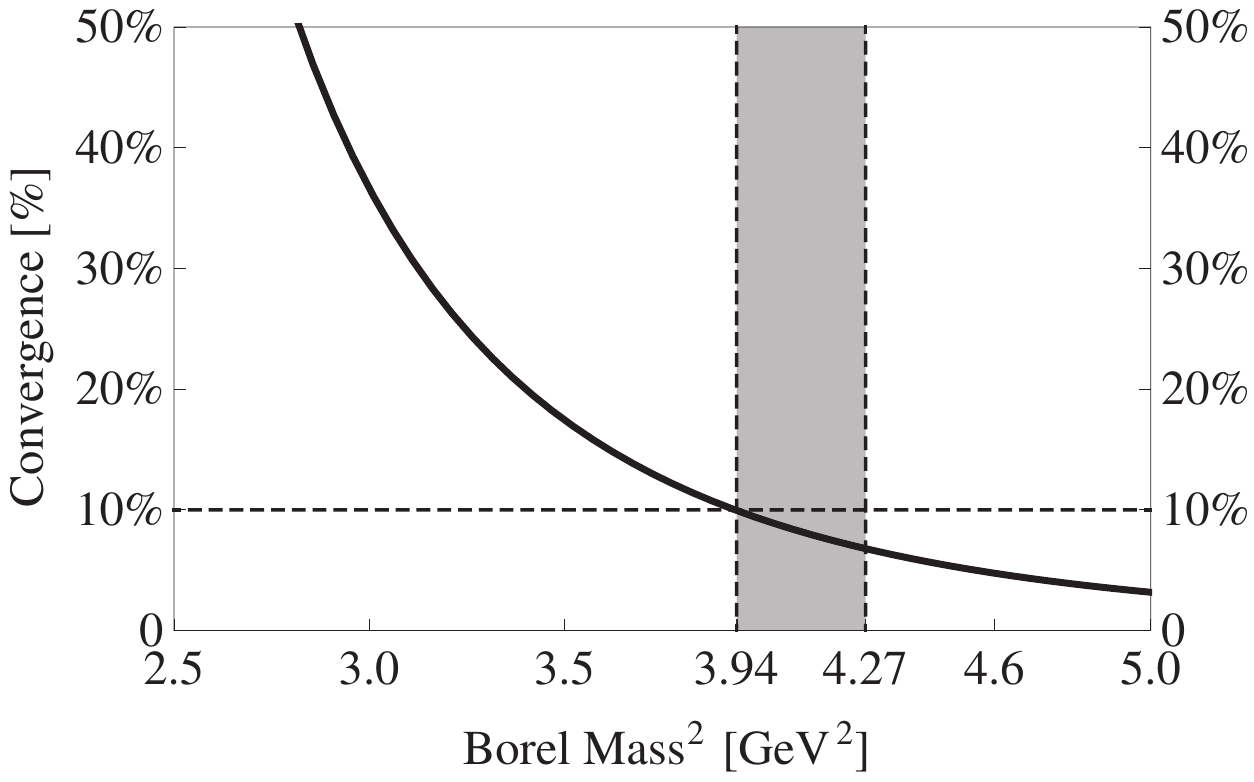}}
\scalebox{0.575}{\includegraphics{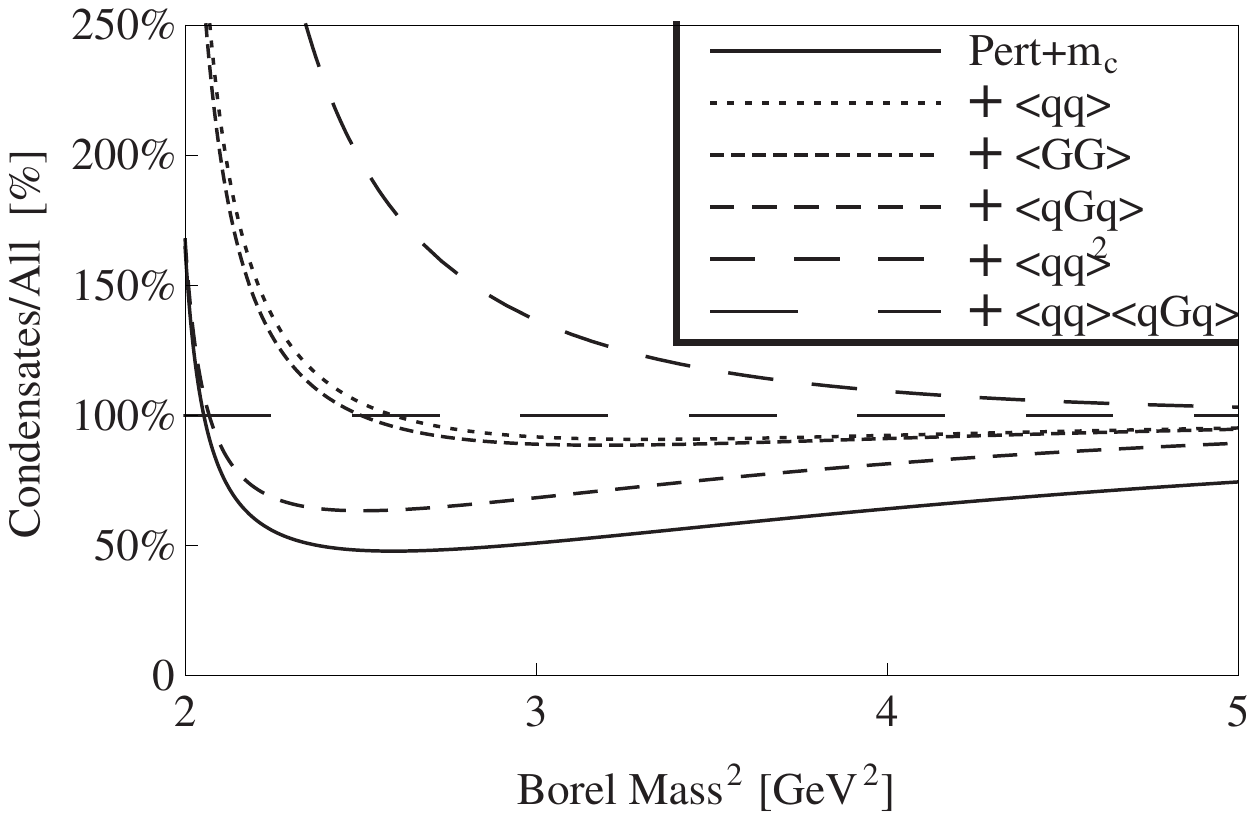}}
\caption{In the left panel we show CVG, defined in Eq.~(\ref{eq_convergence}), as a function of the Borel mass $M_B$.
In the right panel we show the relative contribution of each term on the OPE expansion, as a function of the Borel mass $M_B$.
The current $\psi_{9\mu}$ of $J^P = 3/2^-$ ($J^{\bar D^* \Sigma_c}_\mu$ in Ref.~\cite{Chen:2015moa}) is used here.}
\label{fig:cvg}
\end{center}
\end{figure*}

There are two free parameters in Eq.~(\ref{eq:mass}): the Borel mass $M_B$ and the threshold value $s_0$. We use two criteria to constrain the Borel mass $M_B$. In order to insure the convergence of the OPE series, the first criterion is to require that the dimension eight term be less than 10\% to determine its lower limit $M_B^{min}$:
%
\begin{equation}
\label{eq_convergence}
\mbox{Convergence (CVG)} \equiv \left|\frac{ \Pi_{\langle \bar q q \rangle\langle g_s \bar q \sigma G q \rangle}(\infty, M_B) }{ \Pi(\infty, M_B) }\right| \leq 10\% \, .
\end{equation}
%
We show this function obtained using $\psi_{9\mu}$ in the left panel of Fig.~\ref{fig:cvg}. We find that the OPE convergence improves with the increase of $M_B$. This criterion has a limitation
on the Borel mass that $M_B^2 \geq 3.94$ GeV$^2$.
Actually, the convergence can be even better if there is a clear trend of convergence with the higher order terms giving a progressively smaller contribution. Accordingly, we also show the relative
contribution of each term on the OPE expansion in the right panel of Fig.~\ref{fig:cvg}. We find that a good convergence can be achieved in the same region $M_B^2 \geq 3.94$ GeV$^2$.
While, we shall still use the previous criterion to determine the lower limit of the Borel mass, which can be applied more easily.

\begin{figure*}[hbt]
\begin{center}
\scalebox{0.6}{\includegraphics{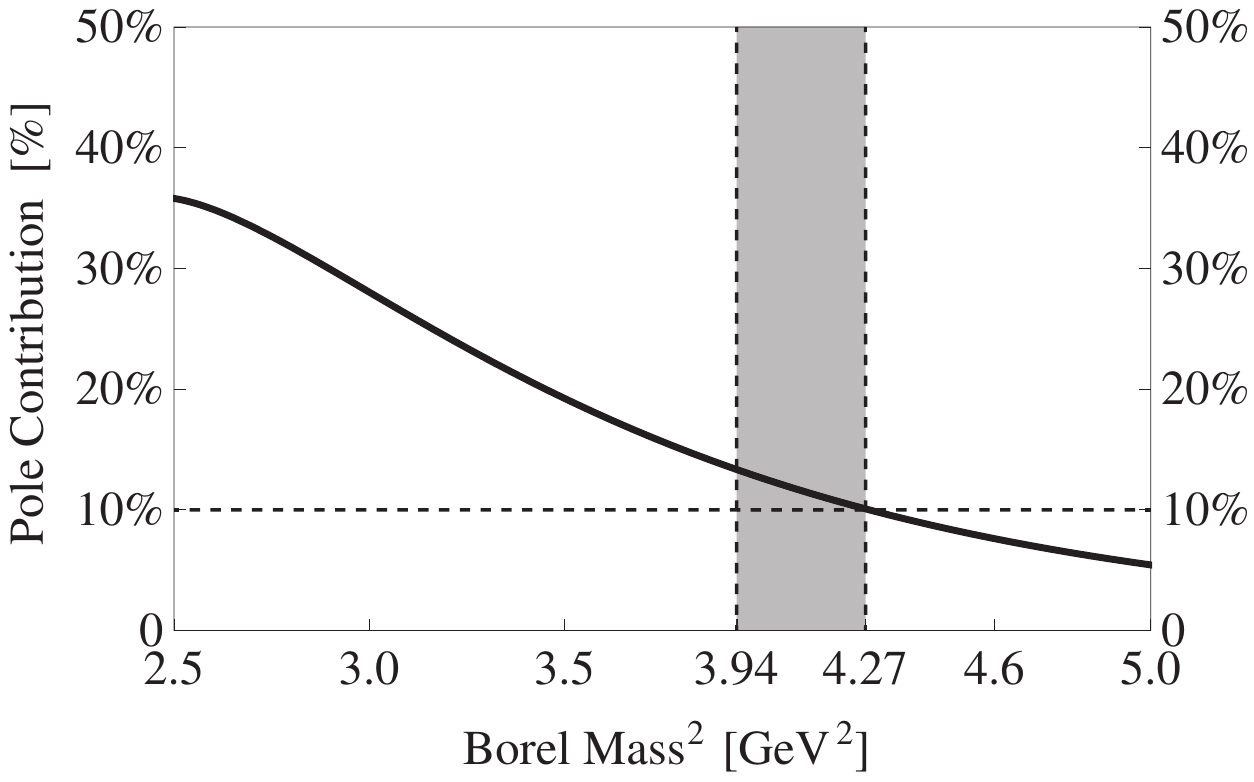}}
\caption{The variation of PC, defined in Eq.~(\ref{eq_pole}), as a function of the Borel mass $M_B$.
The current $\psi_{9\mu}$ of $J^P = 3/2^-$ ($J^{\bar D^* \Sigma_c}_\mu$ in Ref.~\cite{Chen:2015moa}) is used here and the threshold value is chosen to be $s_0$ = 21 GeV$^2$.}
\label{fig:pole}
\end{center}
\end{figure*}

While, to insure that the one-pole parametrization in Eq.~(\ref{eq:rho}) is valid, the second criterion is to require that the pole contribution (PC) be larger than 10\% to determine the upper limit on $M_B^2$:
%
\begin{equation}
\label{eq_pole} \mbox{PC} \equiv \frac{ \Pi(s_0, M_B) }{ \Pi(\infty, M_B) } \geq 10\% \, .
\end{equation}
%
We show the variation of the pole contribution obtained using $\psi_{9\mu}$ in Fig.~\ref{fig:pole}, with respect to the Borel mass $M_B$ and when $s_0$ is chosen to be 21 GeV$^2$.
We find that the PC decreases with the increase of $M_B$. This criterion has a limitation on the Borel mass that $M_B^2 \leq 4.27$ GeV$^2$.
Together we obtain the working region of Borel mass $3.94$ GeV$^2< M_B^2 < 4.27$ GeV$^2$ for the current $\psi_{9\mu}$ with the continuum threshold $s_0 = 21$ GeV$^2$.
The most important drawback of this current, as well as other pentaquark currents used in this paper, is that
their pole contributions are small. One reason is that the two $P_c$'s poles are both mixed with the $J/\psi p$ continuum, so the continuum contribution may not be well suppressed by the Borel transformation.
Another reason is due to the large powers of $s$ in the spectral function; see other sum rule analyses for the six-quark state $d^*(2380)$~\cite{Chen:2014vha} and the $F$-wave heavy mesons~\cite{Zhou:2015ywa}.

Anyway, the Borel mass $M_B$ is just one of the two free parameters. We should also pay attention to the other one, that is the threshold value $s_0$,
and try to find a balance between them.
Actually, we can increase $s_0$ to achieve a large enough pole contribution (then the obtained mass would also be increased and not so reasonable),
but there is another criterion more important, that is the $s_0$ stability.
We note that in Ref.~\cite{Narison:1996fm} the author used the requirements on $M_B$ (called the $\tau$ stability) to extract the upper bound on the $0^{++}$ glueball mass,
and used the requirement on $s_0$ (called the $t_c$ stability) to evaluate its optimal mass. In this paper we shall use a similar requirement on $s_0$, as discussed in the following.

\begin{figure*}[hbt]
\begin{center}
\scalebox{0.6}{\includegraphics{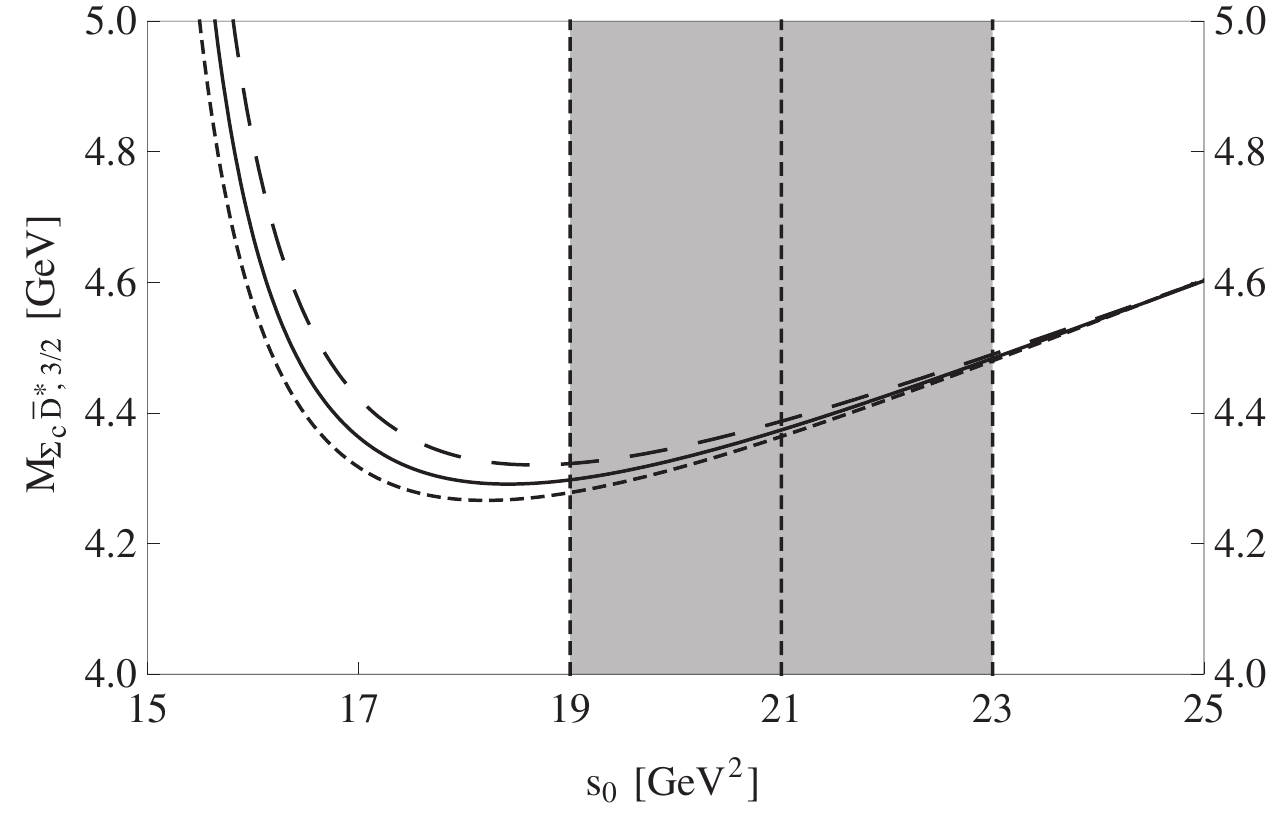}}
\scalebox{0.6}{\includegraphics{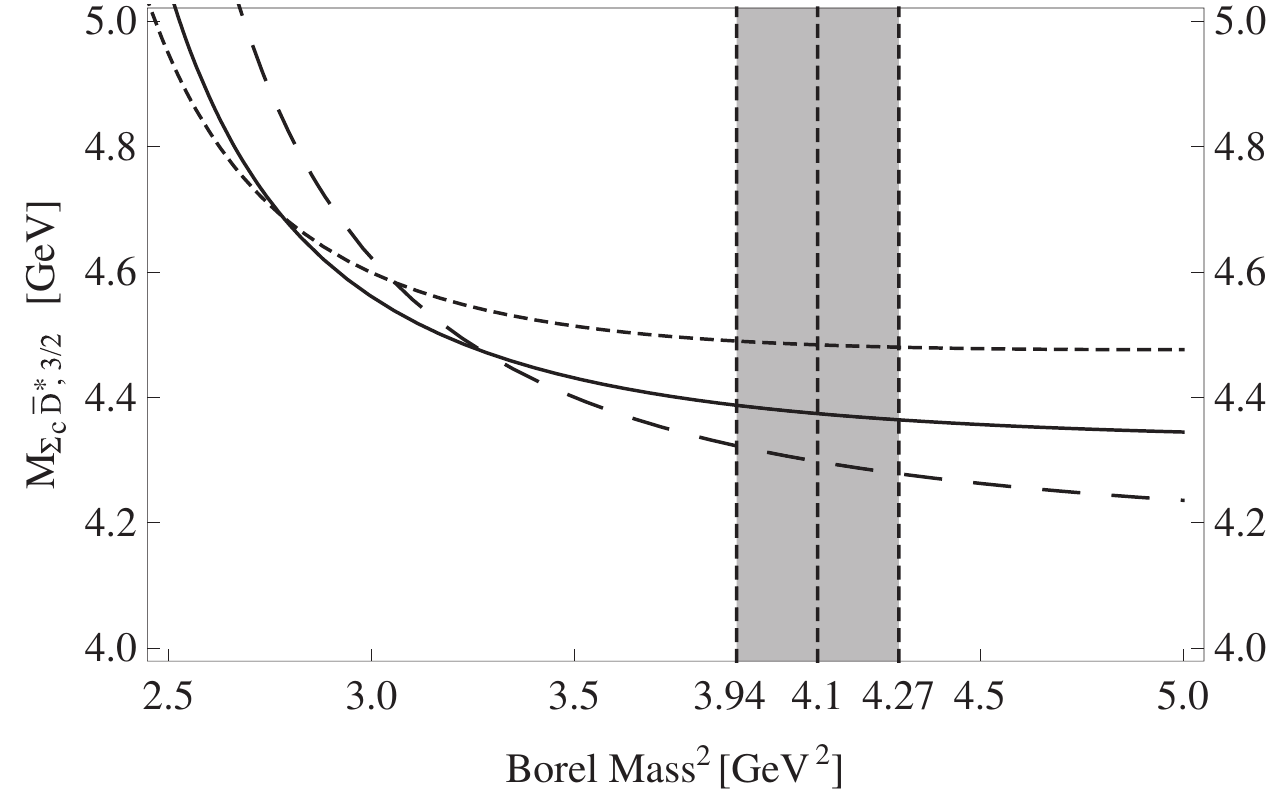}}
\caption{The variation of $M_{[\Sigma_c \bar D^*],3/2^-}$ with respect to the threshold value $s_0$ (left) and the Borel mass $M_B$ (right),
calculated using the current ($J^{\bar D^* \Sigma_c}_\mu$ in Ref.~\cite{Chen:2015moa}) $\psi_{9\mu}$ of $J^P = 3/2^-$.
In the left figure, the long-dashed, solid and short-dashed curves are obtained by fixing $M_B^2 = 3.9$, $4.1$ and $4.3$ GeV$^2$, respectively.
In the right figure, the long-dashed, solid and short-dashed curves are obtained for $s_0 = 19$, $21$ and $23$ GeV$^2$, respectively.}
\label{fig:psi9}
\end{center}
\end{figure*}

To determine $s_0$, we require that both the $s_0$ dependence and the $M_B$ dependence of the mass prediction be the weakest in order to obtain reliable mass prediction.
We show the variation of $M_X$ with respect to the threshold value $s_0$ in the left panel of Fig.~\ref{fig:psi9}, in a large region 15 GeV$^2 < s_0 < 25$ GeV$^2$.
We find that the mass curves have a minimum against $s_0$ when $s_0$ is around 18 GeV$^2$. Hence, the $s_0$ dependence of the mass prediction is the weakest at this point.
However, this value of $s_0$ is too small to give a reasonable working region of $M_B$. A working region can be obtained as long as $s_0>19$ GeV$^2$.
Consequently, we choose the region $19$ GeV$^2\leq s_0\leq 23$ GeV$^2$ as our working region, where the $s_0$ dependence is still weak and the mass curves are flat enough.

Hence, our working regions for the current $\psi_{9\mu}$ are $19$ GeV$^2\leq s_0\leq 23$ GeV$^2$ and $3.94$ GeV$^2\leq M_B^2 \leq 4.27$ GeV$^2$, where the following numerical results can be obtained~\cite{Chen:2015moa}:
\begin{eqnarray}
M_{[\Sigma_c \bar D^*]} = 4.37^{+0.18}_{-0.13} \mbox{ GeV} \, .
\end{eqnarray}
Here the central value corresponds to $M_B=4.10$ GeV$^2$ and $s_0 = 21$ GeV$^2$, and the uncertainty comes
from the Borel mass $M_B$, the threshold value $s_0$, the charm quark mass and the various condensates~\cite{Chen:2015ata}.
We also show the variation of $M_X$ with respect to the Borel mass $M_B$ in the right panel of Fig.~\ref{fig:psi9}, in a broader region $2.5$ GeV$^2\leq M_B^2 \leq 5.0$ GeV$^2$.
We find that these curves are more stable inside the Borel window $3.94$ GeV$^2\leq M_B^2 \leq 4.27$ GeV$^2$.
We note that the threshold value used here, $\sqrt{s_0} \approx 4.58$ GeV, is not far from the obtained mass of 4.37 GeV (but still acceptable), indicating it is not easy to separate the pole and the continuum.

As we have found in Sec.~\ref{sec:sumrule} for the current $\psi_{9\mu}$ that the terms proportional to $q\!\!\!\slash~\times g_{\mu\nu}$ are quite similar to those proportional to $\mathbf{1} \times g_{\mu\nu}$,
suggesting that $X$ has the same parity as $\psi_{9\mu}$, that is negative~\cite{Chen:2015moa}:
\begin{eqnarray}
M_{[\Sigma_c \bar D^*],3/2^-} = 4.37^{+0.18}_{-0.13} \mbox{ GeV} \, .
\label{Pc4380}
\end{eqnarray}
This value is consistent with the experimental mass of the $P_c(4380)$~\cite{lhcb}, and supports it as a $[\Sigma_c \bar D^*]$ hidden-charm pentaquark with the quantum number $J^P=3/2^-$.

\section{Results and discussions}
\label{sec:summary}

\renewcommand{\arraystretch}{1.5}
\begin{table*}[hbtp]
\begin{center}
\caption{Numerical results for the spin $J=1/2$ hidden-charm pentaquark states.}
\begin{tabular}{ccc|cc|cc}
\toprule[1pt]\toprule[1pt]
~~\mbox{Current}~~ & ~~\mbox{Defined in}~~ & ~~\mbox{Structure}~~ & \mbox{$s_0$ [GeV$^2$]} & \mbox{Borel Mass [GeV$^2$]} & ~~\mbox{Mass [GeV]}~~ & ~~\mbox{($J$, $P$)}~~
\\ \midrule[1pt]
$\eta_2 - \eta_4$ & Eq.~(\ref{def:eta24}) & $[p \eta_c]$             & -- & -- & -- & --
\\
$\eta_5 - \eta_7$ & Eq.~(\ref{def:eta57}) & $[p J/\psi]$             & -- & -- & -- & --
\\
$\eta_{13}$       & Eq.~(\ref{def:eta13}) & $[N^* J/\psi]$           & -- & -- & -- & --
\\ \midrule[1pt]
$\xi_2 - \xi_4$   & Eq.~(\ref{def:xi24})  & $[\Lambda_c \bar D]$     & -- & -- & -- & --
\\
$\xi_5 - \xi_7$   & Eq.~(\ref{def:xi57})  & $[\Lambda_c \bar D^*]$   & -- & -- & -- & --
\\
$\xi_{14}$        & Eq.~(\ref{def:xi14})  & $[\Sigma_c \bar D]$      & $20 - 24$ & $4.12 - 4.52$ & $4.45^{+0.17}_{-0.13}$ & ($1/2,-$)
\\
$\xi_{16}$        & Eq.~(\ref{def:xi16})  & $[\Lambda_c^* \bar D]$   & $25 - 29$ & $4.40 - 4.76$ & $4.86^{+0.16}_{-0.19}$ & ($1/2,+$)
\\
$\xi_{17}$        & Eq.~(\ref{def:xi17})  & $[\Sigma_c^* \bar D^*]$  & $22 - 26$ & $3.64 - 4.25$ & $4.73^{+0.19}_{-0.12}$ & ($1/2,-$)
\\
$\xi_{19}$        & Eq.~(\ref{def:xi19})  & $[\Lambda_c^* \bar D^*]$ & $23 - 27$ & $3.70 - 4.22$ & $4.67^{+0.16}_{-0.20}$ & ($1/2,+$)
\\ \midrule[1pt]
$\psi_2$          & Eq.~(\ref{def:psi2})  & $[\Sigma_c^* \bar D]$    & $19 - 23$ & $3.95 - 4.47$ & $4.33^{+0.17}_{-0.13}$ & ($1/2,-$)
\\
$\psi_3$          & Eq.~(\ref{def:psi3})  & $[\Sigma_c^* \bar D^*]$  & $21 - 25$ & $3.50 - 4.11$ & $4.59^{+0.17}_{-0.12}$ & ($1/2,-$)
\\ \bottomrule[1pt]\bottomrule[1pt]
\end{tabular}
\label{tab:spin12}
\end{center}
\end{table*}

\renewcommand{\arraystretch}{1.5}
\begin{table*}[hbtp]
\begin{center}
\caption{Numerical results for the spin $J=3/2$ hidden-charm pentaquark states. $\psi_{9\mu}$ is denoted as $J^{\bar D^* \Sigma_c}_\mu$ in Ref.~\cite{Chen:2015moa}.}
\begin{tabular}{ccc|cc|cc}
\toprule[1pt]\toprule[1pt]
~~\mbox{Current}~~ & ~~\mbox{Defined in}~~ & ~~\mbox{Structure}~~ & \mbox{$s_0$ [GeV$^2$]} & \mbox{Borel Mass [GeV$^2$]} & ~~\mbox{Mass [GeV]}~~ & ~~\mbox{($J$, $P$)}~~
\\ \midrule[1pt]
$\eta_{5\mu} - \eta_{7\mu}$ & Eq.~(\ref{def:eta57mu}) & $[p J/\psi]$             & -- & -- & -- & --
\\
$\eta_{18\mu}$              & Eq.~(\ref{def:eta18mu}) & $[N^* \eta_c]$           & -- & -- & -- & --
\\
$\eta_{19\mu}$              & Eq.~(\ref{def:eta19mu}) & $[N^* J/\psi]$           & -- & -- & -- & --
\\ \midrule[1pt]
$\xi_{5\mu} - \xi_{7\mu}$   & Eq.~(\ref{def:xi57mu})  & $[\Lambda_c \bar D^*]$   & -- & -- & -- & --
\\
$\xi_{18\mu}$               & Eq.~(\ref{def:xi18mu})  & $[\Sigma_c^* \bar D]$    & $21 - 25$ & $3.93 - 4.51$ & $4.56^{+0.16}_{-0.13}$ & ($3/2,-$)
\\
$\xi_{20\mu}$               & Eq.~(\ref{def:xi20mu})  & $[\Lambda_c^* \bar D]$   & $23 - 27$ & $4.12 - 4.63$ & $4.56^{+0.18}_{-0.22}$ & ($3/2,+$)
\\
$\xi_{25\mu}$               & Eq.~(\ref{def:xi25mu})  & $[\Sigma_c^* \bar D^*]$  & $21 - 25$ & $3.85 - 4.30$ & $4.67^{+0.21}_{-0.12}$ & ($3/2,-$)
\\
$\xi_{27\mu}$               & Eq.~(\ref{def:xi27mu})  & $[\Lambda_c^* \bar D^*]$ & $23 - 27$ & $4.07 - 4.50$ & $4.68^{+0.15}_{-0.18}$ & ($3/2,+$)
\\
$\xi_{33\mu}$               & Eq.~(\ref{def:xi33mu})  & $[\Sigma_c \bar D^*]$    & $20 - 24$ & $3.97 - 4.41$ & $4.46^{+0.18}_{-0.13}$ & ($3/2,-$)
\\
$\xi_{35\mu}$               & Eq.~(\ref{def:xi35mu})  & $[\Lambda_c \bar D^*]$   & $27 - 31$ & $4.32 - 5.11$ & $5.18^{+0.16}_{-0.12}$ & ($3/2,+$)
\\ \midrule[1pt]
$\psi_{2\mu}$               & Eq.~(\ref{def:psi2mu})  & $[\Sigma_c^* \bar D]$    & $20 - 24$ & $3.88 - 4.41$ & $4.45^{+0.16}_{-0.13}$ & ($3/2,-$)
\\
$\psi_{5\mu}$               & Eq.~(\ref{def:psi5mu})  & $[\Sigma_c^* \bar D^*]$  & $21 - 25$ & $3.86 - 4.46$ & $4.61^{+0.18}_{-0.12}$ & ($3/2,-$)
\\
$\psi_{9\mu}$               & Eq.~(\ref{def:psi9mu})  & $[\Sigma_c \bar D^*]$    & $19 - 23$ & $3.94 - 4.27$ & $4.37^{+0.18}_{-0.13}$ & ($3/2,-$)
\\ \bottomrule[1pt]\bottomrule[1pt]
\end{tabular}
\label{tab:spin32}
\end{center}
\end{table*}

\renewcommand{\arraystretch}{1.5}
\begin{table*}[hbtp]
\begin{center}
\caption{Numerical results for the spin $J=5/2$ hidden-charm pentaquark states. $\xi_{15\mu\nu}$, $\psi_{4\mu\nu}$ and $J^{\rm mix}_{\mu\nu}$ are denoted as $J^{\bar D^* \Lambda_c}_{\{\mu\nu\}}$, $J^{\bar D \Sigma_c^*}_{\{\mu\nu\}}$ and $J^{\bar D \Sigma_c^*\&\bar D^* \Lambda_c}_{\{\mu\nu\}}$ in Ref.~\cite{Chen:2015moa}, respectively.}
\begin{tabular}{ccc|cc|cc}
\toprule[1pt]\toprule[1pt]
~~\mbox{Current}~~ & ~~\mbox{Defined in}~~ & ~~\mbox{Structure}~~ & \mbox{$s_0$ [GeV$^2$]} & \mbox{Borel Mass [GeV$^2$]} & ~~\mbox{Mass [GeV]}~~ & ~~\mbox{($J$, $P$)}~~
\\ \midrule[1pt]
$\eta_{11\mu\nu}$    & Eq.~(\ref{def:eta11munu}) & $[N^* J/\psi]$           & -- & -- & -- & --
\\ \midrule[1pt]
$\xi_{13\mu\nu}$     & Eq.~(\ref{def:xi13munu})  & $[\Sigma_c^* \bar D^*]$  & $20 - 24$ & $3.51 - 4.00$ & $4.50^{+0.18}_{-0.12}$ & ($5/2,-$)
\\
$\xi_{15\mu\nu}$     & Eq.~(\ref{def:xi15munu})  & $[\Lambda_c^* \bar D^*]$ & $24 - 28$ & $4.09 - 4.59$ & $4.76^{+0.15}_{-0.19}$ & ($5/2,+$)
\\ \midrule[1pt]
$\psi_{3\mu\nu}$     & Eq.~(\ref{def:psi3munu})  & $[\Sigma_c^* \bar D^*]$  & $21 - 25$ & $3.88 - 4.40$ & $4.59^{+0.17}_{-0.12}$ & ($5/2,-$)
\\ \midrule[1pt]
$\psi_{4\mu\nu}$     & Eq.~(\ref{def:psi4munu})  & $P$-wave $[\Sigma_c^* \bar D]$  & $25 - 29$ & $4.30 - 4.73$ & $4.82^{+0.15}_{-0.14}$ & ($5/2,+$)
\\ \midrule[1pt]
$J^{\rm mix}_{\mu\nu}$ & Eq.~(\ref{def:mix}) & $P$-wave $[\Lambda_c \bar D^* \& \Sigma_c^* \bar D]$  & $20 - 24$ & $3.22 - 3.50$ & $4.47^{+0.18}_{-0.13}$ & ($5/2,+$)
\\ \bottomrule[1pt]\bottomrule[1pt]
\end{tabular}
\label{tab:spin52}
\end{center}
\end{table*}

We use the currents selected in Sec.~\ref{sec:current} and in Appendixes~\ref{app:spin32} and \ref{app:spin52} to perform QCD sum rule analyses.
Some of them lead to the OPE series from which the parity can be well determined. We further use these currents to perform numerical analyses. The masses obtained using the $J^P=1/2^+$ currents $\xi_{14}$, $\xi_{16}$, $\xi_{17}$, $\xi_{19}$, $\psi_2$, and $\psi_3$ are listed in Table~\ref{tab:spin12}, those obtained using the $J^P=3/2^-$ currents $\xi_{18\mu}$, $\xi_{20\mu}$, $\xi_{25\mu}$, $\xi_{27\mu}$, $\xi_{33\mu}$, $\xi_{35\mu}$, $\psi_{2\mu}$, $\psi_{5\mu}$, and $\psi_{9\mu}$ are listed in Table~\ref{tab:spin32}, and those obtained using the $J^P=5/2^+$ currents $\xi_{13\mu\nu}$, $\xi_{15\mu\nu}$, and $\psi_{3\mu\nu}$ are listed in Table~\ref{tab:spin52}.

\begin{figure*}[hbt]
\begin{center}
\scalebox{0.6}{\includegraphics{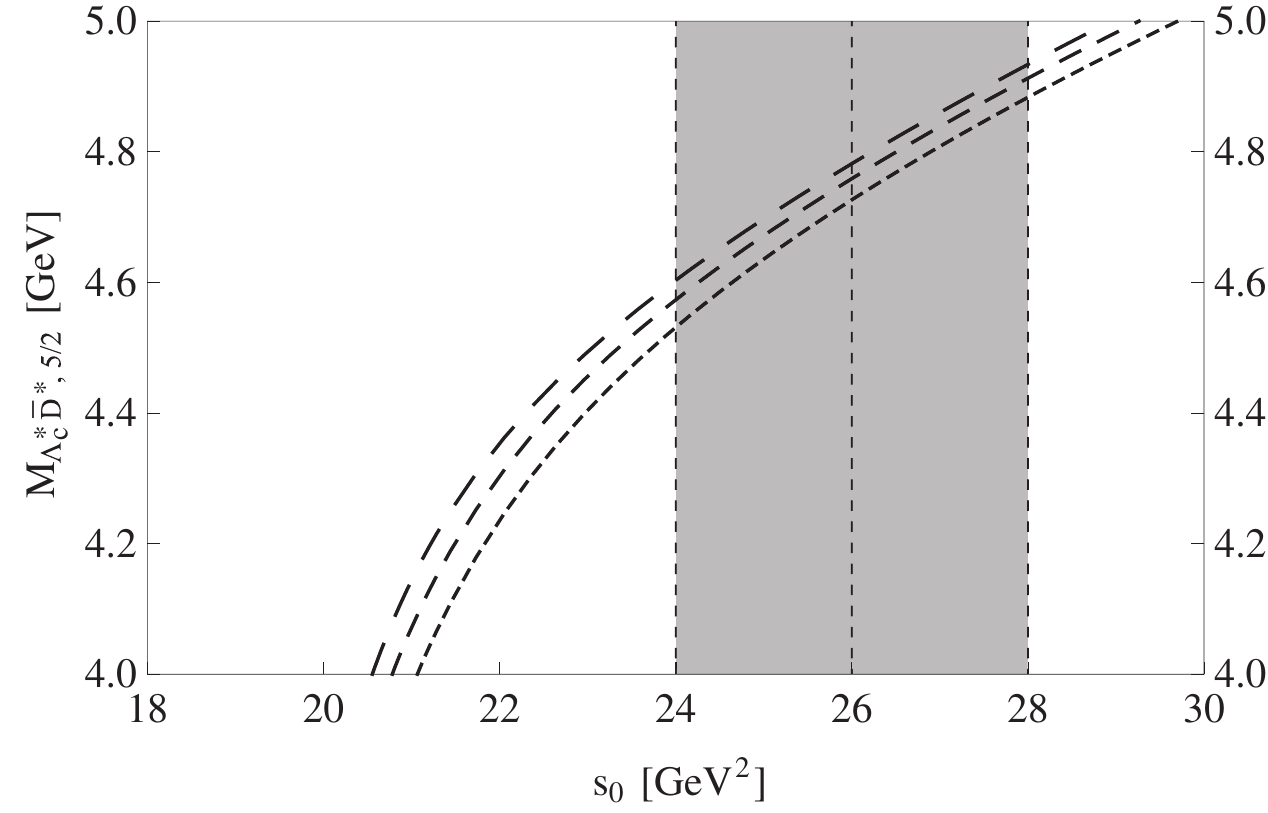}}
\scalebox{0.6}{\includegraphics{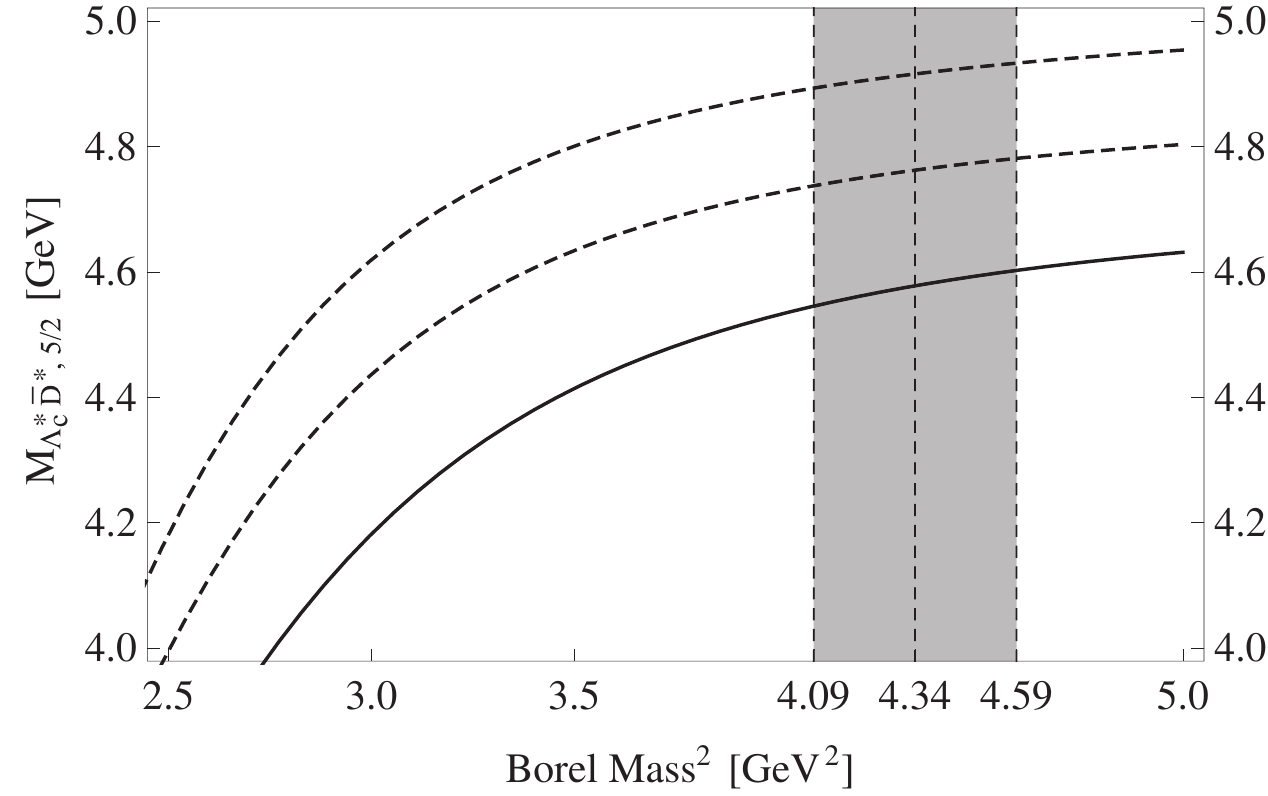}}
\caption{The variation of $M_{[\Lambda_c^* \bar D^*],5/2^+}$ with respect to the threshold value $s_0$ (left) and the Borel mass $M_B$ (right),
calculated using the current $\xi_{15\mu\nu}$ ($J^{\bar D^* \Lambda_c}_{\{\mu\nu\}}$ in Ref.~\cite{Chen:2015moa}) of $J^P = 5/2^+$.
In the left figure, the long-dashed, solid and short-dashed curves are obtained by fixing $M_B^2 = 4.0$, $4.3$ and $4.6$ GeV$^2$, respectively.
In the right figure, the long-dashed, solid and short-dashed curves are obtained for $s_0 = 24$, $26$ and $28$ GeV$^2$, respectively.}
\label{fig:xi15}
\end{center}
\end{figure*}

\begin{figure*}[hbt]
\begin{center}
\scalebox{0.6}{\includegraphics{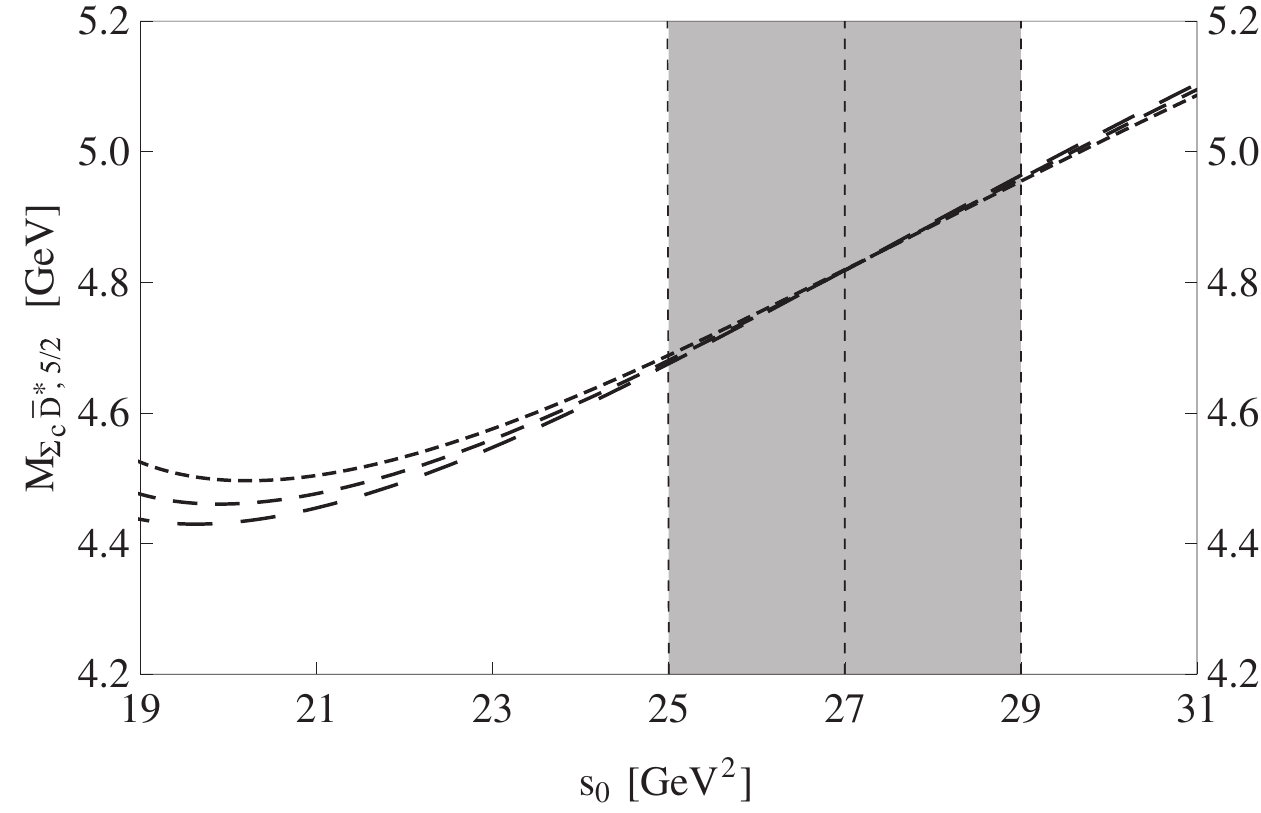}}
\scalebox{0.6}{\includegraphics{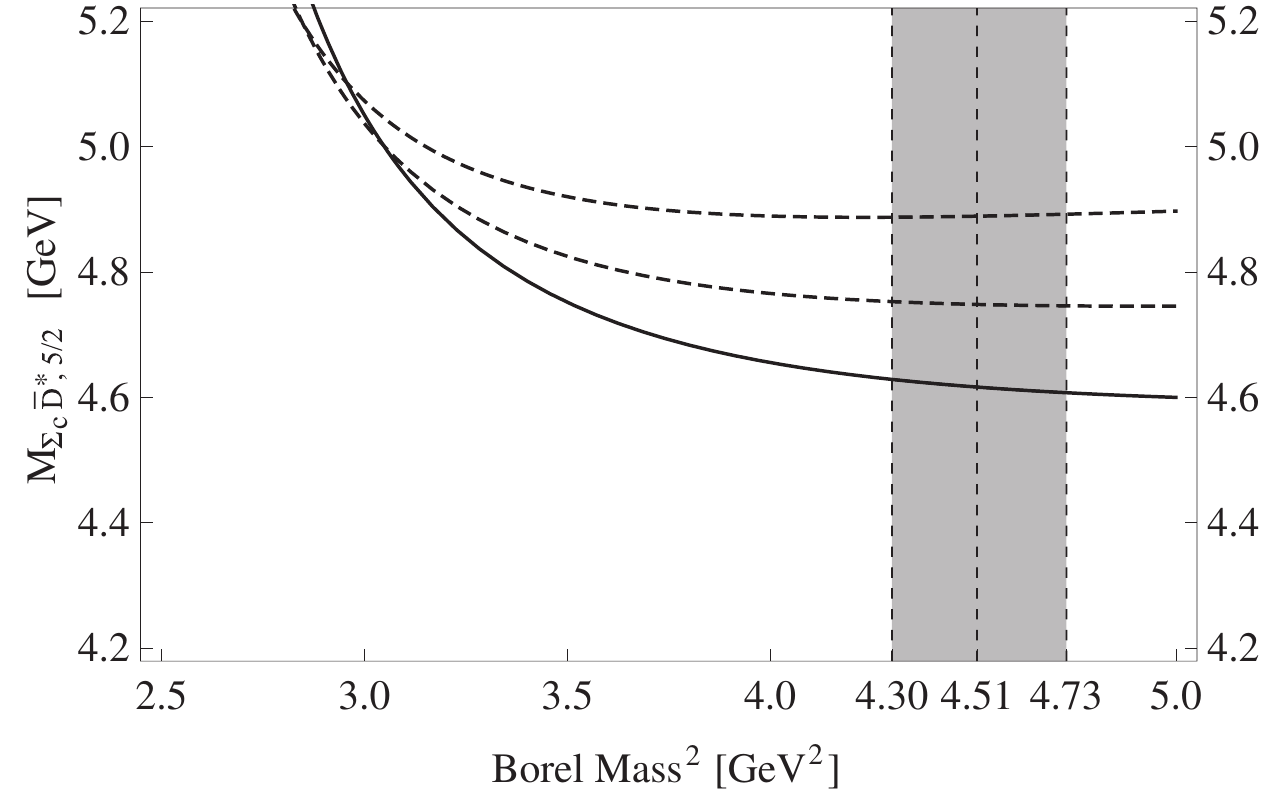}}
\caption{The variations of $M_{[\Sigma_c^* \bar D],5/2^+}$ with respect to the threshold value $s_0$ (left) and the Borel mass $M_B$ (right),
calculated using the current $\psi_{4\mu\nu}$ ($J^{\bar D \Sigma_c^*}_{\{\mu\nu\}}$ in Ref.~\cite{Chen:2015moa}) of $J^P = 5/2^+$.
In the left figure, the long-dashed, solid and short-dashed curves are obtained by fixing $M_B^2 = 4.3$, $4.5$ and $4.7$ GeV$^2$, respectively.
In the right figure, the long-dashed, solid and short-dashed curves are obtained for $s_0 = 25$, $27$ and $29$ GeV$^2$, respectively.}
\label{fig:psi4}
\end{center}
\end{figure*}

\begin{figure*}[hbt]
\begin{center}
\scalebox{0.6}{\includegraphics{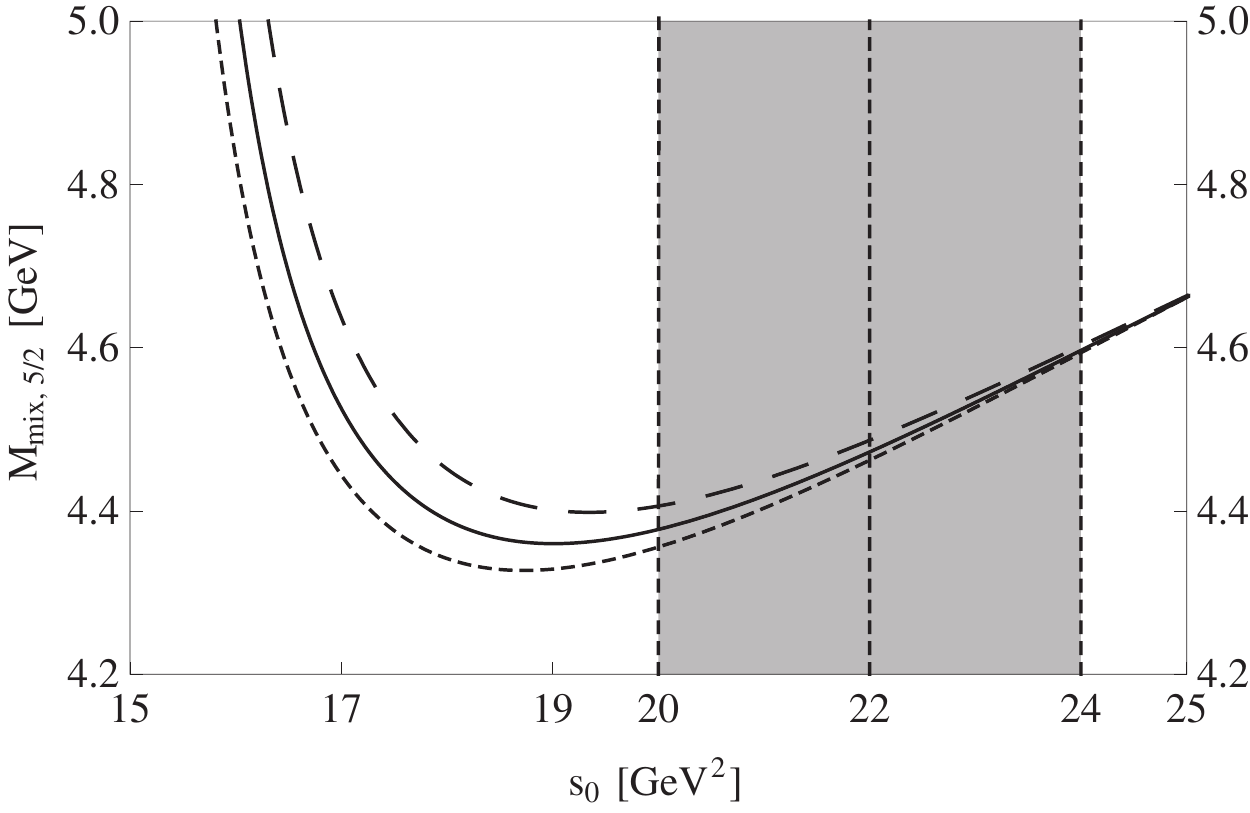}}
\scalebox{0.6}{\includegraphics{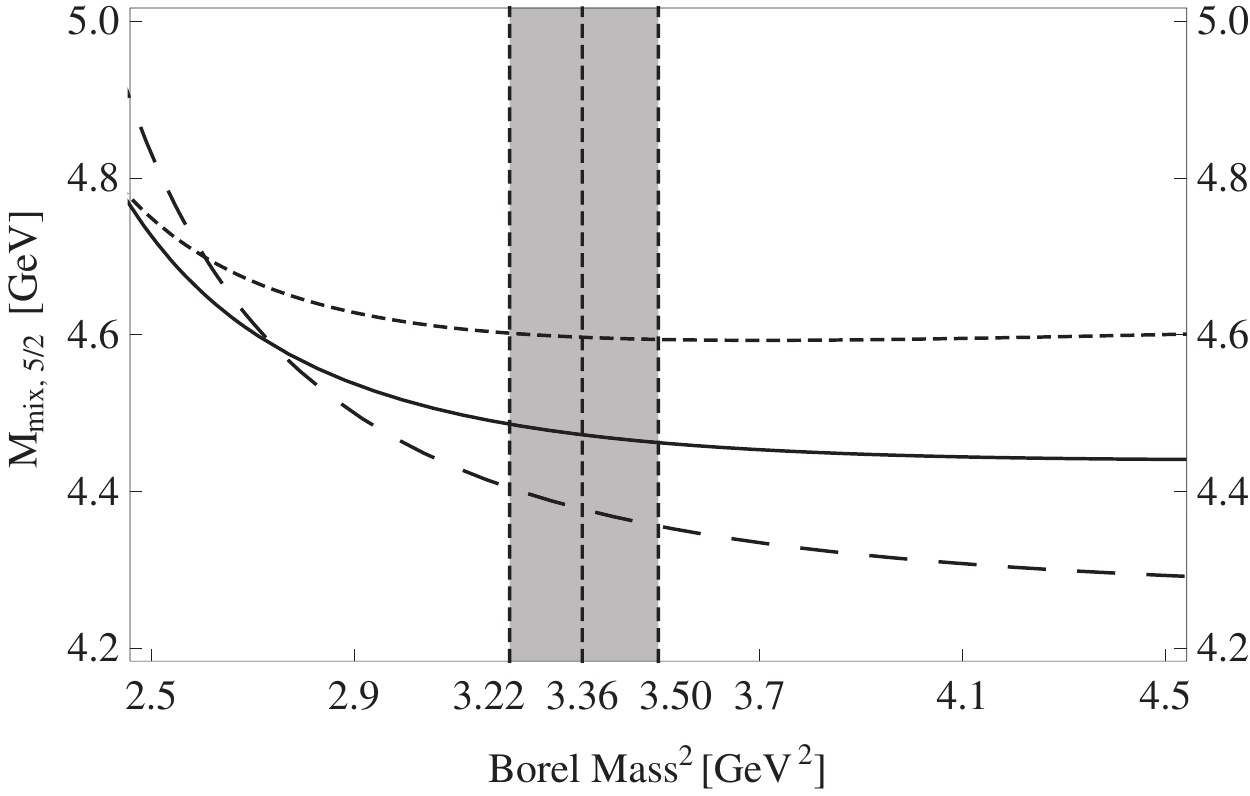}}
\caption{The variations of $M_{{\rm mix},5/2^+}$ with respect to the threshold value $s_0$ (left) and the Borel mass $M_B$ (right),
calculated using the mixied current $J^{\rm mix}_{\mu\nu}$ ($J^{\bar D\Sigma_c^*\&\bar D^*\Lambda_c}_{\{\mu\nu\}}$ in Ref.~\cite{Chen:2015moa}) of $J^P = 5/2^+$.
In the left figure, the long-dashed, solid and short-dashed curves are obtained by fixing $M_B^2 = 3.2$, $3.35$ and $3.5$ GeV$^2$, respectively.
In the right figure, the long-dashed, solid and short-dashed curves are obtained for $s_0 = 20$, $22$ and $24$ GeV$^2$, respectively.}
\label{fig:mix}
\end{center}
\end{figure*}

Especially, in the previous sections we have used the current $\psi_{9\mu}$ and obtained the mass $M_{[\Sigma_c \bar D^*],3/2^-} = 4.37^{+0.18}_{-0.13}$ GeV~\cite{Chen:2015moa}. This value is consistent with the experimental mass of the $P_c(4380)$~\cite{lhcb}, and supports it as a $[\Sigma_c \bar D^*]$ hidden-charm pentaquark with the quantum number $J^P=3/2^-$.

We also use the current $\xi_{15\mu\nu}$ and obtain the mass $M_{[\Lambda_c^* \bar D^*],5/2^+} = 4.76^{+0.15}_{-0.19}$ GeV. This value is significantly larger than the experimental mass of the $P_c(4450)$~\cite{lhcb}. Moreover, we show the mass as a function of the threshold value $s_0$ in the left panel of Fig.~\ref{fig:xi15}, and find that the mass curves do not have a minimum against $s_0$, which is quite different from the current $\psi_{9\mu}$. We also show the mass as a function of the Borel mass $M_B$ in the right panel of Fig.~\ref{fig:xi15}.

To find a good solution consistent with the experiment, we found the mixed current consisting of $\xi_{15\mu\nu}$ and $\psi_{4\mu\nu}$~\cite{Chen:2015moa}:
\begin{eqnarray}
J^{\rm mix}_{\mu\nu} = \cos\theta \times \xi_{15\mu\nu} + \sin\theta \times \psi_{4\mu\nu} \, .
\label{def:mix}
\end{eqnarray}
We note that $\psi_{4\mu\nu}$ is defined in Eq.~(\ref{def:psi4munu}) and well couples to the $P$-wave $[\Sigma_c^* \bar D]$ channel. However, it contains the axial-vector ($\bar c_d \gamma_\mu \gamma_5 c_d$) component, so not our first choice in this paper. We show the mass obtained using this current $\psi_{4\mu\nu}$, as a function of the threshold value $s_0$ and the Borel mass $M_B$ in Fig.~\ref{fig:psi4}. We find that the mass curves have a minimum against $s_0$  when $s_0$ is around 20 GeV$^2$. Moreover, this mass minimum is just around 4.45 GeV, similar to the mass of the $P_c(4450)$~\cite{lhcb}. However, a working region can be obtained as long as $s_0>25$ GeV$^2$, in which region the mass prediction is $4.82^{+0.15}_{-0.14}$ GeV, significantly larger than the mass of the $P_c(4450)$~\cite{lhcb}.

To solve this problem, we further use the mixed current $J^{\rm mix}_{\mu\nu}$ to perform QCD sum rule analysis. We find that it gives a reliable mass sum rule, when the mixing angle $\theta$ is fine-tuned to be $-51\pm5^\circ$, and the hadron mass is extracted as~\cite{Chen:2015moa}
\begin{eqnarray}
 M_{{\rm mix},{5/2^+}} = 4.47^{+0.18}_{-0.13} \mbox{ GeV} \, ,
 \label{Pc4450}
\end{eqnarray}
with $20$ GeV$^2$ $\leq s_0 \leq 24$ GeV$^2$ and $3.22$ GeV$^2$ $\leq M_B^2 \leq 3.50$ GeV$^2$. This value is consistent with the experimental mass of the $P_c(4450)$~\cite{lhcb}, and supports it as an admixture of $P$-wave $[\Lambda_c\bar D^*]$ and $[\Sigma_c^* \bar D]$ with the quantum number $J^P=5/2^+$. We show the mass as a function of the threshold value $s_0$ and the Borel mass $M_B$ in Fig.~\ref{fig:mix}.

In summary, in this paper we adopt the QCD sum rule approach to study the mass spectrum of hidden-charm pentaquarks.
We systematically construct the local pentaquark interpolating currents having spin $J = {1\over2}/{3\over2}/{5\over2}$ and quark contents $uud c \bar c$, and select those currents containing pseudoscalar ($\bar c_d \gamma_5 c_d$) and vector ($\bar c_d \gamma_\mu c_d$) components to perform QCD sum rule analyses. We find some of them lead to the OPE series from which the parity can be well determined, and further use these currents to perform numerical analyses. The results are listed in Tables~\ref{tab:spin12}, \ref{tab:spin32}, and \ref{tab:spin52}.

We find that the $P_c(4380)$ and $P_c(4450)$ can be identified as hidden-charm pentaquark states composed of an anti-charmed meson and a charmed baryon.
We use $\psi_{9\mu}$ to perform QCD sum rule analysis and the result supports the $P_c(4380)$ as a $S$-wave $[\Sigma_c\bar D^*]$ hidden-charm pentaquark with the quantum number $J^P=3/2^-$. We use the mixed current $J^{\rm mix}_{\mu\nu}$ to perform QCD sum rule analysis, and the result supports the $P_c(4450)$ as an admixture of $P$-wave $[\Lambda_c \bar D^*]$ and $[\Sigma_c^* \bar D]$ with the quantum number $J^P=5/2^+$.
Besides them, our results suggest that
\begin{enumerate}

\item The lowest-lying hidden-charm pentaquark state of $J^P = 1/2^-$ has the mass $4.33^{+0.17}_{-0.13}$ GeV.
This result is obtained by using the current $\psi_2$, which is defined in Eq.~(\ref{def:psi2}), and it well couples to the $S$-wave $[\Sigma_c^* \bar D]$ channel.
While, the one of $J^P = 1/2^+$ is significantly higher, that is around $4.7-4.9$ GeV;

\item The lowest-lying hidden-charm pentaquark state of $J^P = 3/2^-$ has the mass $4.37^{+0.18}_{-0.13}$ GeV, consistent with the experimental mass of the $P_c(4380)$ of $J^P = 3/2^-$~\cite{lhcb}.
This result is obtained by using the current $\psi_{9\mu}$, which is defined in Eq.~(\ref{def:psi9mu}), and it well couples to the $S$-wave $[\Sigma_c \bar D^*]$ channel.
While, the one of $J^P = 3/2^+$ is also significantly higher, that is above 4.6 GeV;

\item However, the hidden-charm pentaquark state of $J^P = 5/2^-$ has a mass around $4.5-4.6$ GeV, that is just slightly larger than the experimental mass of the $P_c(4450)$ of $J^P = 5/2^+$~\cite{lhcb}.

\end{enumerate}

The discovery of the $P_c(4380)$ and $P_c(4450)$ opens a new page on the
exotic hadron states. In the near future, further theoretical and experimental efforts
are required to study these hidden-charm pentaquark states.

\section*{Acknowledgments}
\begin{acknowledgement}
This project is supported by
the National Natural Science Foundation of China under Grants No. 11205011, No. 11475015, No. 11375024, No. 11222547, No. 11175073, and No. 11261130311,
the Ministry of Education of China (SRFDP under Grant No. 20120211110002 and the Fundamental Research Funds for the Central Universities),
and the Natural Sciences and Engineering Research Council of Canada (NSERC).
\end{acknowledgement}

\appendix

\section{Local Pentaquark Currents of Spin 3/2}
\label{app:spin32}

\subsection{Currents of $[\bar c_d c_d][\epsilon^{abc}u_a d_b u_c]$}

In this subsection, we construct the currents of the color configuration $[\bar c_d c_d][\epsilon^{abc}u_a d_b u_c]$.
We find the following currents having $J^P=3/2^-$ and quark contents $uud c \bar c$:
\begin{eqnarray*}
\eta_{1\mu} &=& [\epsilon^{abc} (u^T_a C d_b) \gamma_\mu \gamma_5 u_c] [\bar c_d c_d] \, ,
\\ \eta_{2\mu} &=& [\epsilon^{abc} (u^T_a C d_b) \gamma_\mu u_c] [\bar c_d \gamma_5 c_d] \, ,
\\ \eta_{3\mu} &=& [\epsilon^{abc} (u^T_a C \gamma_5 d_b) \gamma_\mu u_c] [\bar c_d c_d] \, ,
\\ \eta_{4\mu} &=& [\epsilon^{abc} (u^T_a C \gamma_5 d_b) \gamma_\mu \gamma_5 u_c] [\bar c_d \gamma_5 c_d] \, ,
\\ \eta_{5\mu} &=& [\epsilon^{abc} (u^T_a C d_b) \gamma_5 u_c] [\bar c_d \gamma_\mu c_d] \, ,
\\ \eta_{6\mu} &=& [\epsilon^{abc} (u^T_a C d_b) u_c] [\bar c_d \gamma_\mu \gamma_5 c_d] \, ,
\\ \eta_{7\mu} &=& [\epsilon^{abc} (u^T_a C \gamma_5 d_b) u_c] [\bar c_d \gamma_\mu c_d] \, ,
\\ \eta_{8\mu} &=& [\epsilon^{abc} (u^T_a C \gamma_5 d_b) \gamma_5 u_c] [\bar c_d \gamma_\mu \gamma_5 c_d] \, ,
\\ \eta_{9\mu} &=& [\epsilon^{abc} (u^T_a C d_b) \sigma_{\mu\nu} \gamma_5 u_c] [\bar c_d \gamma_\nu c_d] \, ,
\\ \eta_{10\mu} &=& [\epsilon^{abc} (u^T_a C d_b) \sigma_{\mu\nu} u_c] [\bar c_d \gamma_\nu \gamma_5 c_d] \, ,
\\ \eta_{11\mu} &=& [\epsilon^{abc} (u^T_a C \gamma_5 d_b) \sigma_{\mu\nu} u_c] [\bar c_d \gamma_\nu c_d] \, ,
\\ \eta_{12\mu} &=& [\epsilon^{abc} (u^T_a C \gamma_5 d_b) \sigma_{\mu\nu} \gamma_5 u_c] [\bar c_d \gamma_\nu \gamma_5 c_d] \, ,
\\ \eta_{13\mu} &=& [\epsilon^{abc} (u^T_a C d_b) \gamma_\nu \gamma_5 u_c] [\bar c_d \sigma_{\mu\nu} c_d] \, ,
\\ \eta_{14\mu} &=& [\epsilon^{abc} (u^T_a C d_b) \gamma_\nu u_c] [\bar c_d \sigma_{\mu\nu} \gamma_5 c_d] \, ,
\\ \eta_{15\mu} &=& [\epsilon^{abc} (u^T_a C \gamma_5 d_b) \gamma_\nu u_c] [\bar c_d \sigma_{\mu\nu} c_d] \, ,
\\ \eta_{16\mu} &=& [\epsilon^{abc} (u^T_a C \gamma_5 d_b) \gamma_\nu \gamma_5 u_c] [\bar c_d \sigma_{\mu\nu} \gamma_5 c_d] \, ,
\\ \eta_{17\mu} &=& [\epsilon^{abc} (u^T_a C \gamma_\mu \gamma_5 d_b) u_c] [\bar c_d c_d] \, ,
\\ \eta_{18\mu} &=& [\epsilon^{abc} (u^T_a C \gamma_\mu \gamma_5 d_b) \gamma_5 u_c] [\bar c_d \gamma_5 c_d] \, ,
\\ \eta_{19\mu} &=& [\epsilon^{abc} (u^T_a C \gamma_\mu \gamma_5 d_b) \gamma_\nu u_c] [\bar c_d \gamma_\nu c_d] \, ,
\\ \eta_{20\mu} &=& [\epsilon^{abc} (u^T_a C \gamma_\mu \gamma_5 d_b) \gamma_\nu \gamma_5 u_c] [\bar c_d \gamma_\nu \gamma_5 c_d] \, ,
\\ \eta_{21\mu} &=& [\epsilon^{abc} (u^T_a C \gamma_\nu \gamma_5 d_b) \gamma_\mu u_c] [\bar c_d \gamma_\nu c_d] \, ,
\\ \eta_{22\mu} &=& [\epsilon^{abc} (u^T_a C \gamma_\nu \gamma_5 d_b) \gamma_\mu \gamma_5 u_c] [\bar c_d \gamma_\nu \gamma_5 c_d] \, ,
\\ \eta_{23\mu} &=& [\epsilon^{abc} (u^T_a C \gamma_\nu \gamma_5 d_b) u_c] [\bar c_d \sigma_{\mu\nu} c_d] \, ,
\\ \eta_{24\mu} &=& [\epsilon^{abc} (u^T_a C \gamma_\nu \gamma_5 d_b) \gamma_5 u_c] [\bar c_d \sigma_{\mu\nu} \gamma_5 c_d] \, ,
\\ \eta_{25\mu} &=& [\epsilon^{abc} (u^T_a C \gamma_\mu \gamma_5 d_b) \sigma_{\nu\rho} u_c] [\bar c_d \sigma_{\nu\rho} c_d] \, ,
\\ \eta_{26\mu} &=& [\epsilon^{abc} (u^T_a C \gamma_\mu \gamma_5 d_b) \sigma_{\nu\rho} \gamma_5 u_c] [\bar c_d \sigma_{\nu\rho} \gamma_5 c_d] \, ,
\\ \eta_{27\mu} &=& [\epsilon^{abc} (u^T_a C \gamma_\rho \gamma_5 d_b) \sigma_{\mu\nu} u_c] [\bar c_d \sigma_{\nu\rho} c_d] \, ,
\\ \eta_{28\mu} &=& [\epsilon^{abc} (u^T_a C \gamma_\rho \gamma_5 d_b) \sigma_{\mu\nu} \gamma_5 u_c] [\bar c_d \sigma_{\nu\rho} \gamma_5 c_d] \, .
\end{eqnarray*}
We can verify the following relation
\begin{eqnarray*}
\eta_{25\mu} &=& \eta_{26\mu} \, .
\end{eqnarray*}
To perform QCD sum rule analyses, we shall use
\begin{eqnarray}
\eta_{5\mu} - \eta_{7\mu} &=& [\epsilon^{abc} (u^T_a C d_b) \gamma_5 u_c] [\bar c_d \gamma_\mu c_d]
\label{def:eta57mu}
\\ \nonumber && ~~~~~~~~~~ - [\epsilon^{abc} (u^T_a C \gamma_5 d_b) u_c] [\bar c_d \gamma_\mu c_d] \, ,
\\ \eta_{18\mu} &=& [\epsilon^{abc} (u^T_a C \gamma_\mu \gamma_5 d_b) \gamma_5 u_c] [\bar c_d \gamma_5 c_d] \, ,
\label{def:eta18mu}
\\ \eta_{19\mu} &=& [\epsilon^{abc} (u^T_a C \gamma_\mu \gamma_5 d_b) \gamma_\nu u_c] [\bar c_d \gamma_\nu c_d] \, ,
\label{def:eta19mu}
\end{eqnarray}
which well couple to the $[p J/\psi]$, $[N^* \eta_c]$, and $[N^* J/\psi]$ channels, respectively.

We note that the following currents actually have spin $J=1/2$:
\begin{eqnarray*}
\eta_{2\mu} - \eta_{4\mu} &=& [\epsilon^{abc} (u^T_a C d_b) \gamma_\mu u_c] [\bar c_d \gamma_5 c_d]
\\ \nonumber && ~~~~~~~~~~ - [\epsilon^{abc} (u^T_a C \gamma_5 d_b) \gamma_\mu \gamma_5 u_c] [\bar c_d \gamma_5 c_d] \, ,
\\ \eta_{9\mu} - \eta_{11\mu} &=& [\epsilon^{abc} (u^T_a C d_b) \sigma_{\mu\nu} \gamma_5 u_c] [\bar c_d \gamma_\nu c_d]
\\ \nonumber && ~~~~~~~~~~ - [\epsilon^{abc} (u^T_a C \gamma_5 d_b) \sigma_{\mu\nu} u_c] [\bar c_d \gamma_\nu c_d] \, ,
\\ \eta_{21\mu} &=& [\epsilon^{abc} (u^T_a C \gamma_\nu \gamma_5 d_b) \gamma_\mu u_c] [\bar c_d \gamma_\nu c_d] \, .
\end{eqnarray*}

\subsection{Currents of $[\bar c_d u_d][\epsilon^{abc} u_a d_b c_c]$}

In this subsection, we construct the currents of the color configuration $[\bar c_d u_d][\epsilon^{abc} u_a d_b c_c]$.
We find the following currents having $J^P=3/2^-$ and quark contents $uud c \bar c$:
\begin{eqnarray*}
\xi_{1\mu} &=& [\epsilon^{abc} (u^T_a C d_b) \gamma_\mu \gamma_5 c_c] [\bar c_d u_d] \, ,
\\ \xi_{2\mu} &=& [\epsilon^{abc} (u^T_a C d_b) \gamma_\mu c_c] [\bar c_d \gamma_5 u_d] \, ,
\\ \xi_{3\mu} &=& [\epsilon^{abc} (u^T_a C \gamma_5 d_b) \gamma_\mu c_c] [\bar c_d u_d] \, ,
\\ \xi_{4\mu} &=& [\epsilon^{abc} (u^T_a C \gamma_5 d_b) \gamma_\mu \gamma_5 c_c] [\bar c_d \gamma_5 u_d] \, ,
\\ \xi_{5\mu} &=& [\epsilon^{abc} (u^T_a C d_b) \gamma_5 c_c] [\bar c_d \gamma_\mu u_d] \, ,
\\ \xi_{6\mu} &=& [\epsilon^{abc} (u^T_a C d_b) c_c] [\bar c_d \gamma_\mu \gamma_5 u_d] \, ,
\\ \xi_{7\mu} &=& [\epsilon^{abc} (u^T_a C \gamma_5 d_b) c_c] [\bar c_d \gamma_\mu u_d] \, ,
\\ \xi_{8\mu} &=& [\epsilon^{abc} (u^T_a C \gamma_5 d_b) \gamma_5 c_c] [\bar c_d \gamma_\mu \gamma_5 u_d] \, ,
\\ \xi_{9\mu} &=& [\epsilon^{abc} (u^T_a C d_b) \sigma_{\mu\nu} \gamma_5 c_c] [\bar c_d \gamma_\nu u_d] \, ,
\\ \xi_{10\mu} &=& [\epsilon^{abc} (u^T_a C d_b) \sigma_{\mu\nu} c_c] [\bar c_d \gamma_\nu \gamma_5 u_d] \, ,
\\ \xi_{11\mu} &=& [\epsilon^{abc} (u^T_a C \gamma_5 d_b) \sigma_{\mu\nu} c_c] [\bar c_d \gamma_\nu u_d] \, ,
\\ \xi_{12\mu} &=& [\epsilon^{abc} (u^T_a C \gamma_5 d_b) \sigma_{\mu\nu} \gamma_5 c_c] [\bar c_d \gamma_\nu \gamma_5 u_d] \, ,
\\ \xi_{13\mu} &=& [\epsilon^{abc} (u^T_a C d_b) \gamma_\nu \gamma_5 c_c] [\bar c_d \sigma_{\mu\nu} u_d] \, ,
\\ \xi_{14\mu} &=& [\epsilon^{abc} (u^T_a C d_b) \gamma_\nu c_c] [\bar c_d \sigma_{\mu\nu} \gamma_5 u_d] \, ,
\\ \xi_{15\mu} &=& [\epsilon^{abc} (u^T_a C \gamma_5 d_b) \gamma_\nu c_c] [\bar c_d \sigma_{\mu\nu} u_d] \, ,
\\ \xi_{16\mu} &=& [\epsilon^{abc} (u^T_a C \gamma_5 d_b) \gamma_\nu \gamma_5 c_c] [\bar c_d \sigma_{\mu\nu} \gamma_5 u_d] \, ,
\\ \xi_{17\mu} &=& [\epsilon^{abc} (u^T_a C \gamma_\mu d_b) \gamma_5 c_c] [\bar c_d u_d] \, ,
\\ \xi_{18\mu} &=& [\epsilon^{abc} (u^T_a C \gamma_\mu d_b) c_c] [\bar c_d \gamma_5 u_d] \, ,
\\ \xi_{19\mu} &=& [\epsilon^{abc} (u^T_a C \gamma_\mu \gamma_5 d_b) c_c] [\bar c_d u_d] \, ,
\\ \xi_{20\mu} &=& [\epsilon^{abc} (u^T_a C \gamma_\mu \gamma_5 d_b) \gamma_5 c_c] [\bar c_d \gamma_5 u_d] \, ,
\\ \xi_{21\mu} &=& [\epsilon^{abc} (u^T_a C \gamma_\nu d_b) \sigma_{\mu\nu} \gamma_5 c_c] [\bar c_d u_d] \, ,
\\ \xi_{22\mu} &=& [\epsilon^{abc} (u^T_a C \gamma_\nu d_b) \sigma_{\mu\nu} c_c] [\bar c_d \gamma_5 u_d] \, ,
\\ \xi_{23\mu} &=& [\epsilon^{abc} (u^T_a C \gamma_\nu \gamma_5 d_b) \sigma_{\mu\nu} c_c] [\bar c_d u_d] \, ,
\\ \xi_{24\mu} &=& [\epsilon^{abc} (u^T_a C \gamma_\nu \gamma_5 d_b) \sigma_{\mu\nu} \gamma_5 c_c] [\bar c_d \gamma_5 u_d] \, ,
\\ \xi_{25\mu} &=& [\epsilon^{abc} (u^T_a C \gamma_\mu d_b) \gamma_\nu \gamma_5 c_c] [\bar c_d \gamma_\nu u_d] \, ,
\\ \xi_{26\mu} &=& [\epsilon^{abc} (u^T_a C \gamma_\mu d_b) \gamma_\nu c_c] [\bar c_d \gamma_\nu \gamma_5 u_d] \, ,
\\ \xi_{27\mu} &=& [\epsilon^{abc} (u^T_a C \gamma_\mu \gamma_5 d_b) \gamma_\nu c_c] [\bar c_d \gamma_\nu u_d] \, ,
\\ \xi_{28\mu} &=& [\epsilon^{abc} (u^T_a C \gamma_\mu \gamma_5 d_b) \gamma_\nu \gamma_5 c_c] [\bar c_d \gamma_\nu \gamma_5 u_d] \, ,
\\ \xi_{29\mu} &=& [\epsilon^{abc} (u^T_a C \gamma_\nu d_b) \gamma_\mu \gamma_5 c_c] [\bar c_d \gamma_\nu u_d] \, ,
\\ \xi_{30\mu} &=& [\epsilon^{abc} (u^T_a C \gamma_\nu d_b) \gamma_\mu c_c] [\bar c_d \gamma_\nu \gamma_5 u_d] \, ,
\\ \xi_{31\mu} &=& [\epsilon^{abc} (u^T_a C \gamma_\nu \gamma_5 d_b) \gamma_\mu c_c] [\bar c_d \gamma_\nu u_d] \, ,
\\ \xi_{32\mu} &=& [\epsilon^{abc} (u^T_a C \gamma_\nu \gamma_5 d_b) \gamma_\mu \gamma_5 c_c] [\bar c_d \gamma_\nu \gamma_5 u_d] \, ,
\\ \xi_{33\mu} &=& [\epsilon^{abc} (u^T_a C \gamma_\nu d_b) \gamma_\nu \gamma_5 c_c] [\bar c_d \gamma_\mu u_d] \, ,
\\ \xi_{34\mu} &=& [\epsilon^{abc} (u^T_a C \gamma_\nu d_b) \gamma_\nu c_c] [\bar c_d \gamma_\mu \gamma_5 u_d] \, ,
\\ \xi_{35\mu} &=& [\epsilon^{abc} (u^T_a C \gamma_\nu \gamma_5 d_b) \gamma_\nu c_c] [\bar c_d \gamma_\mu u_d] \, ,
\\ \xi_{36\mu} &=& [\epsilon^{abc} (u^T_a C \gamma_\nu \gamma_5 d_b) \gamma_\nu \gamma_5 c_c] [\bar c_d \gamma_\mu \gamma_5 u_d] \, ,
\\ \xi_{37\mu} &=& [\epsilon^{abc} (u^T_a C \gamma_\nu d_b) \gamma_5 c_c] [\bar c_d \sigma_{\mu\nu} u_d] \, ,
\\ \xi_{38\mu} &=& [\epsilon^{abc} (u^T_a C \gamma_\nu d_b) c_c] [\bar c_d \sigma_{\mu\nu} \gamma_5 u_d] \, ,
\\ \xi_{39\mu} &=& [\epsilon^{abc} (u^T_a C \gamma_\nu \gamma_5 d_b) c_c] [\bar c_d \sigma_{\mu\nu} u_d] \, ,
\\ \xi_{40\mu} &=& [\epsilon^{abc} (u^T_a C \gamma_\nu \gamma_5 d_b) \gamma_5 c_c] [\bar c_d \sigma_{\mu\nu} \gamma_5 u_d] \, ,
\\ \xi_{41\mu} &=& [\epsilon^{abc} (u^T_a C \gamma_\mu d_b) \sigma_{\nu\rho} \gamma_5 c_c] [\bar c_d \sigma_{\nu\rho} u_d] \, ,
\\ \xi_{42\mu} &=& [\epsilon^{abc} (u^T_a C \gamma_\mu d_b) \sigma_{\nu\rho} c_c] [\bar c_d \sigma_{\nu\rho} \gamma_5 u_d] \, ,
\\ \xi_{43\mu} &=& [\epsilon^{abc} (u^T_a C \gamma_\mu \gamma_5 d_b) \sigma_{\nu\rho} c_c] [\bar c_d \sigma_{\nu\rho} u_d] \, ,
\\ \xi_{44\mu} &=& [\epsilon^{abc} (u^T_a C \gamma_\mu \gamma_5 d_b) \sigma_{\nu\rho} \gamma_5 c_c] [\bar c_d \sigma_{\nu\rho} \gamma_5 u_d] \, ,
\\ \xi_{45\mu} &=& [\epsilon^{abc} (u^T_a C \gamma_\rho d_b) \sigma_{\mu\nu} \gamma_5 c_c] [\bar c_d \sigma_{\nu\rho} u_d] \, ,
\\ \xi_{46\mu} &=& [\epsilon^{abc} (u^T_a C \gamma_\rho d_b) \sigma_{\mu\nu} c_c] [\bar c_d \sigma_{\nu\rho} \gamma_5 u_d] \, ,
\\ \xi_{47\mu} &=& [\epsilon^{abc} (u^T_a C \gamma_\rho \gamma_5 d_b) \sigma_{\mu\nu} c_c] [\bar c_d \sigma_{\nu\rho} u_d] \, ,
\\ \xi_{48\mu} &=& [\epsilon^{abc} (u^T_a C \gamma_\rho \gamma_5 d_b) \sigma_{\mu\nu} \gamma_5 c_c] [\bar c_d \sigma_{\nu\rho} \gamma_5 u_d] \, ,
\\ \xi_{49\mu} &=& [\epsilon^{abc} (u^T_a C \gamma_\rho d_b) \sigma_{\nu\rho} \gamma_5 c_c] [\bar c_d \sigma_{\mu\nu} u_d] \, ,
\\ \xi_{50\mu} &=& [\epsilon^{abc} (u^T_a C \gamma_\rho d_b) \sigma_{\nu\rho} c_c] [\bar c_d \sigma_{\mu\nu} \gamma_5 u_d] \, ,
\\ \xi_{51\mu} &=& [\epsilon^{abc} (u^T_a C \gamma_\rho \gamma_5 d_b) \sigma_{\nu\rho} c_c] [\bar c_d \sigma_{\mu\nu} u_d] \, ,
\\ \xi_{52\mu} &=& [\epsilon^{abc} (u^T_a C \gamma_\rho \gamma_5 d_b) \sigma_{\nu\rho} \gamma_5 c_c] [\bar c_d \sigma_{\mu\nu} \gamma_5 u_d] \, ,
\\ \xi_{53\mu} &=& [\epsilon^{abc} (u^T_a C \sigma_{\mu\nu} d_b) \gamma_\nu \gamma_5 c_c] [\bar c_d u_d] \, ,
\\ \xi_{54\mu} &=& [\epsilon^{abc} (u^T_a C \sigma_{\mu\nu} d_b) \gamma_\nu c_c] [\bar c_d \gamma_5 u_d] \, ,
\\ \xi_{55\mu} &=& [\epsilon^{abc} (u^T_a C \sigma_{\mu\nu} \gamma_5 d_b) \gamma_\nu c_c] [\bar c_d u_d] \, ,
\\ \xi_{56\mu} &=& [\epsilon^{abc} (u^T_a C \sigma_{\mu\nu} \gamma_5 d_b) \gamma_\nu \gamma_5 c_c] [\bar c_d \gamma_5 u_d] \, ,
\\ \xi_{57\mu} &=& [\epsilon^{abc} (u^T_a C \sigma_{\mu\nu} d_b) \gamma_5 c_c] [\bar c_d \gamma_\nu u_d] \, ,
\\ \xi_{58\mu} &=& [\epsilon^{abc} (u^T_a C \sigma_{\mu\nu} d_b) c_c] [\bar c_d \gamma_\nu \gamma_5 u_d] \, ,
\\ \xi_{59\mu} &=& [\epsilon^{abc} (u^T_a C \sigma_{\mu\nu} \gamma_5 d_b) c_c] [\bar c_d \gamma_\nu u_d] \, ,
\\ \xi_{60\mu} &=& [\epsilon^{abc} (u^T_a C \sigma_{\mu\nu} \gamma_5 d_b) \gamma_5 c_c] [\bar c_d \gamma_\nu \gamma_5 u_d] \, ,
\\ \xi_{61\mu} &=& [\epsilon^{abc} (u^T_a C \sigma_{\mu\nu} d_b) \sigma_{\nu\rho} \gamma_5 c_c] [\bar c_d \gamma_\rho u_d] \, ,
\\ \xi_{62\mu} &=& [\epsilon^{abc} (u^T_a C \sigma_{\mu\nu} d_b) \sigma_{\nu\rho} c_c] [\bar c_d \gamma_\rho \gamma_5 u_d] \, ,
\\ \xi_{63\mu} &=& [\epsilon^{abc} (u^T_a C \sigma_{\mu\nu} \gamma_5 d_b) \sigma_{\nu\rho} c_c] [\bar c_d \gamma_\rho u_d] \, ,
\\ \xi_{64\mu} &=& [\epsilon^{abc} (u^T_a C \sigma_{\mu\nu} \gamma_5 d_b) \sigma_{\nu\rho} \gamma_5 c_c] [\bar c_d \gamma_\rho \gamma_5 u_d] \, ,
\\ \xi_{65\mu} &=& [\epsilon^{abc} (u^T_a C \sigma_{\nu\rho} d_b) \sigma_{\mu\nu} \gamma_5 c_c] [\bar c_d \gamma_\rho u_d] \, ,
\\ \xi_{66\mu} &=& [\epsilon^{abc} (u^T_a C \sigma_{\nu\rho} d_b) \sigma_{\mu\nu} c_c] [\bar c_d \gamma_\rho \gamma_5 u_d] \, ,
\\ \xi_{67\mu} &=& [\epsilon^{abc} (u^T_a C \sigma_{\nu\rho} \gamma_5 d_b) \sigma_{\mu\nu} c_c] [\bar c_d \gamma_\rho u_d] \, ,
\\ \xi_{68\mu} &=& [\epsilon^{abc} (u^T_a C \sigma_{\nu\rho} \gamma_5 d_b) \sigma_{\mu\nu} \gamma_5 c_c] [\bar c_d \gamma_\rho \gamma_5 u_d] \, ,
\\ \xi_{69\mu} &=& [\epsilon^{abc} (u^T_a C \sigma_{\nu\rho} d_b) \sigma_{\nu\rho} \gamma_5 c_c] [\bar c_d \gamma_\mu u_d] \, ,
\\ \xi_{70\mu} &=& [\epsilon^{abc} (u^T_a C \sigma_{\nu\rho} d_b) \sigma_{\nu\rho} c_c] [\bar c_d \gamma_\mu \gamma_5 u_d] \, ,
\\ \xi_{71\mu} &=& [\epsilon^{abc} (u^T_a C \sigma_{\nu\rho} \gamma_5 d_b) \sigma_{\nu\rho} c_c] [\bar c_d \gamma_\mu u_d] \, ,
\\ \xi_{72\mu} &=& [\epsilon^{abc} (u^T_a C \sigma_{\nu\rho} \gamma_5 d_b) \sigma_{\nu\rho} \gamma_5 c_c] [\bar c_d \gamma_\mu \gamma_5 u_d] \, ,
\\ \xi_{73\mu} &=& [\epsilon^{abc} (u^T_a C \sigma_{\mu\nu} d_b) \gamma_\rho \gamma_5 c_c] [\bar c_d \sigma_{\nu\rho} u_d] \, ,
\\ \xi_{74\mu} &=& [\epsilon^{abc} (u^T_a C \sigma_{\mu\nu} d_b) \gamma_\rho c_c] [\bar c_d \sigma_{\nu\rho} \gamma_5 u_d] \, ,
\\ \xi_{75\mu} &=& [\epsilon^{abc} (u^T_a C \sigma_{\mu\nu} \gamma_5 d_b) \gamma_\rho c_c] [\bar c_d \sigma_{\nu\rho} u_d] \, ,
\\ \xi_{76\mu} &=& [\epsilon^{abc} (u^T_a C \sigma_{\mu\nu} \gamma_5 d_b) \gamma_\rho \gamma_5 c_c] [\bar c_d \sigma_{\nu\rho} \gamma_5 u_d] \, ,
\\ \xi_{77\mu} &=& [\epsilon^{abc} (u^T_a C \sigma_{\nu\rho} d_b) \gamma_\mu \gamma_5 c_c] [\bar c_d \sigma_{\nu\rho} u_d] \, ,
\\ \xi_{78\mu} &=& [\epsilon^{abc} (u^T_a C \sigma_{\nu\rho} d_b) \gamma_\mu c_c] [\bar c_d \sigma_{\nu\rho} \gamma_5 u_d] \, ,
\\ \xi_{79\mu} &=& [\epsilon^{abc} (u^T_a C \sigma_{\nu\rho} \gamma_5 d_b) \gamma_\mu c_c] [\bar c_d \sigma_{\nu\rho} u_d] \, ,
\\ \xi_{80\mu} &=& [\epsilon^{abc} (u^T_a C \sigma_{\nu\rho} \gamma_5 d_b) \gamma_\mu \gamma_5 c_c] [\bar c_d \sigma_{\nu\rho} \gamma_5 u_d] \, ,
\\ \xi_{81\mu} &=& [\epsilon^{abc} (u^T_a C \sigma_{\nu\rho} d_b) \gamma_\rho \gamma_5 c_c] [\bar c_d \sigma_{\mu\nu} u_d] \, ,
\\ \xi_{82\mu} &=& [\epsilon^{abc} (u^T_a C \sigma_{\nu\rho} d_b) \gamma_\rho c_c] [\bar c_d \sigma_{\mu\nu} \gamma_5 u_d] \, ,
\\ \xi_{83\mu} &=& [\epsilon^{abc} (u^T_a C \sigma_{\nu\rho} \gamma_5 d_b) \gamma_\rho c_c] [\bar c_d \sigma_{\mu\nu} u_d] \, ,
\\ \xi_{84\mu} &=& [\epsilon^{abc} (u^T_a C \sigma_{\nu\rho} \gamma_5 d_b) \gamma_\rho \gamma_5 c_c] [\bar c_d \sigma_{\mu\nu} \gamma_5 u_d] \, .
\end{eqnarray*}
We can verify the following relations
\begin{eqnarray*}
\xi_{41\mu} &=& \xi_{42\mu} \, ,
\\ \xi_{43\mu} &=& \xi_{44\mu} \, ,
\\ \xi_{69\mu} &=& \xi_{71\mu} \, ,
\\ \xi_{70\mu} &=& \xi_{72\mu} \, ,
\\ \xi_{77\mu} &=& \xi_{80\mu} \, ,
\\ \xi_{78\mu} &=& \xi_{79\mu} \, .
\end{eqnarray*}
To perform QCD sum rule analyses, we shall use
\begin{eqnarray}
\xi_{5\mu} - \xi_{7\mu} &=& [\epsilon^{abc} (u^T_a C d_b) \gamma_5 c_c] [\bar c_d \gamma_\mu u_d]
\label{def:xi57mu}
\\ \nonumber && ~~~~~~~~~~ - [\epsilon^{abc} (u^T_a C \gamma_5 d_b) c_c] [\bar c_d \gamma_\mu u_d] \, ,
\\ \xi_{18\mu} &=& [\epsilon^{abc} (u^T_a C \gamma_\mu d_b) c_c] [\bar c_d \gamma_5 u_d] \, ,
\label{def:xi18mu}
\\ \xi_{20\mu} &=& [\epsilon^{abc} (u^T_a C \gamma_\mu \gamma_5 d_b) \gamma_5 c_c] [\bar c_d \gamma_5 u_d] \, ,
\label{def:xi20mu}
\\ \xi_{25\mu} &=& [\epsilon^{abc} (u^T_a C \gamma_\mu d_b) \gamma_\nu \gamma_5 c_c] [\bar c_d \gamma_\nu u_d] \, ,
\label{def:xi25mu}
\\ \xi_{27\mu} &=& [\epsilon^{abc} (u^T_a C \gamma_\mu \gamma_5 d_b) \gamma_\nu c_c] [\bar c_d \gamma_\nu u_d] \, ,
\label{def:xi27mu}
\\ \xi_{33\mu} &=& [\epsilon^{abc} (u^T_a C \gamma_\nu d_b) \gamma_\nu \gamma_5 c_c] [\bar c_d \gamma_\mu u_d] \, ,
\label{def:xi33mu}
\\ \xi_{35\mu} &=& [\epsilon^{abc} (u^T_a C \gamma_\nu \gamma_5 d_b) \gamma_\nu c_c] [\bar c_d \gamma_\mu u_d] \, ,
\label{def:xi35mu}
\end{eqnarray}
which well couple to the $[\Lambda_c \bar D^*]$, $[\Sigma_c^* \bar D]$, $[\Lambda_c^* \bar D]$, $[\Sigma_c^* \bar D^*]$, $[\Lambda_c^* \bar D^*]$, $[\Sigma_c \bar D^*]$ and $[\Lambda_c \bar D^*]$ channels, respectively.

We note that the following currents actually have spin $J=1/2$:
\begin{eqnarray*}
\xi_{2\mu} - \xi_{4\mu} &=& [\epsilon^{abc} (u^T_a C d_b) \gamma_\mu c_c] [\bar c_d \gamma_5 u_d]
\\ \nonumber && ~~~~~~~~~~ - [\epsilon^{abc} (u^T_a C \gamma_5 d_b) \gamma_\mu \gamma_5 c_c] [\bar c_d \gamma_5 u_d] \, ,
\\ \xi_{9\mu} - \xi_{11\mu} &=& [\epsilon^{abc} (u^T_a C d_b) \sigma_{\mu\nu} \gamma_5 c_c] [\bar c_d \gamma_\nu u_d]
\\ \nonumber && ~~~~~~~~~~ - [\epsilon^{abc} (u^T_a C \gamma_5 d_b) \sigma_{\mu\nu} c_c] [\bar c_d \gamma_\nu u_d] \, ,
\\ \xi_{22\mu} &=& [\epsilon^{abc} (u^T_a C \gamma_\nu d_b) \sigma_{\mu\nu} c_c] [\bar c_d \gamma_5 u_d] \, ,
\\ \xi_{24\mu} &=& [\epsilon^{abc} (u^T_a C \gamma_\nu \gamma_5 d_b) \sigma_{\mu\nu} \gamma_5 c_c] [\bar c_d \gamma_5 u_d] \, ,
\\ \xi_{29\mu} &=& [\epsilon^{abc} (u^T_a C \gamma_\nu d_b) \gamma_\mu \gamma_5 c_c] [\bar c_d \gamma_\nu u_d] \, ,
\\ \xi_{31\mu} &=& [\epsilon^{abc} (u^T_a C \gamma_\nu \gamma_5 d_b) \gamma_\mu c_c] [\bar c_d \gamma_\nu u_d] \, .
\end{eqnarray*}

\subsection{Currents of $[\bar c_d d_d][\epsilon^{abc} u_a u_b c_c]$}

In this subsection, we construct the currents of the color configuration $[\bar c_d d_d][\epsilon^{abc} u_a u_b c_c]$.
We find the following currents having $J^P=3/2^-$ and quark contents $uud c \bar c$:
\begin{eqnarray*}
\psi_{1\mu} &=& [\epsilon^{abc} (u^T_a C \gamma_\mu u_b) \gamma_5 c_c] [\bar c_d d_d] \, ,
\\ \psi_{2\mu} &=& [\epsilon^{abc} (u^T_a C \gamma_\mu u_b) c_c] [\bar c_d \gamma_5 d_d] \, ,
\\ \psi_{3\mu} &=& [\epsilon^{abc} (u^T_a C \gamma_\nu u_b) \sigma_{\mu\nu} \gamma_5 c_c] [\bar c_d d_d] \, ,
\\ \psi_{4\mu} &=& [\epsilon^{abc} (u^T_a C \gamma_\nu u_b) \sigma_{\mu\nu} c_c] [\bar c_d \gamma_5 d_d] \, ,
\\ \psi_{5\mu} &=& [\epsilon^{abc} (u^T_a C \gamma_\mu u_b) \gamma_\nu \gamma_5 c_c] [\bar c_d \gamma_\nu d_d] \, ,
\\ \psi_{6\mu} &=& [\epsilon^{abc} (u^T_a C \gamma_\mu u_b) \gamma_\nu c_c] [\bar c_d \gamma_\nu \gamma_5 d_d] \, ,
\\ \psi_{7\mu} &=& [\epsilon^{abc} (u^T_a C \gamma_\nu u_b) \gamma_\mu \gamma_5 c_c] [\bar c_d \gamma_\nu d_d] \, ,
\\ \psi_{8\mu} &=& [\epsilon^{abc} (u^T_a C \gamma_\nu u_b) \gamma_\mu c_c] [\bar c_d \gamma_\nu \gamma_5 d_d] \, ,
\\ \psi_{9\mu} &=& [\epsilon^{abc} (u^T_a C \gamma_\nu u_b) \gamma_\nu \gamma_5 c_c] [\bar c_d \gamma_\mu d_d] \, ,
\\ \psi_{10\mu} &=& [\epsilon^{abc} (u^T_a C \gamma_\nu u_b) \gamma_\nu c_c] [\bar c_d \gamma_\mu \gamma_5 d_d] \, ,
\\ \psi_{11\mu} &=& [\epsilon^{abc} (u^T_a C \gamma_\nu u_b) \gamma_5 c_c] [\bar c_d \sigma_{\mu\nu} d_d] \, ,
\\ \psi_{12\mu} &=& [\epsilon^{abc} (u^T_a C \gamma_\nu u_b) c_c] [\bar c_d \sigma_{\mu\nu} \gamma_5 d_d] \, ,
\\ \psi_{13\mu} &=& [\epsilon^{abc} (u^T_a C \gamma_\mu u_b) \sigma_{\nu\rho} \gamma_5 c_c] [\bar c_d \sigma_{\nu\rho} d_d] \, ,
\\ \psi_{14\mu} &=& [\epsilon^{abc} (u^T_a C \gamma_\mu u_b) \sigma_{\nu\rho} c_c] [\bar c_d \sigma_{\nu\rho} \gamma_5 d_d] \, ,
\\ \psi_{15\mu} &=& [\epsilon^{abc} (u^T_a C \gamma_\rho u_b) \sigma_{\mu\nu} \gamma_5 c_c] [\bar c_d \sigma_{\nu\rho} d_d] \, ,
\\ \psi_{16\mu} &=& [\epsilon^{abc} (u^T_a C \gamma_\rho u_b) \sigma_{\mu\nu} c_c] [\bar c_d \sigma_{\nu\rho} \gamma_5 d_d] \, ,
\\ \psi_{17\mu} &=& [\epsilon^{abc} (u^T_a C \gamma_\rho u_b) \sigma_{\nu\rho} \gamma_5 c_c] [\bar c_d \sigma_{\mu\nu} d_d] \, ,
\\ \psi_{18\mu} &=& [\epsilon^{abc} (u^T_a C \gamma_\rho u_b) \sigma_{\nu\rho} c_c] [\bar c_d \sigma_{\mu\nu} \gamma_5 d_d] \, ,
\\ \psi_{19\mu} &=& [\epsilon^{abc} (u^T_a C \sigma_{\mu\nu} u_b) \gamma_\nu \gamma_5 c_c] [\bar c_d d_d] \, ,
\\ \psi_{20\mu} &=& [\epsilon^{abc} (u^T_a C \sigma_{\mu\nu} u_b) \gamma_\nu c_c] [\bar c_d \gamma_5 d_d] \, ,
\\ \psi_{21\mu} &=& [\epsilon^{abc} (u^T_a C \sigma_{\mu\nu} \gamma_5 u_b) \gamma_\nu c_c] [\bar c_d d_d] \, ,
\\ \psi_{22\mu} &=& [\epsilon^{abc} (u^T_a C \sigma_{\mu\nu} \gamma_5 u_b) \gamma_\nu \gamma_5 c_c] [\bar c_d \gamma_5 d_d] \, ,
\\ \psi_{23\mu} &=& [\epsilon^{abc} (u^T_a C \sigma_{\mu\nu} u_b) \gamma_5 c_c] [\bar c_d \gamma_\nu d_d] \, ,
\\ \psi_{24\mu} &=& [\epsilon^{abc} (u^T_a C \sigma_{\mu\nu} u_b) c_c] [\bar c_d \gamma_\nu \gamma_5 d_d] \, ,
\\ \psi_{25\mu} &=& [\epsilon^{abc} (u^T_a C \sigma_{\mu\nu} \gamma_5 u_b) c_c] [\bar c_d \gamma_\nu d_d] \, ,
\\ \psi_{26\mu} &=& [\epsilon^{abc} (u^T_a C \sigma_{\mu\nu} \gamma_5 u_b) \gamma_5 c_c] [\bar c_d \gamma_\nu \gamma_5 d_d] \, ,
\\ \psi_{27\mu} &=& [\epsilon^{abc} (u^T_a C \sigma_{\mu\nu} u_b) \sigma_{\nu\rho} \gamma_5 c_c] [\bar c_d \gamma_\rho d_d] \, ,
\\ \psi_{28\mu} &=& [\epsilon^{abc} (u^T_a C \sigma_{\mu\nu} u_b) \sigma_{\nu\rho} c_c] [\bar c_d \gamma_\rho \gamma_5 d_d] \, ,
\\ \psi_{29\mu} &=& [\epsilon^{abc} (u^T_a C \sigma_{\mu\nu} \gamma_5 u_b) \sigma_{\nu\rho} c_c] [\bar c_d \gamma_\rho d_d] \, ,
\\ \psi_{30\mu} &=& [\epsilon^{abc} (u^T_a C \sigma_{\mu\nu} \gamma_5 u_b) \sigma_{\nu\rho} \gamma_5 c_c] [\bar c_d \gamma_\rho \gamma_5 d_d] \, ,
\\ \psi_{31\mu} &=& [\epsilon^{abc} (u^T_a C \sigma_{\nu\rho} u_b) \sigma_{\mu\nu} \gamma_5 c_c] [\bar c_d \gamma_\rho d_d] \, ,
\\ \psi_{32\mu} &=& [\epsilon^{abc} (u^T_a C \sigma_{\nu\rho} u_b) \sigma_{\mu\nu} c_c] [\bar c_d \gamma_\rho \gamma_5 d_d] \, ,
\\ \psi_{33\mu} &=& [\epsilon^{abc} (u^T_a C \sigma_{\nu\rho} \gamma_5 u_b) \sigma_{\mu\nu} c_c] [\bar c_d \gamma_\rho d_d] \, ,
\\ \psi_{34\mu} &=& [\epsilon^{abc} (u^T_a C \sigma_{\nu\rho} \gamma_5 u_b) \sigma_{\mu\nu} \gamma_5 c_c] [\bar c_d \gamma_\rho \gamma_5 d_d] \, ,
\\ \psi_{35\mu} &=& [\epsilon^{abc} (u^T_a C \sigma_{\nu\rho} u_b) \sigma_{\nu\rho} \gamma_5 c_c] [\bar c_d \gamma_\mu d_d] \, ,
\\ \psi_{36\mu} &=& [\epsilon^{abc} (u^T_a C \sigma_{\nu\rho} u_b) \sigma_{\nu\rho} c_c] [\bar c_d \gamma_\mu \gamma_5 d_d] \, ,
\\ \psi_{37\mu} &=& [\epsilon^{abc} (u^T_a C \sigma_{\nu\rho} \gamma_5 u_b) \sigma_{\nu\rho} c_c] [\bar c_d \gamma_\mu d_d] \, ,
\\ \psi_{38\mu} &=& [\epsilon^{abc} (u^T_a C \sigma_{\nu\rho} \gamma_5 u_b) \sigma_{\nu\rho} \gamma_5 c_c] [\bar c_d \gamma_\mu \gamma_5 d_d] \, ,
\\ \psi_{39\mu} &=& [\epsilon^{abc} (u^T_a C \sigma_{\mu\nu} u_b) \gamma_\rho \gamma_5 c_c] [\bar c_d \sigma_{\nu\rho} d_d] \, ,
\\ \psi_{40\mu} &=& [\epsilon^{abc} (u^T_a C \sigma_{\mu\nu} u_b) \gamma_\rho c_c] [\bar c_d \sigma_{\nu\rho} \gamma_5 d_d] \, ,
\\ \psi_{41\mu} &=& [\epsilon^{abc} (u^T_a C \sigma_{\mu\nu} \gamma_5 u_b) \gamma_\rho c_c] [\bar c_d \sigma_{\nu\rho} d_d] \, ,
\\ \psi_{42\mu} &=& [\epsilon^{abc} (u^T_a C \sigma_{\mu\nu} \gamma_5 u_b) \gamma_\rho \gamma_5 c_c] [\bar c_d \sigma_{\nu\rho} \gamma_5 d_d] \, ,
\\ \psi_{43\mu} &=& [\epsilon^{abc} (u^T_a C \sigma_{\nu\rho} u_b) \gamma_\mu \gamma_5 c_c] [\bar c_d \sigma_{\nu\rho} d_d] \, ,
\\ \psi_{44\mu} &=& [\epsilon^{abc} (u^T_a C \sigma_{\nu\rho} u_b) \gamma_\mu c_c] [\bar c_d \sigma_{\nu\rho} \gamma_5 d_d] \, ,
\\ \psi_{45\mu} &=& [\epsilon^{abc} (u^T_a C \sigma_{\nu\rho} \gamma_5 u_b) \gamma_\mu c_c] [\bar c_d \sigma_{\nu\rho} d_d] \, ,
\\ \psi_{46\mu} &=& [\epsilon^{abc} (u^T_a C \sigma_{\nu\rho} \gamma_5 u_b) \gamma_\mu \gamma_5 c_c] [\bar c_d \sigma_{\nu\rho} \gamma_5 d_d] \, ,
\\ \psi_{47\mu} &=& [\epsilon^{abc} (u^T_a C \sigma_{\nu\rho} u_b) \gamma_\rho \gamma_5 c_c] [\bar c_d \sigma_{\mu\nu} d_d] \, ,
\\ \psi_{48\mu} &=& [\epsilon^{abc} (u^T_a C \sigma_{\nu\rho} u_b) \gamma_\rho c_c] [\bar c_d \sigma_{\mu\nu} \gamma_5 d_d] \, ,
\\ \psi_{49\mu} &=& [\epsilon^{abc} (u^T_a C \sigma_{\nu\rho} \gamma_5 u_b) \gamma_\rho c_c] [\bar c_d \sigma_{\mu\nu} d_d] \, ,
\\ \psi_{50\mu} &=& [\epsilon^{abc} (u^T_a C \sigma_{\nu\rho} \gamma_5 u_b) \gamma_\rho \gamma_5 c_c] [\bar c_d \sigma_{\mu\nu} \gamma_5 d_d] \, .
\end{eqnarray*}
We can verify the following relations
\begin{eqnarray*}
\psi_{13\mu} &=& \psi_{14\mu} \, ,
\\ \psi_{35\mu} &=& \psi_{37\mu} \, ,
\\ \psi_{36\mu} &=& \psi_{38\mu} \, ,
\\ \psi_{43\mu} &=& \psi_{46\mu} \, .
\\ \psi_{44\mu} &=& \psi_{45\mu} \, .
\end{eqnarray*}
To perform QCD sum rule analyses, we shall use
\begin{eqnarray}
\psi_{2\mu} &=& [\epsilon^{abc} (u^T_a C \gamma_\mu u_b) c_c] [\bar c_d \gamma_5 d_d] \, ,
\label{def:psi2mu}
\\ \psi_{5\mu} &=& [\epsilon^{abc} (u^T_a C \gamma_\mu u_b) \gamma_\nu \gamma_5 c_c] [\bar c_d \gamma_\nu d_d] \, ,
\label{def:psi5mu}
\\ \psi_{9\mu} &=& [\epsilon^{abc} (u^T_a C \gamma_\nu u_b) \gamma_\nu \gamma_5 c_c] [\bar c_d \gamma_\mu d_d] \, ,
\label{def:psi9mu}
\end{eqnarray}
which well couple to the $[\Sigma^*_c \bar D]$, $[\Sigma_c^* \bar D^*]$ and $[\Sigma_c \bar D^*]$ channels, respectively.

We note that the following currents actually have spin $J=1/2$:
\begin{eqnarray*}
\psi_{4\mu} &=& [\epsilon^{abc} (u^T_a C \gamma_\nu u_b) \sigma_{\mu\nu} c_c] [\bar c_d \gamma_5 d_d] \, ,
\\ \psi_{7\mu} &=& [\epsilon^{abc} (u^T_a C \gamma_\nu u_b) \gamma_\mu \gamma_5 c_c] [\bar c_d \gamma_\nu d_d] \, .
\end{eqnarray*}

\section{Local Pentaquark Currents of Spin 5/2}
\label{app:spin52}

\subsection{Currents of $[\bar c_d c_d][\epsilon^{abc}u_a d_b u_c]$}

In this subsection, we construct the currents of the color configuration $[\bar c_d c_d][\epsilon^{abc}u_a d_b u_c]$.
We find the following currents having $J^P=5/2^+$ and quark contents $uud c \bar c$:
\begin{eqnarray*}
\eta_{1\mu\nu} &=& [\epsilon^{abc} (u^T_a C d_b) \gamma_\mu \gamma_5 u_c] [\bar c_d \gamma_\nu c_d] + \{ \mu \leftrightarrow \nu \} \, ,
\\ \eta_{2\mu\nu} &=& [\epsilon^{abc} (u^T_a C d_b) \gamma_\mu u_c] [\bar c_d \gamma_\nu \gamma_5 c_d] + \{ \mu \leftrightarrow \nu \} \, ,
\\ \eta_{3\mu\nu} &=& [\epsilon^{abc} (u^T_a C \gamma_5 d_b) \gamma_\mu u_c] [\bar c_d \gamma_\nu c_d] + \{ \mu \leftrightarrow \nu \} \, ,
\\ \eta_{4\mu\nu} &=& [\epsilon^{abc} (u^T_a C \gamma_5 d_b) \gamma_\mu \gamma_5 u_c] [\bar c_d \gamma_\nu \gamma_5 c_d] + \{ \mu \leftrightarrow \nu \} \, ,
\\ \eta_{5\mu\nu} &=& [\epsilon^{abc} (u^T_a C d_b) \sigma_{\mu\rho} \gamma_5 u_c] [\bar c_d \sigma_{\nu\rho} c_d] + \{ \mu \leftrightarrow \nu \} \, ,
\\ \eta_{6\mu\nu} &=& [\epsilon^{abc} (u^T_a C d_b) \sigma_{\mu\rho} u_c] [\bar c_d \sigma_{\nu\rho} \gamma_5 c_d] + \{ \mu \leftrightarrow \nu \} \, ,
\\ \eta_{7\mu\nu} &=& [\epsilon^{abc} (u^T_a C \gamma_5 d_b) \sigma_{\mu\rho} u_c] [\bar c_d \sigma_{\nu\rho} c_d] + \{ \mu \leftrightarrow \nu \} \, ,
\\ \eta_{8\mu\nu} &=& [\epsilon^{abc} (u^T_a C \gamma_5 d_b) \sigma_{\mu\rho} \gamma_5 u_c] [\bar c_d \sigma_{\nu\rho} \gamma_5 c_d] + \{ \mu \leftrightarrow \nu \} \, ,
\\ \eta_{9\mu\nu} &=& [\epsilon^{abc} (u^T_a C \gamma_\mu \gamma_5 d_b) \gamma_\nu u_c] [\bar c_d c_d] + \{ \mu \leftrightarrow \nu \} \, ,
\\ \eta_{10\mu\nu} &=& [\epsilon^{abc} (u^T_a C \gamma_\mu \gamma_5 d_b) \gamma_\nu \gamma_5 u_c] [\bar c_d \gamma_5 c_d] + \{ \mu \leftrightarrow \nu \} \, ,
\\ \eta_{11\mu\nu} &=& [\epsilon^{abc} (u^T_a C \gamma_\mu \gamma_5 d_b) u_c] [\bar c_d \gamma_\nu c_d] + \{ \mu \leftrightarrow \nu \} \, ,
\\ \eta_{12\mu\nu} &=& [\epsilon^{abc} (u^T_a C \gamma_\mu \gamma_5 d_b) \gamma_5 u_c] [\bar c_d \gamma_\nu \gamma_5 c_d] + \{ \mu \leftrightarrow \nu \} \, ,
\\ \eta_{13\mu\nu} &=& [\epsilon^{abc} (u^T_a C \gamma_\mu \gamma_5 d_b) \sigma_{\nu\rho} u_c] [\bar c_d \gamma_\rho c_d] + \{ \mu \leftrightarrow \nu \} \, ,
\\ \eta_{14\mu\nu} &=& [\epsilon^{abc} (u^T_a C \gamma_\mu \gamma_5 d_b) \sigma_{\nu\rho} \gamma_5 u_c] [\bar c_d \gamma_\rho \gamma_5 c_d] + \{ \mu \leftrightarrow \nu \} \, ,
\\ \eta_{15\mu\nu} &=& [\epsilon^{abc} (u^T_a C \gamma_\mu \gamma_5 d_b) \gamma_\rho u_c] [\bar c_d \sigma_{\nu\rho} c_d] + \{ \mu \leftrightarrow \nu \} \, ,
\\ \eta_{16\mu\nu} &=& [\epsilon^{abc} (u^T_a C \gamma_\mu \gamma_5 d_b) \gamma_\rho \gamma_5 u_c] [\bar c_d \sigma_{\nu\rho} \gamma_5 c_d] + \{ \mu \leftrightarrow \nu \} \, .
\\ \eta_{17\mu\nu} &=& [\epsilon^{abc} (u^T_a C \gamma_\rho \gamma_5 d_b) \gamma_\mu u_c] [\bar c_d \sigma_{\nu\rho} c_d] + \{ \mu \leftrightarrow \nu \} \, ,
\\ \eta_{18\mu\nu} &=& [\epsilon^{abc} (u^T_a C \gamma_\rho \gamma_5 d_b) \gamma_\mu \gamma_5 u_c] [\bar c_d \sigma_{\nu\rho} \gamma_5 c_d] + \{ \mu \leftrightarrow \nu \} \, .
\end{eqnarray*}
To perform QCD sum rule analyses, we shall use
\begin{eqnarray}
\eta_{11\mu\nu} &=& [\epsilon^{abc} (u^T_a C \gamma_\mu \gamma_5 d_b) u_c] [\bar c_d \gamma_\nu c_d] + \{ \mu \leftrightarrow \nu \} \, ,
\label{def:eta11munu}
\end{eqnarray}
which well couples to the $[N^* J/\psi]$ channel.

We note that the following currents actually have spin $J=3/2$:
\begin{eqnarray*}
\eta_{1\mu\nu} - \eta_{3\mu\nu} &=& [\epsilon^{abc} (u^T_a C d_b) \gamma_\mu \gamma_5 u_c] [\bar c_d \gamma_\nu c_d]
\\ \nonumber && - [\epsilon^{abc} (u^T_a C \gamma_5 d_b) \gamma_\mu u_c] [\bar c_d \gamma_\nu c_d] + \{ \mu \leftrightarrow \nu \} \, ,
\\ \eta_{10\mu\nu} &=& [\epsilon^{abc} (u^T_a C \gamma_\mu \gamma_5 d_b) \gamma_\nu \gamma_5 u_c] [\bar c_d \gamma_5 c_d] + \{ \mu \leftrightarrow \nu \} \, ,
\\ \eta_{13\mu\nu} &=& [\epsilon^{abc} (u^T_a C \gamma_\mu \gamma_5 d_b) \sigma_{\nu\rho} u_c] [\bar c_d \gamma_\rho c_d] + \{ \mu \leftrightarrow \nu \} \, .
\end{eqnarray*}

\subsection{Currents of $[\bar c_d u_d][\epsilon^{abc} u_a d_b c_c]$}

In this subsection, we construct the currents of the color configuration $[\bar c_d u_d][\epsilon^{abc} u_a d_b c_c]$.
We find the following currents having $J^P=5/2^+$ and quark contents $uud c \bar c$:
\begin{eqnarray*}
\xi_{1\mu\nu} &=& [\epsilon^{abc} (u^T_a C d_b) \gamma_\mu \gamma_5 c_c] [\bar c_d \gamma_\nu u_d] + \{ \mu \leftrightarrow \nu \} \, ,
\\ \xi_{2\mu\nu} &=& [\epsilon^{abc} (u^T_a C d_b) \gamma_\mu c_c] [\bar c_d \gamma_\nu \gamma_5 u_d] + \{ \mu \leftrightarrow \nu \} \, ,
\\ \xi_{3\mu\nu} &=& [\epsilon^{abc} (u^T_a C \gamma_5 d_b) \gamma_\mu c_c] [\bar c_d \gamma_\nu u_d] + \{ \mu \leftrightarrow \nu \} \, ,
\\ \xi_{4\mu\nu} &=& [\epsilon^{abc} (u^T_a C \gamma_5 d_b) \gamma_\mu \gamma_5 c_c] [\bar c_d \gamma_\nu \gamma_5 u_d] + \{ \mu \leftrightarrow \nu \} \, ,
\\ \xi_{5\mu\nu} &=& [\epsilon^{abc} (u^T_a C d_b) \sigma_{\mu\rho} \gamma_5 c_c] [\bar c_d \sigma_{\nu\rho} u_d] + \{ \mu \leftrightarrow \nu \} \, ,
\\ \xi_{6\mu\nu} &=& [\epsilon^{abc} (u^T_a C d_b) \sigma_{\mu\rho} c_c] [\bar c_d \sigma_{\nu\rho} \gamma_5 u_d] + \{ \mu \leftrightarrow \nu \} \, ,
\\ \xi_{7\mu\nu} &=& [\epsilon^{abc} (u^T_a C \gamma_5 d_b) \sigma_{\mu\rho} c_c] [\bar c_d \sigma_{\nu\rho} u_d] + \{ \mu \leftrightarrow \nu \} \, ,
\\ \xi_{8\mu\nu} &=& [\epsilon^{abc} (u^T_a C \gamma_5 d_b) \sigma_{\mu\rho} \gamma_5 c_c] [\bar c_d \sigma_{\nu\rho} \gamma_5 u_d] + \{ \mu \leftrightarrow \nu \} \, ,
\\ \xi_{9\mu\nu} &=& [\epsilon^{abc} (u^T_a C \gamma_\mu d_b) \gamma_\nu \gamma_5 c_c] [\bar c_d u_d] + \{ \mu \leftrightarrow \nu \} \, ,
\\ \xi_{10\mu\nu} &=& [\epsilon^{abc} (u^T_a C \gamma_\mu d_b) \gamma_\nu c_c] [\bar c_d \gamma_5 u_d] + \{ \mu \leftrightarrow \nu \} \, ,
\\ \xi_{11\mu\nu} &=& [\epsilon^{abc} (u^T_a C \gamma_\mu \gamma_5 d_b) \gamma_\nu c_c] [\bar c_d u_d] + \{ \mu \leftrightarrow \nu \} \, ,
\\ \xi_{12\mu\nu} &=& [\epsilon^{abc} (u^T_a C \gamma_\mu \gamma_5 d_b) \gamma_\nu \gamma_5 c_c] [\bar c_d \gamma_5 u_d] + \{ \mu \leftrightarrow \nu \} \, ,
\\ \xi_{13\mu\nu} &=& [\epsilon^{abc} (u^T_a C \gamma_\mu d_b) \gamma_5 c_c] [\bar c_d \gamma_\nu u_d] + \{ \mu \leftrightarrow \nu \} \, ,
\\ \xi_{14\mu\nu} &=& [\epsilon^{abc} (u^T_a C \gamma_\mu d_b) c_c] [\bar c_d \gamma_\nu \gamma_5 u_d] + \{ \mu \leftrightarrow \nu \} \, ,
\\ \xi_{15\mu\nu} &=& [\epsilon^{abc} (u^T_a C \gamma_\mu \gamma_5 d_b) c_c] [\bar c_d \gamma_\nu u_d] + \{ \mu \leftrightarrow \nu \} \, ,
\\ \xi_{16\mu\nu} &=& [\epsilon^{abc} (u^T_a C \gamma_\mu \gamma_5 d_b) \gamma_5 c_c] [\bar c_d \gamma_\nu \gamma_5 u_d] + \{ \mu \leftrightarrow \nu \} \, ,
\\ \xi_{17\mu\nu} &=& [\epsilon^{abc} (u^T_a C \gamma_\mu d_b) \sigma_{\nu\rho} \gamma_5 c_c] [\bar c_d \gamma_\rho u_d] + \{ \mu \leftrightarrow \nu \} \, ,
\\ \xi_{18\mu\nu} &=& [\epsilon^{abc} (u^T_a C \gamma_\mu d_b) \sigma_{\nu\rho} c_c] [\bar c_d \gamma_\rho \gamma_5 u_d] + \{ \mu \leftrightarrow \nu \} \, ,
\\ \xi_{19\mu\nu} &=& [\epsilon^{abc} (u^T_a C \gamma_\mu \gamma_5 d_b) \sigma_{\nu\rho} c_c] [\bar c_d \gamma_\rho u_d] + \{ \mu \leftrightarrow \nu \} \, ,
\\ \xi_{20\mu\nu} &=& [\epsilon^{abc} (u^T_a C \gamma_\mu \gamma_5 d_b) \sigma_{\nu\rho} \gamma_5 c_c] [\bar c_d \gamma_\rho \gamma_5 u_d] + \{ \mu \leftrightarrow \nu \} \, ,
\\ \xi_{21\mu\nu} &=& [\epsilon^{abc} (u^T_a C \gamma_\rho d_b) \sigma_{\mu\rho} \gamma_5 c_c] [\bar c_d \gamma_\nu u_d] + \{ \mu \leftrightarrow \nu \} \, ,
\\ \xi_{22\mu\nu} &=& [\epsilon^{abc} (u^T_a C \gamma_\rho d_b) \sigma_{\mu\rho} c_c] [\bar c_d \gamma_\nu \gamma_5 u_d] + \{ \mu \leftrightarrow \nu \} \, ,
\\ \xi_{23\mu\nu} &=& [\epsilon^{abc} (u^T_a C \gamma_\rho \gamma_5 d_b) \sigma_{\mu\rho} c_c] [\bar c_d \gamma_\nu u_d] + \{ \mu \leftrightarrow \nu \} \, ,
\\ \xi_{24\mu\nu} &=& [\epsilon^{abc} (u^T_a C \gamma_\rho \gamma_5 d_b) \sigma_{\mu\rho} \gamma_5 c_c] [\bar c_d \gamma_\nu \gamma_5 u_d] + \{ \mu \leftrightarrow \nu \} \, ,
\\ \xi_{25\mu\nu} &=& [\epsilon^{abc} (u^T_a C \gamma_\mu d_b) \gamma_\rho \gamma_5 c_c] [\bar c_d \sigma_{\nu\rho} u_d] + \{ \mu \leftrightarrow \nu \} \, ,
\\ \xi_{26\mu\nu} &=& [\epsilon^{abc} (u^T_a C \gamma_\mu d_b) \gamma_\rho c_c] [\bar c_d \sigma_{\nu\rho} \gamma_5 u_d] + \{ \mu \leftrightarrow \nu \} \, ,
\\ \xi_{27\mu\nu} &=& [\epsilon^{abc} (u^T_a C \gamma_\mu \gamma_5 d_b) \gamma_\rho c_c] [\bar c_d \sigma_{\nu\rho} u_d] + \{ \mu \leftrightarrow \nu \} \, ,
\\ \xi_{28\mu\nu} &=& [\epsilon^{abc} (u^T_a C \gamma_\mu \gamma_5 d_b) \gamma_\rho \gamma_5 c_c] [\bar c_d \sigma_{\nu\rho} \gamma_5 u_d] + \{ \mu \leftrightarrow \nu \} \, ,
\\ \xi_{29\mu\nu} &=& [\epsilon^{abc} (u^T_a C \gamma_\rho d_b) \gamma_\mu \gamma_5 c_c] [\bar c_d \sigma_{\nu\rho} u_d] + \{ \mu \leftrightarrow \nu \} \, ,
\\ \xi_{30\mu\nu} &=& [\epsilon^{abc} (u^T_a C \gamma_\rho d_b) \gamma_\mu c_c] [\bar c_d \sigma_{\nu\rho} \gamma_5 u_d] + \{ \mu \leftrightarrow \nu \} \, ,
\\ \xi_{31\mu\nu} &=& [\epsilon^{abc} (u^T_a C \gamma_\rho \gamma_5 d_b) \gamma_\mu c_c] [\bar c_d \sigma_{\nu\rho} u_d] + \{ \mu \leftrightarrow \nu \} \, ,
\\ \xi_{32\mu\nu} &=& [\epsilon^{abc} (u^T_a C \gamma_\rho \gamma_5 d_b) \gamma_\mu \gamma_5 c_c] [\bar c_d \sigma_{\nu\rho} \gamma_5 u_d] + \{ \mu \leftrightarrow \nu \} \, ,
\\ \xi_{33\mu\nu} &=& [\epsilon^{abc} (u^T_a C \sigma_{\mu\rho} d_b) \sigma_{\nu\rho} \gamma_5 c_c] [\bar c_d u_d] + \{ \mu \leftrightarrow \nu \} \, ,
\\ \xi_{34\mu\nu} &=& [\epsilon^{abc} (u^T_a C \sigma_{\mu\rho} d_b) \sigma_{\nu\rho} c_c] [\bar c_d \gamma_5 u_d] + \{ \mu \leftrightarrow \nu \} \, ,
\\ \xi_{35\mu\nu} &=& [\epsilon^{abc} (u^T_a C \sigma_{\mu\rho} \gamma_5 d_b) \sigma_{\nu\rho} c_c] [\bar c_d u_d] + \{ \mu \leftrightarrow \nu \} \, ,
\\ \xi_{36\mu\nu} &=& [\epsilon^{abc} (u^T_a C \sigma_{\mu\rho} \gamma_5 d_b) \sigma_{\nu\rho} \gamma_5 c_c] [\bar c_d \gamma_5 u_d] + \{ \mu \leftrightarrow \nu \} \, ,
\\ \xi_{37\mu\nu} &=& [\epsilon^{abc} (u^T_a C \sigma_{\mu\rho} d_b) \gamma_\nu \gamma_5 c_c] [\bar c_d \gamma_\rho u_d] + \{ \mu \leftrightarrow \nu \} \, ,
\\ \xi_{38\mu\nu} &=& [\epsilon^{abc} (u^T_a C \sigma_{\mu\rho} d_b) \gamma_\nu c_c] [\bar c_d \gamma_\rho \gamma_5 u_d] + \{ \mu \leftrightarrow \nu \} \, ,
\\ \xi_{39\mu\nu} &=& [\epsilon^{abc} (u^T_a C \sigma_{\mu\rho} \gamma_5 d_b) \gamma_\nu c_c] [\bar c_d \gamma_\rho u_d] + \{ \mu \leftrightarrow \nu \} \, ,
\\ \xi_{40\mu\nu} &=& [\epsilon^{abc} (u^T_a C \sigma_{\mu\rho} \gamma_5 d_b) \gamma_\nu \gamma_5 c_c] [\bar c_d \gamma_\rho \gamma_5 u_d] + \{ \mu \leftrightarrow \nu \} \, ,
\\ \xi_{41\mu\nu} &=& [\epsilon^{abc} (u^T_a C \sigma_{\mu\rho} d_b) \gamma_\rho \gamma_5 c_c] [\bar c_d \gamma_\nu u_d] + \{ \mu \leftrightarrow \nu \} \, ,
\\ \xi_{42\mu\nu} &=& [\epsilon^{abc} (u^T_a C \sigma_{\mu\rho} d_b) \gamma_\rho c_c] [\bar c_d \gamma_\nu \gamma_5 u_d] + \{ \mu \leftrightarrow \nu \} \, ,
\\ \xi_{43\mu\nu} &=& [\epsilon^{abc} (u^T_a C \sigma_{\mu\rho} \gamma_5 d_b) \gamma_\rho c_c] [\bar c_d \gamma_\nu u_d] + \{ \mu \leftrightarrow \nu \} \, ,
\\ \xi_{44\mu\nu} &=& [\epsilon^{abc} (u^T_a C \sigma_{\mu\rho} \gamma_5 d_b) \gamma_\rho \gamma_5 c_c] [\bar c_d \gamma_\nu \gamma_5 u_d] + \{ \mu \leftrightarrow \nu \} \, ,
\\ \xi_{45\mu\nu} &=& [\epsilon^{abc} (u^T_a C \sigma_{\mu\rho} d_b) \gamma_5 c_c] [\bar c_d \sigma_{\nu\rho} u_d] + \{ \mu \leftrightarrow \nu \} \, ,
\\ \xi_{46\mu\nu} &=& [\epsilon^{abc} (u^T_a C \sigma_{\mu\rho} d_b) c_c] [\bar c_d \sigma_{\nu\rho} \gamma_5 u_d] + \{ \mu \leftrightarrow \nu \} \, ,
\\ \xi_{47\mu\nu} &=& [\epsilon^{abc} (u^T_a C \sigma_{\mu\rho} \gamma_5 d_b) c_c] [\bar c_d \sigma_{\nu\rho} u_d] + \{ \mu \leftrightarrow \nu \} \, ,
\\ \xi_{48\mu\nu} &=& [\epsilon^{abc} (u^T_a C \sigma_{\mu\rho} \gamma_5 d_b) \gamma_5 c_c] [\bar c_d \sigma_{\nu\rho} \gamma_5 u_d] + \{ \mu \leftrightarrow \nu \} \, ,
\\ \xi_{49\mu\nu} &=& [\epsilon^{abc} (u^T_a C \sigma_{\mu\rho} d_b) \sigma_{\nu\sigma} \gamma_5 c_c] [\bar c_d \sigma_{\rho\sigma} u_d] + \{ \mu \leftrightarrow \nu \} \, ,
\\ \xi_{50\mu\nu} &=& [\epsilon^{abc} (u^T_a C \sigma_{\mu\rho} d_b) \sigma_{\nu\sigma} c_c] [\bar c_d \sigma_{\rho\sigma} \gamma_5 u_d] + \{ \mu \leftrightarrow \nu \} \, ,
\\ \xi_{51\mu\nu} &=& [\epsilon^{abc} (u^T_a C \sigma_{\mu\rho} \gamma_5 d_b) \sigma_{\nu\sigma} c_c] [\bar c_d \sigma_{\rho\sigma} u_d] + \{ \mu \leftrightarrow \nu \} \, ,
\\ \xi_{52\mu\nu} &=& [\epsilon^{abc} (u^T_a C \sigma_{\mu\rho} \gamma_5 d_b) \sigma_{\nu\sigma} \gamma_5 c_c] [\bar c_d \sigma_{\rho\sigma} \gamma_5 u_d]
\\ \nonumber && ~~~~~~~~~~ + \{ \mu \leftrightarrow \nu \} \, ,
\\ \xi_{53\mu\nu} &=& [\epsilon^{abc} (u^T_a C \sigma_{\mu\rho} d_b) \sigma_{\rho\sigma} \gamma_5 c_c] [\bar c_d \sigma_{\nu\sigma} u_d] + \{ \mu \leftrightarrow \nu \} \, ,
\\ \xi_{54\mu\nu} &=& [\epsilon^{abc} (u^T_a C \sigma_{\mu\rho} d_b) \sigma_{\rho\sigma} c_c] [\bar c_d \sigma_{\nu\sigma} \gamma_5 u_d] + \{ \mu \leftrightarrow \nu \} \, ,
\\ \xi_{55\mu\nu} &=& [\epsilon^{abc} (u^T_a C \sigma_{\mu\rho} \gamma_5 d_b) \sigma_{\rho\sigma} c_c] [\bar c_d \sigma_{\nu\sigma} u_d] + \{ \mu \leftrightarrow \nu \} \, ,
\\ \xi_{56\mu\nu} &=& [\epsilon^{abc} (u^T_a C \sigma_{\mu\rho} \gamma_5 d_b) \sigma_{\rho\sigma} \gamma_5 c_c] [\bar c_d \sigma_{\nu\sigma} \gamma_5 u_d]
\\ \nonumber && ~~~~~~~~~~ + \{ \mu \leftrightarrow \nu \} \, ,
\\ \xi_{57\mu\nu} &=& [\epsilon^{abc} (u^T_a C \sigma_{\rho\sigma} d_b) \sigma_{\mu\rho} \gamma_5 c_c] [\bar c_d \sigma_{\nu\sigma} u_d] + \{ \mu \leftrightarrow \nu \} \, ,
\\ \xi_{58\mu\nu} &=& [\epsilon^{abc} (u^T_a C \sigma_{\rho\sigma} d_b) \sigma_{\mu\rho} c_c] [\bar c_d \sigma_{\nu\sigma} \gamma_5 u_d] + \{ \mu \leftrightarrow \nu \} \, ,
\\ \xi_{59\mu\nu} &=& [\epsilon^{abc} (u^T_a C \sigma_{\rho\sigma} \gamma_5 d_b) \sigma_{\mu\rho} c_c] [\bar c_d \sigma_{\nu\sigma} u_d] + \{ \mu \leftrightarrow \nu \} \, ,
\\ \xi_{60\mu\nu} &=& [\epsilon^{abc} (u^T_a C \sigma_{\rho\sigma} \gamma_5 d_b) \sigma_{\mu\rho} \gamma_5 c_c] [\bar c_d \sigma_{\nu\sigma} \gamma_5 u_d]
\\ \nonumber && ~~~~~~~~~~ + \{ \mu \leftrightarrow \nu \} \, .
\end{eqnarray*}
To perform QCD sum rule analyses, we shall use
\begin{eqnarray}
\xi_{13\mu\nu} &=& [\epsilon^{abc} (u^T_a C \gamma_\mu d_b) \gamma_5 c_c] [\bar c_d \gamma_\nu u_d] + \{ \mu \leftrightarrow \nu \} \, ,
\label{def:xi13munu}
\\ \xi_{15\mu\nu} &=& [\epsilon^{abc} (u^T_a C \gamma_\mu \gamma_5 d_b) c_c] [\bar c_d \gamma_\nu u_d] + \{ \mu \leftrightarrow \nu \} \, ,
\label{def:xi15munu}
\end{eqnarray}
which well couple to the $[\Sigma_c^* \bar D^*]$ and $[\Lambda_c^* \bar D^*]$ channels, respectively.

We note that the following currents actually have spin $J=3/2$:
\begin{eqnarray*}
\xi_{1\mu\nu} - \xi_{3\mu\nu} &=& [\epsilon^{abc} (u^T_a C d_b) \gamma_\mu \gamma_5 c_c] [\bar c_d \gamma_\nu u_d]
\\ \nonumber &&  - [\epsilon^{abc} (u^T_a C \gamma_5 d_b) \gamma_\mu c_c] [\bar c_d \gamma_\nu u_d] + \{ \mu \leftrightarrow \nu \} \, ,
\\ \xi_{10\mu\nu} &=& [\epsilon^{abc} (u^T_a C \gamma_\mu d_b) \gamma_\nu c_c] [\bar c_d \gamma_5 u_d] + \{ \mu \leftrightarrow \nu \} \, ,
\\ \xi_{12\mu\nu} &=& [\epsilon^{abc} (u^T_a C \gamma_\mu \gamma_5 d_b) \gamma_\nu \gamma_5 c_c] [\bar c_d \gamma_5 u_d] + \{ \mu \leftrightarrow \nu \} \, ,
\\ \xi_{17\mu\nu} &=& [\epsilon^{abc} (u^T_a C \gamma_\mu d_b) \sigma_{\nu\rho} \gamma_5 c_c] [\bar c_d \gamma_\rho u_d] + \{ \mu \leftrightarrow \nu \} \, ,
\\ \xi_{19\mu\nu} &=& [\epsilon^{abc} (u^T_a C \gamma_\mu \gamma_5 d_b) \sigma_{\nu\rho} c_c] [\bar c_d \gamma_\rho u_d] + \{ \mu \leftrightarrow \nu \} \, ,
\\ \xi_{21\mu\nu} &=& [\epsilon^{abc} (u^T_a C \gamma_\rho d_b) \sigma_{\mu\rho} \gamma_5 c_c] [\bar c_d \gamma_\nu u_d] + \{ \mu \leftrightarrow \nu \} \, ,
\\ \xi_{23\mu\nu} &=& [\epsilon^{abc} (u^T_a C \gamma_\rho \gamma_5 d_b) \sigma_{\mu\rho} c_c] [\bar c_d \gamma_\nu u_d] + \{ \mu \leftrightarrow \nu \} \, .
\end{eqnarray*}

\subsection{Currents of $[\bar c_d d_d][\epsilon^{abc} u_a u_b c_c]$}

In this subsection, we construct the currents of the color configuration $[\bar c_d d_d][\epsilon^{abc} u_a u_b c_c]$.
We find the following currents having $J^P=5/2^+$ and quark contents $uud c \bar c$:
\begin{eqnarray*}
\psi_{1\mu\nu} &=& [\epsilon^{abc} (u^T_a C \gamma_\mu u_b) \gamma_\nu \gamma_5 c_c] [\bar c_d d_d] + \{ \mu \leftrightarrow \nu \} \, ,
\\ \psi_{2\mu\nu} &=& [\epsilon^{abc} (u^T_a C \gamma_\mu u_b) \gamma_\nu c_c] [\bar c_d \gamma_5 d_d] + \{ \mu \leftrightarrow \nu \} \, ,
\\ \psi_{3\mu\nu} &=& [\epsilon^{abc} (u^T_a C \gamma_\mu u_b) \gamma_5 c_c] [\bar c_d \gamma_\nu d_d] + \{ \mu \leftrightarrow \nu \} \, ,
\\ \psi_{4\mu\nu} &=& [\epsilon^{abc} (u^T_a C \gamma_\mu u_b) c_c] [\bar c_d \gamma_\nu \gamma_5 d_d] + \{ \mu \leftrightarrow \nu \} \, ,
\\ \psi_{5\mu\nu} &=& [\epsilon^{abc} (u^T_a C \gamma_\mu u_b) \sigma_{\nu\rho} \gamma_5 c_c] [\bar c_d \gamma_\rho d_d] + \{ \mu \leftrightarrow \nu \} \, ,
\\ \psi_{6\mu\nu} &=& [\epsilon^{abc} (u^T_a C \gamma_\mu u_b) \sigma_{\nu\rho} c_c] [\bar c_d \gamma_\rho \gamma_5 d_d] + \{ \mu \leftrightarrow \nu \} \, ,
\\ \psi_{7\mu\nu} &=& [\epsilon^{abc} (u^T_a C \gamma_\rho u_b) \sigma_{\mu\rho} \gamma_5 c_c] [\bar c_d \gamma_\nu d_d] + \{ \mu \leftrightarrow \nu \} \, ,
\\ \psi_{8\mu\nu} &=& [\epsilon^{abc} (u^T_a C \gamma_\rho u_b) \sigma_{\mu\rho} c_c] [\bar c_d \gamma_\nu \gamma_5 d_d] + \{ \mu \leftrightarrow \nu \} \, ,
\\ \psi_{9\mu\nu} &=& [\epsilon^{abc} (u^T_a C \gamma_\mu u_b) \gamma_\rho \gamma_5 c_c] [\bar c_d \sigma_{\nu\rho} d_d] + \{ \mu \leftrightarrow \nu \} \, ,
\\ \psi_{10\mu\nu} &=& [\epsilon^{abc} (u^T_a C \gamma_\mu u_b) \gamma_\rho c_c] [\bar c_d \sigma_{\nu\rho} \gamma_5 d_d] + \{ \mu \leftrightarrow \nu \} \, ,
\\ \psi_{11\mu\nu} &=& [\epsilon^{abc} (u^T_a C \gamma_\rho u_b) \gamma_\mu \gamma_5 c_c] [\bar c_d \sigma_{\nu\rho} d_d] + \{ \mu \leftrightarrow \nu \} \, ,
\\ \psi_{12\mu\nu} &=& [\epsilon^{abc} (u^T_a C \gamma_\rho u_b) \gamma_\mu c_c] [\bar c_d \sigma_{\nu\rho} \gamma_5 d_d] + \{ \mu \leftrightarrow \nu \} \, ,
\\ \psi_{13\mu\nu} &=& [\epsilon^{abc} (u^T_a C \sigma_{\mu\rho} u_b) \sigma_{\nu\rho} \gamma_5 c_c] [\bar c_d d_d] + \{ \mu \leftrightarrow \nu \} \, ,
\\ \psi_{14\mu\nu} &=& [\epsilon^{abc} (u^T_a C \sigma_{\mu\rho} u_b) \sigma_{\nu\rho} c_c] [\bar c_d \gamma_5 d_d] + \{ \mu \leftrightarrow \nu \} \, ,
\\ \psi_{15\mu\nu} &=& [\epsilon^{abc} (u^T_a C \sigma_{\mu\rho} \gamma_5 u_b) \sigma_{\nu\rho} c_c] [\bar c_d d_d] + \{ \mu \leftrightarrow \nu \} \, ,
\\ \psi_{16\mu\nu} &=& [\epsilon^{abc} (u^T_a C \sigma_{\mu\rho} \gamma_5 u_b) \sigma_{\nu\rho} \gamma_5 c_c] [\bar c_d \gamma_5 d_d] + \{ \mu \leftrightarrow \nu \} \, ,
\\ \psi_{17\mu\nu} &=& [\epsilon^{abc} (u^T_a C \sigma_{\mu\rho} u_b) \gamma_\nu \gamma_5 c_c] [\bar c_d \gamma_\rho d_d] + \{ \mu \leftrightarrow \nu \} \, ,
\\ \psi_{18\mu\nu} &=& [\epsilon^{abc} (u^T_a C \sigma_{\mu\rho} u_b) \gamma_\nu c_c] [\bar c_d \gamma_\rho \gamma_5 d_d] + \{ \mu \leftrightarrow \nu \} \, ,
\\ \psi_{19\mu\nu} &=& [\epsilon^{abc} (u^T_a C \sigma_{\mu\rho} \gamma_5 u_b) \gamma_\nu c_c] [\bar c_d \gamma_\rho d_d] + \{ \mu \leftrightarrow \nu \} \, ,
\\ \psi_{20\mu\nu} &=& [\epsilon^{abc} (u^T_a C \sigma_{\mu\rho} \gamma_5 u_b) \gamma_\nu \gamma_5 c_c] [\bar c_d \gamma_\rho \gamma_5 d_d] + \{ \mu \leftrightarrow \nu \} \, ,
\\ \psi_{21\mu\nu} &=& [\epsilon^{abc} (u^T_a C \sigma_{\mu\rho} u_b) \gamma_\rho \gamma_5 c_c] [\bar c_d \gamma_\nu d_d] + \{ \mu \leftrightarrow \nu \} \, ,
\\ \psi_{22\mu\nu} &=& [\epsilon^{abc} (u^T_a C \sigma_{\mu\rho} u_b) \gamma_\rho c_c] [\bar c_d \gamma_\nu \gamma_5 d_d] + \{ \mu \leftrightarrow \nu \} \, ,
\\ \psi_{23\mu\nu} &=& [\epsilon^{abc} (u^T_a C \sigma_{\mu\rho} \gamma_5 u_b) \gamma_\rho c_c] [\bar c_d \gamma_\nu d_d] + \{ \mu \leftrightarrow \nu \} \, ,
\\ \psi_{24\mu\nu} &=& [\epsilon^{abc} (u^T_a C \sigma_{\mu\rho} \gamma_5 u_b) \gamma_\rho \gamma_5 c_c] [\bar c_d \gamma_\nu \gamma_5 d_d] + \{ \mu \leftrightarrow \nu \} \, ,
\\ \psi_{25\mu\nu} &=& [\epsilon^{abc} (u^T_a C \sigma_{\mu\rho} u_b) \gamma_5 c_c] [\bar c_d \sigma_{\nu\rho} d_d] + \{ \mu \leftrightarrow \nu \} \, ,
\\ \psi_{26\mu\nu} &=& [\epsilon^{abc} (u^T_a C \sigma_{\mu\rho} u_b) c_c] [\bar c_d \sigma_{\nu\rho} \gamma_5 d_d] + \{ \mu \leftrightarrow \nu \} \, ,
\\ \psi_{27\mu\nu} &=& [\epsilon^{abc} (u^T_a C \sigma_{\mu\rho} \gamma_5 u_b) c_c] [\bar c_d \sigma_{\nu\rho} d_d] + \{ \mu \leftrightarrow \nu \} \, ,
\\ \psi_{28\mu\nu} &=& [\epsilon^{abc} (u^T_a C \sigma_{\mu\rho} \gamma_5 u_b) \gamma_5 c_c] [\bar c_d \sigma_{\nu\rho} \gamma_5 d_d] + \{ \mu \leftrightarrow \nu \} \, ,
\\ \psi_{29\mu\nu} &=& [\epsilon^{abc} (u^T_a C \sigma_{\mu\rho} u_b) \sigma_{\nu\sigma} \gamma_5 c_c] [\bar c_d \sigma_{\rho\sigma} d_d] + \{ \mu \leftrightarrow \nu \} \, ,
\\ \psi_{30\mu\nu} &=& [\epsilon^{abc} (u^T_a C \sigma_{\mu\rho} u_b) \sigma_{\nu\sigma} c_c] [\bar c_d \sigma_{\rho\sigma} \gamma_5 d_d] + \{ \mu \leftrightarrow \nu \} \, ,
\\ \psi_{31\mu\nu} &=& [\epsilon^{abc} (u^T_a C \sigma_{\mu\rho} \gamma_5 u_b) \sigma_{\nu\sigma} c_c] [\bar c_d \sigma_{\rho\sigma} d_d] + \{ \mu \leftrightarrow \nu \} \, ,
\\ \psi_{32\mu\nu} &=& [\epsilon^{abc} (u^T_a C \sigma_{\mu\rho} \gamma_5 u_b) \sigma_{\nu\sigma} \gamma_5 c_c] [\bar c_d \sigma_{\rho\sigma} \gamma_5 d_d]
\\ \nonumber && ~~~~~~~~~~ + \{ \mu \leftrightarrow \nu \} \, ,
\\ \psi_{33\mu\nu} &=& [\epsilon^{abc} (u^T_a C \sigma_{\mu\rho} u_b) \sigma_{\rho\sigma} \gamma_5 c_c] [\bar c_d \sigma_{\nu\sigma} d_d] + \{ \mu \leftrightarrow \nu \} \, ,
\\ \psi_{34\mu\nu} &=& [\epsilon^{abc} (u^T_a C \sigma_{\mu\rho} u_b) \sigma_{\rho\sigma} c_c] [\bar c_d \sigma_{\nu\sigma} \gamma_5 d_d] + \{ \mu \leftrightarrow \nu \} \, ,
\\ \psi_{35\mu\nu} &=& [\epsilon^{abc} (u^T_a C \sigma_{\mu\rho} \gamma_5 u_b) \sigma_{\rho\sigma} c_c] [\bar c_d \sigma_{\nu\sigma} d_d] + \{ \mu \leftrightarrow \nu \} \, ,
\\ \psi_{36\mu\nu} &=& [\epsilon^{abc} (u^T_a C \sigma_{\mu\rho} \gamma_5 u_b) \sigma_{\rho\sigma} \gamma_5 c_c] [\bar c_d \sigma_{\nu\sigma} \gamma_5 d_d]
\\ \nonumber && ~~~~~~~~~~ + \{ \mu \leftrightarrow \nu \} \, ,
\\ \psi_{37\mu\nu} &=& [\epsilon^{abc} (u^T_a C \sigma_{\rho\sigma} u_b) \sigma_{\mu\rho} \gamma_5 c_c] [\bar c_d \sigma_{\nu\sigma} d_d] + \{ \mu \leftrightarrow \nu \} \, ,
\\ \psi_{38\mu\nu} &=& [\epsilon^{abc} (u^T_a C \sigma_{\rho\sigma} u_b) \sigma_{\mu\rho} c_c] [\bar c_d \sigma_{\nu\sigma} \gamma_5 d_d] + \{ \mu \leftrightarrow \nu \} \, ,
\\ \psi_{39\mu\nu} &=& [\epsilon^{abc} (u^T_a C \sigma_{\rho\sigma} \gamma_5 u_b) \sigma_{\mu\rho} c_c] [\bar c_d \sigma_{\nu\sigma} d_d] + \{ \mu \leftrightarrow \nu \} \, ,
\\ \psi_{40\mu\nu} &=& [\epsilon^{abc} (u^T_a C \sigma_{\rho\sigma} \gamma_5 u_b) \sigma_{\mu\rho} \gamma_5 c_c] [\bar c_d \sigma_{\nu\sigma} \gamma_5 d_d]
\\ \nonumber && ~~~~~~~~~~ + \{ \mu \leftrightarrow \nu \} \, .
\end{eqnarray*}
To perform QCD sum rule analyses, we shall use
\begin{eqnarray}
\psi_{3\mu\nu} &=& [\epsilon^{abc} (u^T_a C \gamma_\mu u_b) \gamma_5 c_c] [\bar c_d \gamma_\nu d_d] + \{ \mu \leftrightarrow \nu \} \, ,
\label{def:psi3munu}
\end{eqnarray}
which well couples to the $[\Sigma_c^* \bar D^*]$ channel. While, we also need to use
\begin{eqnarray}
\psi_{4\mu\nu} &=& [\epsilon^{abc} (u^T_a C \gamma_\mu u_b) c_c] [\bar c_d \gamma_\nu \gamma_5 d_d] + \{ \mu \leftrightarrow \nu \} \, .
\label{def:psi4munu}
\end{eqnarray}
which well couples to the $P$-wave $[\Sigma_c^* \bar D]$ channel.

We note that the following currents actually have spin $J=3/2$:
\begin{eqnarray*}
\psi_{2\mu\nu} &=& [\epsilon^{abc} (u^T_a C \gamma_\mu u_b) \gamma_\nu c_c] [\bar c_d \gamma_5 d_d] + \{ \mu \leftrightarrow \nu \} \, ,
\\ \psi_{5\mu\nu} &=& [\epsilon^{abc} (u^T_a C \gamma_\mu u_b) \sigma_{\nu\rho} \gamma_5 c_c] [\bar c_d \gamma_\rho d_d] + \{ \mu \leftrightarrow \nu \} \, ,
\\ \psi_{7\mu\nu} &=& [\epsilon^{abc} (u^T_a C \gamma_\rho u_b) \sigma_{\mu\rho} \gamma_5 c_c] [\bar c_d \gamma_\nu d_d] + \{ \mu \leftrightarrow \nu \} \, .
\end{eqnarray*}

\section{One Example}
\label{app:example}

In this appendix, we show one example and express the current $\eta_1$ as a combination of $\xi_i$ and $\psi_i$:
\begin{eqnarray*}
\eta_1 &=& [\epsilon_{abc} (u^T_a C d_b) \gamma_5 u_c] [\bar c_d c_d] \, ,
\\ \nonumber &=& [\epsilon_{abc} (u^T_a C d_b) \gamma_5 u_d] [\bar c_d c_c] + [\epsilon_{abc} (u^T_a C d_d) \gamma_5 u_c] [\bar c_d c_b]
\\ \nonumber && ~~~~~~~~~~ + [\epsilon_{abc} (u^T_d C d_b) \gamma_5 u_c] [\bar c_d c_a]
\\ \nonumber &=& + { 5\over4 } \times [\epsilon_{abc} (u^T_a C d_b) \gamma_5 u_d] [\bar c_d c_c]
\\ \nonumber && + { 1\over4 } \times [\epsilon_{abc} (u^T_a C \gamma_\mu d_b) \gamma_\mu \gamma_5 u_d] [\bar c_d c_c]
\\ \nonumber && - { 1\over8 } \times [\epsilon_{abc} (u^T_a C \sigma_{\mu\nu} d_b) \sigma_{\mu\nu} \gamma_5 u_d] [\bar c_d c_c]
\\ \nonumber && + { 1\over4 } \times [\epsilon_{abc} (u^T_a C \gamma_\mu \gamma_5 d_b) \gamma_\mu u_d] [\bar c_d c_c]
\\ \nonumber && + { 1\over4 } \times [\epsilon_{abc} (u^T_a C \gamma_5 d_b) u_d] [\bar c_d c_c]
\\ \nonumber && - { 1\over4 } \times [\epsilon_{abc} (u^T_a C \gamma_\mu u_b) \gamma_\mu \gamma_5 d_d] [\bar c_d c_c]
\\ \nonumber && + { 1\over8 } \times [\epsilon_{abc} (u^T_a C \sigma_{\mu\nu} u_b) \sigma_{\mu\nu} \gamma_5 d_d] [\bar c_d c_c]
\\ \nonumber &=&
- {5\over16} \xi_1
- {5\over16} \xi_2
- {1\over16} \xi_3
- {1\over16} \xi_4
- {5\over16} \xi_5
\\ \nonumber &&
+ {5\over16} \xi_6
+ {1\over16} \xi_7
- {1\over16} \xi_8
- {5\over32} \xi_{10}
\\ \nonumber &&
- {1\over32} \xi_{11}
- {1\over16} \xi_{13}
- {1\over16} \xi_{14}
- {1\over16} \xi_{15}
\\ \nonumber &&
- {1\over16} \xi_{16}
- {1\over16} \xi_{17}
+ {1\over16} \xi_{18}
- {1\over16} \xi_{19}
+ {1\over16} \xi_{20}
\\ \nonumber &&
+ {i\over16} \xi_{21}
- {i\over16} \xi_{22}
+ {i\over16} \xi_{23}
- {i\over16} \xi_{24}
- {i\over16} \xi_{25}
\\ \nonumber &&
- {i\over16} \xi_{26}
- {i\over16} \xi_{27}
- {i\over16} \xi_{28}
+ {1\over32} \xi_{29}
+ {1\over32} \xi_{30}
\\ \nonumber &&
- {i\over16} \xi_{33}
+ {i\over16} \xi_{34}
+ {i\over16} \xi_{35}
- {i\over16} \xi_{36}
+ {1\over32} \xi_{37}
\\ \nonumber &&
+ {1\over32} \xi_{38}
- {i\over16} \xi_{43}
\\ \nonumber &&
\\ \nonumber &&
+ {1\over16} \psi_1
+ {1\over16} \psi_2
+ {1\over16} \psi_3
- {1\over16} \psi_4
- {i\over16} \psi_5
\\ \nonumber &&
+ {i\over16} \psi_6
+ {i\over16} \psi_7
+ {i\over16} \psi_8
- {1\over32} \psi_9
- {1\over32} \psi_{10}
\\ \nonumber &&
+ {i\over16} \psi_{13}
- {i\over16} \psi_{14}
- {i\over16} \psi_{15}
\\ \nonumber &&
+ {i\over16} \psi_{16}
- {1\over32} \psi_{18}
- {1\over32} \psi_{19}
+ {i\over16} \psi_{23}
\, .
\end{eqnarray*}

\end{document}